\documentclass[12pt,a4paper]{amsart}
\usepackage[dvips,colorlinks=true,linkcolor=blue]{hyperref}
\usepackage{subfigure,caption2}
\usepackage{amssymb}

\usepackage{graphicx}
\usepackage[rflt]{floatflt}
\numberwithin{equation}{section}
\newcommand{\smallpagebreak}{{\par\vspace{2 mm}\noindent}}

\newcommand{\demo}{\par\noindent{\it Proof.\/} \ }
\newcommand{\QED}{\ \qed\smallpagebreak}
\newcommand{\dsize}{\textstyle}
\newcommand{\D}{\displaystyle}

\newcommand{\R}{{\mathbb R}}

\newcommand{\Z}{{\mathbb Z}}
\newcommand{\N}{{\mathbb N}}
\newcommand{\C}{{\mathbb C}}

\newcommand{\re}{{\rm Re}\,}
\newcommand{\im}{{\rm Im}\,}
\newcommand{\ind}{\text{ind}\,}
\newcommand{\res}{{\rm res}\, }

\newcommand{\dist}{{\rm dist}\,}
\newcommand{\mes}{{\rm mes}\,}

\theoremstyle{plain}
\newtheorem{Th}{Theorem}[section]
\newtheorem{Le}{Lemma}[section]
\newtheorem{Pro}{Proposition}[section]
\newtheorem{Cor}{Corollary}[section]
\theoremstyle{definition}
\newtheorem{Rem}{Remark}[section]

\title{Strong resonant tunneling, level repulsion and spectral type
  for one-dimensional adiabatic quasi-periodic {Schr{\"o}dinger} operators}
\author{Alexander Fedotov} \author{Fr{\'e}d{\'e}ric Klopp}
\address[Alexander Fedotov]{Department of Mathematical Physics, St
  Petersburg State University, 1, Ulia\-novskaja, 198904 St
  Petersburg-Petrodvorets, Russia}
\email{\href{mailto:fedotov@svs.ru}{fedotov@svs.ru}}
\address[Fr{\'e}d{\'e}ric Klopp]{LAGA, Institut Galil{\'e}e, U.M.R. 7539 C.N.R.S,
  Universit{\'e} Paris-Nord, Avenue J.-B.  Cl{\'e}ment, F-93430 Villetaneuse,
  France}
\email{\href{mailto:klopp@math.univ-paris13.fr}{klopp@math.univ-paris13.fr}}
\keywords{quasi periodic Schr{\"o}dinger equation, pure point spectrum,
  absolutely continuous spectrum, resonant tunneling, level repulsion}
\subjclass[2000]{34E05, 34E20, 34L05, 34L40}
\thanks{F.K.'s research was partially supported by the program RIAC
  160 at Universit{\'e} Paris 13 and by the FNS 2000 ``Programme Jeunes
  Chercheurs''. A.F. thanks the LAGA, Universit{\'e} Paris 13 for its kind
  hospitality. Both authors thank the Mittag-Leffler Institute, Mc
  Gill University and the Centre de Recherches Math{\'e}matiques where
  part of this work was done.}
\begin{document}
\begin{abstract}
  In this paper, we consider one dimensional adiabatic quasi-periodic
  Schr{\"o}dinger operators in the regime of strong resonant tunneling.
  We show the emergence of a level repulsion phenomenon which is seen
  to be very naturally related to the local spectral type of the
  operator: the more singular the spectrum, the weaker the repulsion.

  \vskip.5cm
  \par\noindent   \textsc{R{\'e}sum{\'e}.}
  Dans cet article, nous {\'e}tudions une famille d'op{\'e}rateurs
  quasi-p{\'e}riodiques adiabatiques dans un cas d'effet tunnel r{\'e}sonant
  fort.  Nous voyons l'apparition d'un ph{\'e}nom{\`e}ne de r{\'e}pulsion de
  niveaux fort qui est reli{\'e} au type spectral local de l'op{\'e}rateur :
  plus le spectre est singulier, plus la r{\'e}pulsion est faible.
\end{abstract}
\setcounter{section}{-1}
\maketitle
\section{Introduction}
\label{sec:intro}
In~\cite{Fe-Kl:04a}, we studied the spectrum of the family of
one-dimensional quasi-periodic Schr{\"o}dinger operators acting on
$L^2(\R)$  and  defined by
\begin{equation}
  \label{family}
  H_{z,\varepsilon}\psi=-\frac{d^2}{dx^2}\psi(x)+
  (V(x-z)+\alpha\cos(\varepsilon x))\psi(x),
\end{equation}
where
\begin{description}
\item[(H1)] $V:\ \R\to\R$ is a non constant, locally square
  integrable, $1$-periodic function;
\item[(H2)] $\varepsilon$ is a small positive number chosen such that
  $2\pi/\varepsilon$ be irrational;
\item[(H3)] $z$ is a real parameter indexing the operators;
\item[(H4)] $\alpha$ is a strictly positive parameter that we will
  keep fixed in most of the paper.
\end{description}
To describe the energy region where we worked, consider the spectrum
of the periodic Schr{\"o}dinger operator (on $L^2(\R)$)
\begin{equation}
  \label{Ho}
  H_0=-\frac{d^2}{dx^2}+V\,(x)
\end{equation}
%
\begin{floatingfigure}[r]{3.5cm}
  \centering
  \includegraphics[bbllx=71,bblly=662,bburx=176,bbury=721,width=3.5cm]{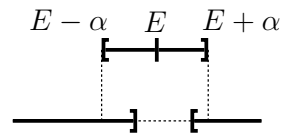}
  \caption{Bands in interaction}\label{interactingfigure}
\end{floatingfigure}
%
\noindent We assumed that two of its spectral bands are interacting
through the perturbation $\alpha\cos$ i.e., that the relative position
of the {\it spectral window} $\mathcal{F}(E):=[E-\alpha,E+\alpha]$ and
the spectrum of the unperturbed operator $H_0$ is that shown in
Fig.~\ref{interactingfigure}. In such an energy region, the spectrum
is localized near two sequences of quantized energy values, say
$(E_0^{(l)})_l$ and $(E_\pi^{(l')})_l'$ (see Theorem~\ref{thr:2});
each of these sequences is ``generated'' by one of the ends of the
neighboring spectral bands of $H_0$. In~\cite{Fe-Kl:04a}, we
restricted our study to neighborhoods of such quantized energy values
that were not resonant i.e. to neighborhoods of the points $E_\mu ^{(l)}$
that were not ``too'' close to the points $(E_\nu^{(l')})_l'$ for
$\{\mu ,\nu\}=\{0,\pi\}$. Already in this case, the distance between the
two sequences influences the nature and location of the spectrum: we
saw a weak level repulsion arise ``due to weak resonant tunneling''.\\
Similarly to what happens in the standard ``double well'' case
(see~\cite{MR85d:35085,He-Sj:84}), the resonant tunneling begins to
play an important role when the two energies, each generated by one of
the quantization conditions, are sufficiently close to each other.
\vskip.1cm\noindent In the present paper, we deal with the resonant
case, i.e. the case when two of the ``interacting energies'' are
``very'' close to each other or even coincide. We find a strong
relationship between the level repulsion and the nature of the
spectrum.  Recall that the latter is determined by the speed of decay
of the solutions to $(H_{z,\varepsilon}-E)\psi=0$
(see~\cite{Gi-Pe:87}).  Expressed in this way, it is very natural that
the two characteristics are related: the slower the decay of the
generalized eigenfunctions, the larger the overlap between generalized
eigenfunction corresponding to close energy levels, hence, the larger
the tunneling between these levels and thus the repulsion between
them.
\par Let us now briefly describe the various situations we encounter.
Therefore, we briefly recall the settings and results
of~\cite{Fe-Kl:04a}. Let $J$ be an interval of energies such that, for
all $E\in J$, the spectral window $\mathcal{F}(E)$ covers the edges of
two neighboring spectral bands of $H_0$ and the gap located between
them (see Fig.~\ref{interactingfigure} and assumption (TIBM)). Under
this assumption, consider the {\it real} and {\it complex iso-energy
  curves} associated to~(\ref{family}).  Denoted respectively by
$\Gamma_\R$ and $\Gamma$, they are defined by
\begin{gather}
  \label{isoenr}
  \Gamma_\R:=\{(\kappa,\zeta)\in\R^2,\
  \mathbf{E}(\kappa)+\alpha\cdot \cos(\zeta)=E\},\\
  \label{isoen}
  \Gamma:=\{(\kappa,\zeta)\in\C^2,\
  \mathbf{E}(\kappa)+\alpha\cdot \cos(\zeta)=E\}.
\end{gather}
where $\mathbf{E}(\kappa)$ be the dispersion relation associated to
$H_0$ (see section~\ref{sec:le-quasi-moment}). These curves are
roughly depicted in Fig.~\ref{TIBMfig:actions}.  They are periodic
both in $\zeta$ and $\kappa$ directions.

\noindent Consider one of the periodicity cells of $\Gamma_\R$. It
contains two tori. They are denoted by $\gamma_0$ and $\gamma_\pi$
%
\begin{floatingfigure}[r]{7.5cm}
  \centering
  \includegraphics[bbllx=71,bblly=566,bburx=247,bbury=721,width=7cm]{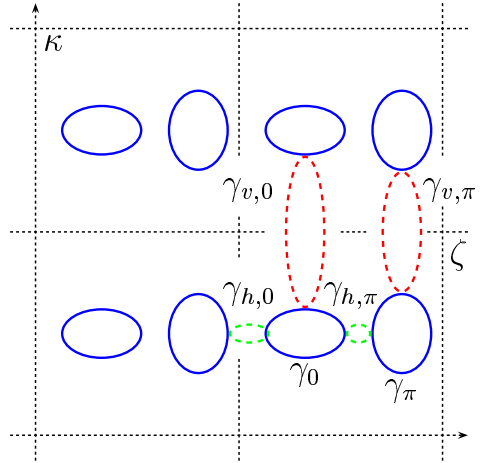}
  \caption{The adiabatic phase space}\label{TIBMfig:actions}
\end{floatingfigure}
%
\noindent and  shown in full lines. To each of them,  one 
associates a phase obtained by integrating $1/2$ times the fundamental
$1$-form on $\Gamma$ along the tori; we denote the phases by
$\Phi_0$ and $\Phi_\pi$ respectively (see
section~\ref{sec:iso-energy-curve}).\\  
Each of the dashed lines in Fig.~\ref{TIBMfig:actions} represents a
loop on $\Gamma$ that connects certain connected components of
$\Gamma_\R$; one can distinguish between the ``horizontal'' loops and
the ``vertical'' loops. There are two special horizontal loops denoted
by $\gamma_{h,0}$ and $\gamma_{h,\pi}$; the loop $\gamma_{h,0}$ (resp.
$\gamma_{h,\pi}$) connects $\gamma_0$ to $\gamma_\pi-(2\pi,0)$ (resp.
$\gamma_0$ to $\gamma_\pi$). In the same way, there are two special
vertical loops denoted by $\gamma_{v,0}$ and $\gamma_{v,\pi}$ ; the
loop $\gamma_{v,0}$ (resp.  $\gamma_{v,\pi}$) connects $\gamma_0$ to
$\gamma_0+(0,2\pi)$ (resp. $\gamma_\pi$ to $\gamma_\pi+(0,2\pi)$). To
each of these complex loops, one associates an action obtained by
integrating $-i/2$ times the fundamental $1$-form on $\Gamma$ along
the loop. For $a\in\{0,\pi\}$ and $b\in\{v,h\}$, we denote by
$S_{a,b}$ the action associated to $\gamma_{a,b}$. For $E$ real, all
these actions are real. One orients the loops so that they all be
positive.  Finally, we define tunneling coefficients as
\begin{equation*}
 t_{a,b}=e^{-S_{a,b}/\varepsilon},\quad a\in\{0,\pi\},\ b\in\{v,h\}.
\end{equation*}
Each of the curves $\gamma_0$ and $\gamma_\pi$ defines a sequence of
``quantized energies'' in $J$ (see Theorem~\ref{thr:2}). They satisfy
the ``quantization'' conditions
\begin{equation}\label{q-c} 
    \frac{1}{\varepsilon}\Phi_0(E_0^{(l)})=\frac\pi2+l\pi+o(1),\quad l\in\Z,\quad\quad
  \frac{1}{\varepsilon}\check\Phi_\pi(E_\pi^{(m)})=\frac\pi2+m\pi+o(1),\quad m\in\Z,     
\end{equation}
where $o(1)$ denote real analytic functions of $E$ that are small for
small $\varepsilon$. By Theorem~\ref{thr:2}, for $a\in\{0,\pi\}$, near
each $E_a^{(l)}$, there is one exponentially small interval
$I_a^{(l)}$ such that the spectrum of $H_{z,\varepsilon}$ in $J$ is
contained in the union of all these intervals. The precise description
of the spectrum in the interval $I_a^{(l)}$ depends on whether it
intersects another such interval or not. Note that an interval of type
$I_0^{(l)}$ can only intersect intervals of type $I_\pi^{(l)}$ and
vice versa (see section~\ref{sec:acti-integr-tunn-1}).
\par The paper~\cite{Fe-Kl:04a} was devoted to the study of the
spectrum in intervals $I_a^{(l)}$ that do not intersect any other
interval. In the present paper, we consider two intervals $I_0^{(l)}$
and $I_\pi^{(l')}$ that do intersect and describe the spectrum in the
union of these two intervals. It is useful to keep in mind that this
union is exponentially small (when $\varepsilon$ goes to $0$).
\par There are two main parameters controlling the spectral type:
\begin{equation*}
   \tau=2\sqrt{\frac{t_{v,0}\,t_{v,\pi}}{t_{h,0}\,t_{h,\pi}}}
   \quad\quad\quad\text{and}\quad\quad\quad
  \rho=2\sqrt{\frac{\max(t_{v,0},t_{v,\pi})^2}{t_{h,0}\,t_{h,\pi}}}
\end{equation*}
Here, as we are working inside an exponentially small interval, the
point inside this interval at which we compute the tunneling
coefficients does not really matter: over this interval, the relative
variation of any of the actions is exponentially small.\\
Noting that $\rho\geq\tau$, we distinguish three regimes:
\begin{itemize}
\item $\tau\gg1$,
\item $\rho\gg1$ and $\tau\ll1$,
\item $\rho\ll1$.
\end{itemize}
Here, the symbol $\gg1$ (resp. $\ll1$) mean that the quantity is
exponentially large (resp. small) in $\frac1{\varepsilon}$ as
$\varepsilon$ goes to $0$, the exponential rate being arbitrary, see
section~\ref{sec:plus-proche-de}.\\
In each of the three cases, we consider two energies, say $E_0$ and
$E_\pi$, satisfying respectively the first and the second relation
in~\eqref{q-c}, and describe the evolution of the spectrum as $E_0$
and $E_\pi$ become closer to each other. As explained in Remark 1.2
in~\cite{Fe-Kl:04a}, this can be achieved by reducing $\varepsilon$
somewhat. As noted above, when moving these two energies, we can
consider that the other parameters in the problem, mainly the
tunneling coefficients, hence, the coupling constants $\tau$ and
$\rho$ stay constant; in particular, we stay in one of the three cases
described above when we move $E_0$ and $E_\pi$ closer together.
\par Assume we are in the case $\tau\gg1$. Then, when $E_0$ and
$E_\pi$ are still ``far'' away from each other, one sees two intervals
containing spectrum (sub-intervals of the corresponding intervals
$I_0$ and $I_\pi$), one located near each energy; they contain the
same amount of spectrum i.e. the measure with respect to the density
of states of each interval is $\varepsilon/2\pi$; and the Lyapunov
exponent is positive on both intervals (see
Fig.~\ref{fig:tau_grand1}). When $E_0$ and $E_\pi$ approach each
other, at some moment these two intervals merge into one, so only a
single interval is seen (see Fig.~\ref{fig:tau_grand2}); its density
of states measure is $\varepsilon/\pi$ and the Lyapunov exponent is
still positive on this interval. There is no gap separating the
intervals of spectrum generated by the two quantization
conditions~\footnote{In the present paper, we do not discuss the gaps
  in the spectrum that are exponentially small with respect to the
  lengths of the two intervals $I_0$ and $I_\pi$).}. This can be
interpreted as a consequence of the positivity of the Lyapunov
exponent : the states are well localized so the overlapping is weak
and there no level repulsion. Nevertheless, the resonance has one
effect: when the two intervals merge, it gives rise to a sharp drop of
the Lyapunov exponent (which still stays positive) in the middle of
the interval containing spectrum. Over a distance exponentially small
in $\varepsilon$, the Lyapunov exponent drops by an amount of order
one.
\par In the other extreme, in the case $\rho\ll1$, the ``starting''
geometry of the spectrum is the same as in the previous case, namely,
two well separated intervals containing each an $\varepsilon/2\pi$
``part'' of spectrum. When these intervals become sufficiently close
one to another most of the spectrum on these intervals is absolutely
continuous (even if the spectrum was singular when the intervals were
``far'' enough). As the energies $E_0$ and $E_\pi$ approach each
other, so do the intervals until they roughly reach an interspacing of
size $\sqrt{t_h}$; during this process, the sizes of the intervals
which, at the start, were roughly of order $t_{v,0}+t_h$ and
$t_{v,\pi}+t_h$ grew to reach the order of $\sqrt{t_h}$ (this number
is much larger than any of the other two as $\rho\ll1$). When $E_0$
and $E_\pi$ move closer to each other, the intervals containing the
spectrum stay ``frozen'' at a distance of size $\sqrt{t_h}$ from each
other, and their sizes do not vary noticeably either (see
Fig.~\ref{fig:rho_petit}). They start moving and changing size again
when $E_0$ and $E_\pi$ again become separated by an interspacing of
size at least $\sqrt{t_h}$. So, in this case, we see a very strong
repulsion preventing the intervals of spectra from intersecting each
other. The spacing between them is quite similar to that observed in
the case of the standard double well problem
(see~\cite{MR85d:35085,He-Sj:84}).
\par In the last case, when $\rho\gg1$ and $\tau\ll1$, we see an
intermediate behavior. For the sake of simplicity, let us assume that
$t_{v,\pi}\gg t_{v,0}$. Starting from the situation when $E_0$ and
$E_\pi$ are ``far'' apart, we see two intervals, one around each point
$E_0$ and $E_\pi$ and this as long as $|E_0-E_\pi|\gg t_{v,\pi}$ (see
Fig.~\ref{fig:rho_grand0}). When $|E_0-E_\pi|$ becomes roughly of size
$ t_{v,\pi}$ or smaller, as in the first case, the Lyapunov exponent
varies by an amount of order one over each of the exponentially small
intervals. The difference is that it need not stay positive: at the
edges of the two intervals that are facing each other, it becomes
small and even can vanish. These are the edges that seem more prone to
interaction. Now, when one moves $E_0$ towards $E_\pi$, the lacuna
separating the two intervals stays open and starts moving with $E_0$;
it becomes roughly centered at $E_0$, stays of fixed size (of order
$t_h/t_{v,\pi}$) and moves along with $E_0$ as $E_0$ crosses $E_\pi$
and up to a distance roughly $t_{v,\pi}$ on the other side of $E_\pi$
(see Fig.~\ref{fig:rho_grand1}).  Then, when $E_0$ moves still further
away from $E_\pi$, it becomes again the center of some interval
containing spectrum that starts moving away from the band centered at
$E_\pi$. We see that, in this case as in the case of strong repulsion,
there always are two intervals separated by a gap; both intervals
contain a $\varepsilon/2\pi$ ``part'' of spectrum. But, now, the two
intervals can become exponentially larger than the gap (in the case of
strong interaction, the length of the gap was at least of the same
order of magnitude as the lengths of the bands). Moreover, on both
intervals, the Lyapunov exponent is positive near the outer edges i.e.
the edges that are not facing each other; it can become small or even
vanish on the inner edges. So, there may be some Anderson transitions
within the intervals. We see here the effects of a weaker form of
resonant tunneling and a weaker repulsion.
\par To complete this introduction, let us note that, though in the
present paper we only considered a perturbation given by a cosine, it
is clear from our techniques that the same phenomena appear as long as
the phase space picture is that given in Fig.~\ref{TIBMfig:actions}.


%
\section{The results}
\label{sec:MainResults}
\noindent We now state our assumptions and results in a precise way.
\subsection{The periodic operator}
\label{sec:periodic-operator}
This section is devoted to the description of elements of the spectral
theory of one-dimensional periodic Schr{\"o}dinger operator $H_0$ that we
need to present our results. For more details and proofs, we refer to
section~\ref{S3} and to~\cite{Eas:73,MR2002f:81151}.
\subsubsection{The spectrum of $H_0$}
\label{sec:son-spectre}
The spectrum of the operator $H_0$ defined in~\eqref{Ho} is a union of
countably many intervals of the real axis, say $[E_{2n+1},\,E_{2n+2}]$
for $n\in\N$ , such that
\begin{gather*}
  E_1<E_2\le E_3<E_4\dots E_{2n}\le E_{2n+1}<E_{2n+2}\le \dots\,,\\
  E_n\to+\infty,\quad n\to+\infty.
\end{gather*}
This spectrum is purely absolutely continuous. The points
$(E_{j})_{j}$ are the eigenvalues of the self-adjoint operator
obtained by considering the differential polynomial~\eqref{Ho} acting
in $L^2([0,2])$ with periodic boundary conditions (see~\cite{Eas:73}).
For $n\in\N$, the intervals $[E_{2n+1},\,E_{2n+2}]$ are the {\it
  spectral bands}, and the intervals $(E_{2n},\,E_{2n+1})$, the {\it
  spectral gaps}. When $E_{2n}<E_{2n+1}$, one says that the $n$-th gap
is {\it open}; when $[E_{2n-1},E_{2n}]$ is separated from the rest of
the spectrum by open gaps, the $n$-th band is said to be {\it
  isolated}. Generically all the gap are open.
\smallpagebreak From now on, to simplify the exposition, we suppose
that 
\begin{description}
\item[(O)] all the gaps of the spectrum of $H_0$ are open.
\end{description}
\subsubsection{The Bloch quasi-momentum}
\label{sec:le-quasi-moment}
Let $x\mapsto\psi(x,E)$ be a non trivial solution to the periodic
Schr{\"o}dinger equation $H_0\psi=E\psi$ such that
$\psi\,(x+1,E)=\mu \,\psi\,(x,E)$, $\forall x\in\R$, for some
$\mu\in\C^*$ independent of $x$. Such a solution is called a {\it
  Bloch solution} to the equation, and $\mu$ is the {\it Floquet
  multiplier} associated to $\psi$. One may write $\mu=\exp(ik)$
where, $k$ is the {\it Bloch quasi-momentum} of the Bloch solution
$\psi$.
\smallpagebreak It appears that the mapping $E\mapsto k(E)$ is
analytic and multi-valued; its branch points are the points $\{E_n;\ 
n\in\N\}$. They are all of ``square root'' type.
\smallpagebreak The dispersion relation $k\mapsto{\mathbf E}(k)$ is
the inverse of the Bloch quasi-momentum.  We refer to
section~\ref{SS3.2} for more details on $k$.
\subsection{A ``geometric'' assumption on the energy region under study}
\label{sec:main-assumption-w}
Let us now describe the energy region where our study is valid.
\smallpagebreak Recall that the spectral window $\mathcal{F}(E)$ is
the range of the mapping $\zeta\in\R\mapsto E-\alpha\cos(\zeta)$.
\smallpagebreak In the sequel, $J$ always denotes a compact interval
such that, for some $n\in\N^*$ and for all $E\in J$, one has
\begin{description}
  \label{TIBMcondition}
\item[(TIBM)]
  $[E_{2n},E_{2n+1}]\subset\dot{\mathcal F}(E)$ and  ${\mathcal
    F}(E)\subset]E_{2n-1},E_{2n+2}[$.
\end{description}
where $\dot{\mathcal F}(E)$ is the interior of $\mathcal{F}(E)$ (see
figure~\ref{interactingfigure}).
\begin{Rem}
  \label{rem:5}
  As all the spectral gaps of $H_0$ are assumed to be open, as their
  length tends to $0$ at infinity, and, as the length of the spectral
  bands goes to infinity at infinity, it is clear that, for any non
  vanishing $\alpha$, assumption (TIBM) is satisfied in any gap at a
  sufficiently high energy; it suffices that this gap be of length
  smaller than $2\alpha$.
\end{Rem}
\subsection{The definitions of the phase integrals and the tunneling
  coefficients}
\label{sec:iso-energy-curve}
We now give the precise definitions of the phase integrals and the
tunneling coefficients introduced in the introduction.
\subsubsection{The complex momentum and its branch points}
\label{sec:complex-momentum-its}
The phase integrals and the tunneling coefficients are expressed in
terms of integrals of the {\it complex momentum}. Fix $E$ in $J$.  The
complex momentum $\zeta\mapsto\kappa(\zeta)$ is defined by
%
\begin{floatingfigure}[r]{7.5cm}
  \centering
  \includegraphics[bbllx=71,bblly=606,bburx=275,bbury=721,width=7cm]{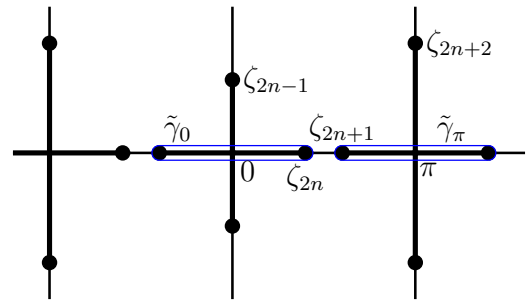}
  \caption{The branch points}\label{bp-tibm}
\end{floatingfigure}
%
\begin{equation}
  \label{complex-mom}
  \hskip-4cm\kappa(\zeta)=k(E-\alpha\cos(\zeta)).
\end{equation}
As $k$, $\kappa$ is analytic and multi-valued. The set $\Gamma$
defined in~\eqref{isoen} is the graph of the function $\kappa$.  As
the branch points of $k$ are the points $(E_i)_{i\in\N}$, the branch
points of $\kappa$ satisfy
\begin{equation}
  \label{BPCM}
  \hskip-4cm E-\alpha\cos(\zeta)=E_j,\ j\in\N^*.
\end{equation}
As $E$ is real, the set of these points is symmetric with respect to
the real axis and to the imaginary axis, and it is\\ $2\pi$-periodic
in $\zeta$. All the branch points of $\kappa$ lie on $\arccos(\R)$.
This set consists of the real axis and all the translates of the
imaginary axis by a multiple of $\pi$. As the branch points of the
Bloch quasi-momentum, the branch points of $\kappa$ are of ``square
root'' type.
\vskip.1cm Due to the symmetries, it suffices to describe the branch
points in the half-strip $\{\zeta;\ \im\zeta\geq0,\ 
0\leq\re\zeta\leq\pi\}$. These branch points are described in detail
in section 7.1.1 of~\cite{Fe-Kl:04a}. In figure~\ref{bp-tibm}, we show
some of them.  The points $\zeta_j$ being defined by~\eqref{BPCM}, one has
$0<\zeta_{2n}<\zeta_{2n+1}<\pi$,
$0<\im\zeta_{2n+2}<\im\zeta_{2n+3}<\cdots$,
$0<\im\zeta_{2n-1}<\cdots<\im\zeta_1$.
\subsubsection{The contours}
\label{sec:contours}
To define the phases and the tunneling coefficients, we introduce some
integration contours in the complex $\zeta$-plane.\\
These loops are shown in figures~\ref{bp-tibm} and~\ref{TIBMfig:2}.
The loops $\tilde\gamma_{0}$, $\tilde\gamma_{\pi}$,
$\tilde\gamma_{h,0}$, $\tilde\gamma_{h,\pi}$, $\tilde\gamma_{v,0}$ and
$\tilde\gamma_{v,\pi}$ are simple loops, respectively, going once
around the intervals $[-\zeta_{2n},\zeta_{2n}]$,
$[\zeta_{2n+1},2\pi-\zeta_{2n+1}]$, $[-\zeta_{2n+1},-\zeta_{2n}]$,
$[\zeta_{2n},\zeta_{2n+1}]$, $[\zeta_{2n-1},\overline{\zeta_{2n-1}}]$
and $[\zeta_{2n+2},\overline{\zeta_{2n+2}}]$.\\
In section 10.1 of~\cite{Fe-Kl:04a}, we have shown that, on each of
the above loops, one can fix a continuous branch of the complex
momentum.\\
Consider $\Gamma$, the complex iso-energy curve defined
by~\eqref{isoen}. Define the projection $\Pi:
(\zeta,\kappa)\in\Gamma\mapsto\zeta\in\C$. As on each of the loops
$\tilde\gamma_{0}$, $\tilde\gamma_{\pi}$, $\tilde\gamma_{h,0}$,
$\tilde\gamma_{h,\pi}$, $\tilde\gamma_{v,0}$ and
$\tilde\gamma_{v,\pi}$, one can fix a continuous branch of the complex
momentum, each of these loops is the projection on the complex plane
of some loop in $\Gamma$. In sections 10.6.1 and 10.6.2
of~\cite{Fe-Kl:04a}, we give the precise definitions of the curves
$\gamma_{0}$, $\gamma_{\pi}$, $\gamma_{h,0}$, $\gamma_{h,\pi}$,
$\gamma_{v,0}$ and $\gamma_{v,\pi}$ represented in
figure~\ref{TIBMfig:actions} and show that they respectively project
onto the curves $\tilde\gamma_{0}$, $\tilde\gamma_{\pi}$,
$\tilde\gamma_{h,0}$, $\tilde\gamma_{h,\pi}$, $\tilde\gamma_{v,0}$ and
$\tilde\gamma_{v,\pi}$.
\subsubsection{The phase integrals, the action integrals and the tunneling
  coefficients}
\label{sec:acti-integr-tunn-1}
The results described below are proved in
section 10 of~\cite{Fe-Kl:04a}.\\
%
\begin{floatingfigure}[r]{7cm}
  \centering
  \includegraphics[bbllx=71,bblly=606,bburx=275,bbury=721,width=7cm]{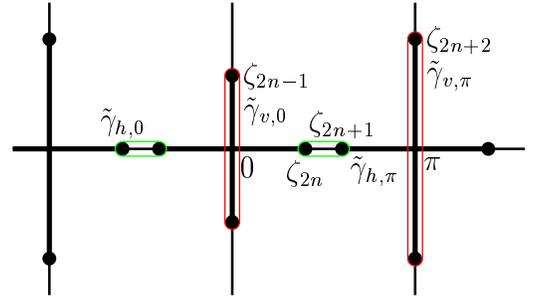}
  \caption{The loops for the action integrals}\label{TIBMfig:2}
\end{floatingfigure}
%
\noindent Let $\nu\in\{0,\pi\}$. To the loop $\gamma_\nu$, we associate
the {\it phase integral} $\Phi_\nu$ defined as
\begin{equation}
  \label{Phi:Gamma}
  \hskip-5cm\Phi_\nu(E)=\frac12\oint_{\tilde\gamma_\nu}\kappa\,d\zeta,
\end{equation}
where $\kappa$ is a branch of the complex momentum that is continuous
on $\tilde\gamma_\nu$. The function $E\mapsto\Phi_\nu(E)$ is real analytic
and does not vanish on $J$. The loop $\tilde\gamma_\nu$ is oriented so that
$\Phi_\nu(E)$ be positive.  One shows that, for all $E\in J$,
\begin{equation}
  \label{eq:21}
  \hskip-5cm \quad\Phi_0'(E)<0\quad\text{ and
  }\quad\Phi_\pi'(E)>0.
\end{equation}
To the loop $\gamma_{v,\nu}$, we associate the {\it vertical action
  integral} $S_{v,\nu}$ defined as
\begin{equation}
  \label{Sv:Gamma}
  S_{v,\nu}(E)=-\frac i2\oint_{\tilde\gamma_{v,\nu}}\kappa d\zeta,
\end{equation}
\vskip.1cm\noindent where $\kappa$ is a branch of the complex momentum
that is continuous on $\tilde\gamma_{v,\nu}$. The function $E\mapsto
S_{v,\nu}(E)$ is real analytic and does not vanish on $J$.  
The loop $\tilde\gamma_{v,\nu}$ is oriented so that $S_{v,\nu}(E)$ be positive.
\\
The {\it vertical tunneling coefficient} is defined to be
\begin{equation}
  \label{tv}
\hspace{-5cm}  t_{v,\nu}(E)=\exp\left(-\frac1\varepsilon S_{v,\nu}(E)\right).
\end{equation}

\smallpagebreak The index $\nu$ being chosen as above, we define
{\it horizontal action integral} $S_{h,\nu}$ by
\begin{equation}
  \label{Sh:Gamma}\hspace{-5cm}
  S_{h,\nu}(E)=-\frac i2\oint_{\tilde\gamma_{h,\nu}} \kappa(\zeta)\,d\zeta,
\end{equation}
where $\kappa$ is a branch of the complex momentum that is continuous
on $\tilde\gamma_{h,\nu}$. The function $E\mapsto S_{h,\nu}(E)$ is real
analytic and does not vanish on $J$.  The loop $\tilde\gamma_{h,\nu}$ is
oriented so that $S_{h,\nu}(E)$ be positive.  \\
The {\it horizontal
  tunneling coefficient} is defined as
\begin{equation}
  \label{th}
  t_{h,\nu}(E)=\exp\left(-\frac1\varepsilon S_{h,\nu}(E)\right).
\end{equation}
As the cosine is even, one has
\begin{equation}
  \label{parity-h}
  S_{h,0}(E)=S_{h,\pi}(E)\quad\text{ and
  }\quad t_{h,0}(E)=t_{h,\pi}(E).
\end{equation}
Finally, one defines
\begin{equation}
  \label{eq:15}
  S_h(E)=S_{h,0}(E)+S_{h,\pi}(E)\quad\text{ and }\quad
  t_h(E)=t_{h,0}(E)\cdot t_{h,\pi}(E).
\end{equation}
In~\eqref{Phi:Gamma},~\eqref{Sv:Gamma}, and~\eqref{Sh:Gamma}, only the
sign of the integral depends on the choice of the branch of $\kappa$;
this sign was fixed by orienting the integration contour.
\subsection{A coarse description of the location of the spectrum in $J$}
\label{sec:une-descr-gross}
Henceforth, we assume that the assumptions (H) and (O) are satisfied
and that $J$ is a compact interval satisfying (TIBM). As
in~\cite{Fe-Kl:04a}, we suppose that
\begin{description}
\item[(T)]\quad $\D 2\pi\cdot\min_{E\in
    J}\min(\im\zeta_{2n-2}(E),\,\im\zeta_{2n+3}(E)) >\max_{E\in
    J}\max(S_h(E),\,S_{v,0}(E),\,S_{v,\pi}(E))$.
\end{description}
Note that (T) is verified if the spectrum of $H_0$ has two successive
bands that are sufficiently close to each other and sufficiently far
away from the remainder of the spectrum (this can be checked
numerically on simple examples, see section~\ref{sec:numer-comp}). In
section 1.9 of~\cite{Fe-Kl:04a}, we discuss this assumption further.
\smallpagebreak Define
  \begin{equation}
    \label{eq:6}
    \delta_0:=\frac12\inf_{E\in J}\min(S_h(E),S_{v,0}(E),S_{v,\pi}(E))>0.
  \end{equation}
One has
\begin{Th}[\cite{Fe-Kl:04a}]
  \label{thr:2} 
  Fix $E_*\in J$. For $\varepsilon$ sufficiently small, there exists
  $V_*\subset \C$, a neighborhood of $E_*$, and two real analytic
  functions $E\mapsto\check \Phi_0(E,\varepsilon)$ and
  $E\mapsto\check\Phi_\pi(E,\varepsilon)$, defined on $V_*$ satisfying
  the uniform asymptotics
  \begin{equation}
    \label{eq:17}
    \check\Phi_0(E,\varepsilon)=\Phi_0(E)+o(\varepsilon),\quad
    \check\Phi_\pi(E,\varepsilon)=\Phi_\pi(E)+o(\varepsilon)\quad\text{when
    }\varepsilon\to0,
  \end{equation}
  such that, if one defines two finite sequences of points in $J\cap
  V_*$, say $(E_{0}^{(l)})_l:=(E_{0}^{(l)}(\varepsilon))_l$ and
  $(E_{\pi}^{(l')})_{l'}:=(E_{\pi}^{(l')}(\varepsilon))_{l'}$, by
  \begin{equation}
    \label{eq:11}
    \frac1{\varepsilon}\check\Phi_0(E_0^{(l)},\varepsilon)=\frac\pi2+\pi
    l\quad\text{ and }\quad\frac1{\varepsilon}\check
    \Phi_\pi(E_\pi^{(l')},\varepsilon)=\frac\pi2+\pi l',\quad
    (l,\,l')\in\N^2,
  \end{equation}
  then, for all real $z$, the spectrum of $H_{z,\varepsilon}$ in
  $J\cap V_*$ is contained in the union of the intervals
  \begin{equation}
    \label{eq:16}
    I_0^{(l)}:=
    E_{0}^{(l)}+[-e^{-\delta_0/\varepsilon},e^{-\delta_0/\varepsilon}]
    \quad\text{ and }\quad
    I_{\pi}^{(l')}:=E_{\pi}^{(l')}+[-e^{-\delta_0/\varepsilon},
    e^{-\delta_0/\varepsilon}]
  \end{equation}
  that is
  \begin{equation*}
    \sigma(H_{z,\varepsilon})\cap J\cap V_*\subset\left(\bigcup_{l}
    I_{0}^{(l)}\right)\bigcup\left(\bigcup_{l'}
    I_{\pi}^{(l')}\right).
  \end{equation*}
\end{Th}
\noindent In the sequel, to alleviate the notations, we omit the
reference to $\varepsilon$ in the functions $\check\Phi_0$ and
$\check\Phi_\pi$.
\smallpagebreak By~\eqref{eq:21} and~\eqref{eq:17},  there exists $C>0$
such that, for $\varepsilon$ sufficiently small, the points defined
in~\eqref{eq:11}  satisfy
\begin{gather}
  \label{eq:7}
  \frac{1}{C}\varepsilon\le E_0^{(l)}-E_0^{(l-1)}\le
  C\varepsilon,\\
  \label{eq:8}
  \frac{1}{C}\varepsilon\le E_{\pi}^{(l)}-E_{\pi}^{(l-1)}\le
  C\varepsilon.
\end{gather}
Moreover, for $\nu\in\{0,\pi\}$, in the interval $J\cap V_*$, the
number of points $E_{\nu}^{(l)}$ is of order $1/\varepsilon$.
\smallpagebreak In the sequel, we refer to the points $E_{0}^{(l)}$
(resp. $E_{\pi}^{(l)}$), and, by extension, to the intervals
$I_{0}^{(l)}$ (resp. $I_{\pi}^{(l)}$) attached to them, as of type $0$
(resp. type $\pi$).
\smallpagebreak By~\eqref{eq:7} and~\eqref{eq:8}, the intervals of
type $0$ (resp. $\pi$) are two by two disjoints; any interval of type
$0$ (resp. $\pi$) intersects at most a single interval of type $\pi$
(resp. $0$).
\subsection{A precise description of the spectrum}
\label{sec:plus-proche-de}
As pointed out in the introduction, the present paper deals with the
resonant case that is we consider two energies, say $E_0^{(l)}$ and
$E_\pi^{(l')}$, that satisfy
\begin{equation}
  \label{eq:20}
  |E_\pi^{(l')}-E_0^{(l)}|\le 2e^{-\frac{\delta_0}{\varepsilon}}.
\end{equation}
This means that the intervals $I_0^{(l)}$ and $I_\pi^{(l')}$
intersect each other. Moreover, by~\eqref{eq:7} and~\eqref{eq:8},
these intervals stay at a distance at least $C^{-1}\varepsilon$ of all the
other intervals of the sequences defined in Theorem~\ref{thr:2}.  We
now describe the spectrum of $H_{z,\varepsilon}$ in the union
$I_0^{(l)}\cup I_\pi^{(l')}$.\\
To simplify the exposition, we set
\begin{equation}
  \label{eq:13}
  E_0:=E_0^{(l)},\quad\quad
  E_\pi:=E_\pi^{(l')},\quad\quad I_0:=I_0^{(l)},\quad\text{ and }\quad
  I_\pi:=I_\pi^{(l')}.
\end{equation}
In the resonant case, the primary parameter controlling the location
and the nature of the spectrum is
\begin{equation}
  \label{tau}
  \tau=2\sqrt{\frac{t_{v,0}(\bar E)\,t_{v,\pi}(\bar E)}{t_h(\bar
      E)}}\quad\text{where}\quad\bar E=\frac{E_\pi+E_0}2.
\end{equation}
As tunneling coefficients are exponentially small,
one typically has either $\tau\gg1$ or $\tau\ll1$.
We will give a detailed analysis of these cases. 
More precisely, we fix $\delta_\tau>0$ arbitrary
and assume that either
\begin{gather}
  \label{eq:28}
  \forall E\in V_*\cap J,\quad
  S_h(E)-S_{v,0}(E)-S_{v,\pi}(E)\geq\delta_\tau,\quad
  (\text{case we denote by }\tau\gg1),\\
  \intertext{or}
  \label{eq:14}
  \forall E\in V_*\cap J,\quad S_h(E)-S_{v,0}(E)-S_{v,\pi}(E)\le
  -\delta_\tau,\quad (\text{case we denote by }\tau\ll1).
\end{gather}
The case $\tau\asymp1$ is more complicated (also less frequent i.e.
satisfied by less energies). We discuss it briefly later.
\smallpagebreak To describe our results, it is convenient to introduce
the following ``local variables''
\begin{equation}
  \label{xi}
  \xi_\nu(E)=\frac{\check \Phi'_\nu(\bar E)}\varepsilon\cdot \frac{E-
    E_\nu}{t_{v,\nu}(\bar E)}\quad\text{where}\quad\nu\in\{0,\pi\}.
\end{equation}
\subsubsection{When $\tau$ is large}
\label{sec:quand-tau-g}
Let us now assume $\tau\gg1$. The location of the spectrum in $I_0\cup
I_\pi$ is described by
\begin{Th}
  \label{th:tib-res:sp:1}
  Assume we are in the case of Theorem~\ref{thr:2}.
  Assume~\eqref{eq:28} is satisfied. Then, there exist
  $\varepsilon_0>0$ and a non-negative function $\varepsilon\mapsto
  f(\varepsilon)$ tending to zero as $\varepsilon\to 0$ such that, for
  $\varepsilon\in (0,\varepsilon_0]$, the spectrum of
  $H_{z,\varepsilon}$ in $I_0\cup I_\pi$ is located in two intervals
  $\check I_0$ and $\check I_\pi$ defined by
  \begin{equation*}
    \check I_0= \{E\in I_0:\ |\xi_0(E)|\leq1+f(\varepsilon)\}
    \quad\text{and}\quad
    \check I_\pi= \{E\in I_\pi:\ |\xi_\pi(E)|\leq1+f(\varepsilon)\}.
  \end{equation*}
  If $dN_{\varepsilon}(E)$ denotes the density of states measure of
  $H_{z,\varepsilon}$, then
  \begin{equation}
    \label{eq:2}
    \int_{\check I_0}dN_{\varepsilon}(E)=\int_{\check
      I_\pi}dN_{\varepsilon}(E) =\frac\varepsilon{2\pi}\
    \text{if}\ \check I_0\cap\check I_\pi=\emptyset,\quad\text{and}
    \quad\int_{\check I_0\cup \check I_\pi}dN_{\varepsilon}(E)=
    \frac\varepsilon{\pi}\ \text{otherwise}.
  \end{equation}
  Moreover, the Lyapunov exponent on $\check I_0\cup\check I_\pi$
  satisfies
  \begin{equation}
    \label{Theta-res}
    \Theta(E,\varepsilon)=\frac\varepsilon{\pi}
    \log\left(\tau\sqrt{1+|\xi_0(E)|+|\xi_\pi(E)|}\right)+o(1),
  \end{equation}
  where $o(1)\to0$ when $\varepsilon\to0$ uniformly in $E$, in $E_0$
  and in $E_\pi$.
\end{Th}
\noindent By~\eqref{eq:2}, if $\check I_0$ and
$\check I_\pi$ are disjoint, they both contain spectrum of
$H_{z,\varepsilon}$; if not, one only knows that their union contains
spectrum.
\smallpagebreak Let us analyze the results of
Theorem~\ref{th:tib-res:sp:1}.
%
\begin{figure}[htbp]
  \centering
  \subfigure[$|E_\pi-E_0|\gg\max(t_{v,\pi},t_{v,0})$]{
    \includegraphics[bbllx=71,bblly=655,bburx=275,bbury=721,width=7cm]{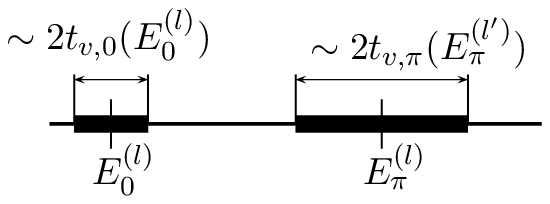}
    \label{fig:tau_grand1}}
  \hskip1cm
  \subfigure[$|E_\pi-E_0|\ll\max(t_{v,\pi},t_{v,0})$]{
    \includegraphics[bbllx=71,bblly=655,bburx=247,bbury=721,width=7cm]{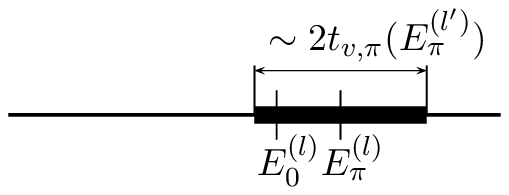}
    \label{fig:tau_grand2}}
  \caption{The location of the spectrum for $\tau$ large}
  \label{fig:tau_grand}
\end{figure}
%
\smallpagebreak{\it The location of the spectrum.\/} By~\eqref{xi},
the intervals $\check I_0$ and $\check I_\pi$ defined in
Theorem~\ref{th:tib-res:sp:1} are respectively ``centered'' at the
points $E_0$ and $E_\pi$. Their lengths are given by
\begin{equation*}
  |\check I_0|=\frac{2\,\varepsilon}{|\check\Phi'_0(E_0)|}\cdot
  {t_{v,0}(E_0)}\cdot(1+o(1))\quad\text{and}\quad
  |\check I_\pi|=\frac{2\,\varepsilon}{|\check\Phi'_\pi(E_\pi)|}\cdot
  {t_{v,\pi}(E_\pi)}\cdot(1+o(1)).
\end{equation*}
where $o(1)$ only depends on $\varepsilon$ and $o(1)\to0$ when
$\varepsilon\to0$. Depending on $|E_\pi-E_0|$, the picture of the
spectrum in $I_0\cup I_\pi$ is given by figure~\ref{fig:tau_grand},
case (a) and (b).\\
The repulsion effect observed in the non resonant case
(see~\cite{Fe-Kl:04a}, section 1.6) is negligible with
respect to the length of the intervals $\check I_0$ and $\check
I_\pi$.
\smallpagebreak{\it The nature of the spectrum.\/} In the intervals
$\check I_0$ and $\check I_\pi$, according to~\eqref{eq:28}
and~\eqref{Theta-res}, the Lyapunov exponent is positive.  Hence, by
the Ishii-Pastur-Kotani Theorem (\cite{Pa-Fi:92}), in both $\check
I_0$ and $\check I_\pi$, the spectrum of $H_{z,\varepsilon}$ is
singular.
\smallpagebreak{\it The Lyapunov exponent $\Theta(E,\varepsilon)$ on
  the spectrum.\/} The general formula~\eqref{Theta-res} can be
  simplified in the following way:
\begin{gather*}
  \Theta(E,\varepsilon)=\frac
  \varepsilon{\pi}\log\left(\tau\sqrt{1+|\xi_0(E)|}\right)+o(1)\quad
  {\rm when }\quad E\in \check I_\pi,\\
  \intertext{and}
  \Theta(E,\varepsilon)=\frac
  \varepsilon{\pi}\log\left(\tau\sqrt{1+|\xi_\pi(E)|}\right)+o(1)\quad
  {\rm when }\quad E\in \check I_0.
\end{gather*}
If $|E_\pi-E_0|\gg\max(t_{v,\pi},t_{v,0})$, then the Lyapunov exponent
stays essentially constant on each of the intervals $\check I_0$ and
$\check I_\pi$. On the other hand, if
$|E_\pi-E_0|\ll\max(t_{v,\pi},t_{v,0})$, then, on these exponentially
small intervals, the Lyapunov exponent may vary by a constant. To see
this, let us take a simple example. Assume that $t_{v,0}\ll
t_{v,\pi}$, or, better said, that there exists $\delta>0$ such that
\begin{equation*}
  \forall E\in V_*\cap J,\quad S_{v,0}(E)>S_{v,\pi}(E)+\delta.
\end{equation*}
If $E_0$ and $E_\pi$ coincide, then $\check I_0\subset \check I_\pi$,
and, near the center of $\check I_\pi$, the Lyapunov exponent assumes
the value
\begin{equation*}
    \Theta(E,\varepsilon)=\frac{\varepsilon}{\pi}\log\tau+o(1)=
    \frac{1}{2\pi}\left(S_h(\bar E)-S_{v,\pi}(\bar E)-S_{v,0}(\bar
    E)\right)+o(1).
\end{equation*}
Near the edges of $\check I_\pi$, its value is given by
\begin{equation*}
  \begin{split}
    \Theta(E,\varepsilon)&=\frac{\varepsilon}{\pi}\log\tau+
    \frac{\varepsilon}{2\pi}\log(t_{v,\pi}(\bar E)/t_{v,0}(\bar
    E))+o(1) \\&=\frac{1}{2\pi}\left(S_h(\bar E)-2S_{v,\pi}(\bar
      E)\right)+o(1).
  \end{split}
\end{equation*}
So the variation of the Lyapunov exponent is given by
$\frac{1}{2\pi}\,(S_{v,0}(\bar E)-S_{v,\pi}(\bar E))$ on an
exponentially small interval. One sees a sharp drop of the Lyapunov
exponent on the interval containing spectrum when going from the edges
of $\check I_\pi$ towards $E_\pi$.
\subsubsection{When $\tau$ is small}
\label{sec:quand-tau-p}
We now assume that $\tau\ll1$, i.e., that~\eqref{eq:14} is satisfied.
Then, the spectral behavior depends on the value of the quantity
$\Lambda_n(V)$ defined and analyzed in section~\ref{sec:Omega}, see
Theorem~\ref{Lambda:prop}.  Here, we only note that $\Lambda_n(V)$
depends solely on $V$ and on the number of the gap separating the two
interacting bands; moreover, it can be considered as a ''measure of
symmetry'' of $V$: taking the value 1 for ``symmetric'' potentials
$V$, this quantity generically satisfies
\begin{equation}
  \label{eq:41}
  \Lambda_n(V)>1.
\end{equation}
Below, we only consider this generic case.\\
There are different possible ``scenarii'' for the spectral behavior
when $\tau\ll1$. Before describing them in detail, we start with a
general description of the spectrum. We prove
\begin{Th}
  \label{thr:5}
  Assume we are in the case of Theorem~\ref{thr:2}. Assume
  that~\eqref{eq:14} and~\eqref{eq:41} are satisfied. Then, there
  exists $\varepsilon_0>0$ and a non negative function
  $\varepsilon\mapsto f(\varepsilon)$ tending to zero when
  $\varepsilon\to0$ such that, for $\varepsilon\in]0,\varepsilon_0[$,
  the spectrum of $H_{z,\varepsilon}$ in $I_0^{(l)}\cup I_\pi^{(l')}$
  is contained in $\Sigma(\varepsilon)$, the set of energies $E$
  satisfying
  \begin{equation}
    \label{eq:37}
    \left|\tau^2\xi_0(E)\xi_\pi(E)+2\Lambda_n(V)\right|\leq
    \left(2+\tau^2|\xi_0(E)|+\tau^2|\xi_\pi(E)|\right)(1+f(\varepsilon)).
  \end{equation}
\end{Th}
\noindent In section~\ref{sec:G-F:37}, we analyze the
    inequality~~(\ref{eq:37}) and prove 
\begin{Pro}
  \label{pro:1}
  For sufficiently small $\varepsilon$, the set $\Sigma(\varepsilon)$
  defined in Theorem~\ref{thr:5} is the union of two disjoint compact
  intervals; both intervals are strictly contained inside the
  $(2e^{-\delta_0/\varepsilon})$-neighborhood of $\bar E$.
\end{Pro}
\noindent The intervals described in Proposition~\ref{pro:1} are
denoted by $I_0$ and $I_\pi$.\\
We check the
\begin{Th}
  \label{thr:6}
  Assume we are in the case of Proposition~\ref{pro:1}. If
  $dN_{\varepsilon}(E)$ denotes the density of states measure of
  $H_{z,\varepsilon}$, then
  \begin{equation*}
    \int_{I_0}dN_{\varepsilon}(E)=\int_{I_\pi}dN_{\varepsilon}(E)
    =\frac\varepsilon{2\pi}.
  \end{equation*}
\end{Th}
\noindent Hence, each of the intervals $I_0$ and $I_\pi$ contains some
spectrum of $H_{z,\varepsilon}$. This 
implies that, when $\tau\ll1$ and $\Lambda_n(V)>1$, there is a ``level
repulsion'' or a ``splitting'' of resonant intervals.
\smallpagebreak As for the nature of the spectrum, one shows the
following results. The behavior of the Lyapunov exponent is given by
\begin{Th}
  \label{thr:7}
  Assume we are in the case of Theorem~\ref{thr:5}. On the set
  $\Sigma(\varepsilon)$,  the Lyapunov exponent of
  $H_{z,\varepsilon}$ satisfies
  \begin{equation}
    \label{eq:40}
    \Theta(E,\varepsilon)=\frac{\varepsilon}{2\pi}
    \log\left(\tau^2(|\xi_0(E)|+|\xi_\pi(E)|)+1\right)+o(1),
  \end{equation}
  where $o(1)\to0$ when $\varepsilon\to0$ uniformly in $E$ and in
  $E_0$ and $E_\pi$.
\end{Th}
\noindent For $c>0$, one defines the set
\begin{equation}
  \label{eq:36}
  I^+_c=\left\{E\in \Sigma(\varepsilon):\ \varepsilon
    \log\left(\tau\sqrt{|\xi_0(E)|+|\xi_\pi(E)|}\right)>c\right\}.
\end{equation}
Theorem~\ref{thr:7} and the Ishii-Pastur-Kotani Theorem imply
\begin{Cor}
  \label{cor:1}
  Assume we are in the case of Theorem~\ref{thr:5}. For $\varepsilon$
  sufficiently small, the set $\Sigma(\varepsilon)\cap I^+_c$ only
  contains singular spectrum.
\end{Cor}
\smallpagebreak Define
\begin{equation}
  \label{eq:34}
  I^-_c=\left\{E\in\Sigma(\varepsilon):\ \varepsilon
    \log\left(\tau\sqrt{|\xi_0(E)|+|\xi_\pi(E)|}\right)<-c\right\}.
\end{equation}
Theorem~\ref{thr:5} implies that, for sufficiently small
$\varepsilon$, the set  $I_c^-$ is contained  in the set
\begin{equation}
  \label{eq:37:a}
  \tilde\Sigma_{\rm ac}(\varepsilon)=
  \{E\in
  \R:\,\left|\tau^2\xi_0(E)\xi_\pi(E)+2\Lambda_n(V)\right|\leq 
  2(1+g(\varepsilon))\},
\end{equation}
where $\varepsilon \mapsto g(\varepsilon)$ is independent of $c$ and
satisfies the estimate $g=o(1)$ as $\varepsilon\to 0$.  The set $
\tilde\Sigma_{\rm ac}(\varepsilon)$ consists of $\tilde I_0$ and
$\tilde I_\pi$, two disjoint intervals, and the distance between these
intervals is greater than or equal to $C\,
\varepsilon\sqrt{t_h(\bar E)}$ (see Lemma~\ref{M(Delta)}). \\
Let $\Sigma_{ac}$ denote the absolutely continuous spectrum of
$H_{z,\varepsilon}$.  One shows
\begin{Th}
  \label{thr:1}
  Assume we are in the case of Theorem~\ref{thr:5}. Pick
  $\nu\in\{0,\pi\}$. There exists
  $\eta>0$ and $D\subset (0,1)$, a set of Diophantine numbers such that
  \begin{itemize}
  \item
    \begin{equation*}
      \frac{\mes(D\cap(0,\varepsilon))}{\varepsilon}=
      1+o\left(e^{-\eta/\varepsilon}\right)\text{
        when }\varepsilon\to0.
    \end{equation*}
  \item for $\varepsilon\in D$ sufficiently small, if $I_c^-\cap
    \tilde I_\nu\ne \emptyset$, then 
    \begin{equation*}
      \mes(\tilde I_\nu\cap \Sigma_{\rm ac})=\mes(I_\nu)\,(1+o(1)),
    \end{equation*}
  where $o(1)\to0$ when
  $\varepsilon\to0$ uniformly in $E_0$ and $E_\pi$.
  \end{itemize}
\end{Th}
\subsubsection{Possible scenarii when $\tau$ is small}
\label{sec:les-scen-princ}
\begin{figure}[htbp]
  \centering
  \subfigure[When $|E_\pi-E_0|\gg\varepsilon t_{v,\pi}(\bar E)$]{
    \includegraphics[bbllx=26,bblly=655,bburx=247,bbury=721,width=7cm]{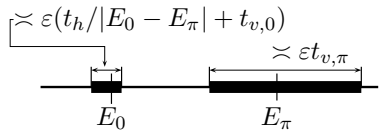}
    \label{fig:rho_grand0}}
  \hskip1cm
  \subfigure[When $|E_\pi-E_0|\ll\varepsilon t_{v,\pi}(\bar E)$]{
    \includegraphics[bbllx=41,bblly=655,bburx=247,bbury=721,width=7cm]{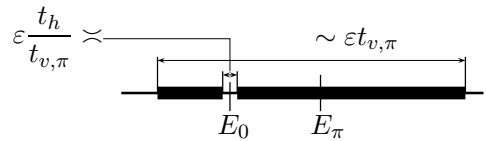}
    \label{fig:rho_grand1}}
  \caption{The locus of the spectrum when $\tau\ll1$ and $\rho\gg1$}
  \label{fig:tau_petit}
\end{figure}

As in section~\ref{sec:quand-tau-p}, we now assume that $\tau\ll1$ and
$\Lambda_n(V)>1$. Essentially, there are two possible cases for the
location and the nature of the spectrum of $H_{z,\varepsilon}$.
Define
\begin{equation}
  \label{eq:1}
  \rho:=\left.\frac{\max(t_{v,\pi},t_{v,0})}{\sqrt{t_h}}\right|_{E=\bar E}
  =\tau\left.\sqrt{\frac{\max(t_{v,\pi},t_{v,0})}{\min(t_{v,\pi},t_{v,0})}}
  \right|_{E=\bar E}.
\end{equation}
Note that $\rho\geq\tau$. We only discuss the cases when $\rho$ is
exponentially small or exponentially large.
\smallpagebreak{\bf 1.} \ Here, following section~\ref{sec:small-rho},
we discuss the case $\rho\ll1$.  If $|E_\pi-E_0|\ll
\varepsilon\sqrt{t_h(\bar E)}$, $\Sigma(\varepsilon)$ is the union of
two intervals of length roughly $\varepsilon\sqrt{t_h}$ ; they are
separated by a gap of length roughly $\varepsilon\sqrt{t_h}$ (see
figure~\ref{fig:rho_petit}); this gap is centered at the point $\bar
E$.  The length of the intervals containing spectrum, as well as the
length and center of the lacuna essentially do not change as the
distance $E_\pi-E_0$ increases up to $\sim\varepsilon\sqrt{t_h(\bar
  E)}$; after that, the intervals containing spectrum begin to move
away from each other.
%
\begin{floatingfigure}[r]{6cm}
  \centering
    \includegraphics[bbllx=71,bblly=655,bburx=275,bbury=721,width=6cm]{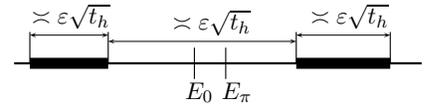}
    \caption{When $\rho\ll1$}\label{fig:rho_petit}
\end{floatingfigure}
%
\noindent As for the nature of the spectrum, when $\rho$ is
exponentially small and $|E_\pi-E_0|\ll\varepsilon\sqrt{t_h(\bar E)}$,
the intervals containing spectrum are contained in the set $I^-_c$;
so, most of the spectrum in these intervals is absolutely continuous
(if $\varepsilon$ satisfies the Diophantine
condition of Theorem~\ref{thr:1}).\\
{\bf 2.} \ Consider he case when $\rho\gg1$. This case is analyzed in
section~\ref{sec:big-rho}. For sake of definiteness, assume that
$t_{v,0}\ll t_{v,\pi}$. Then, there exists an interval, say $I_\pi$,
that is asymptotically centered at $E_\pi$ and that contains spectrum.
The length of this interval is of order $\varepsilon t_{v,\pi}(\bar
E)$.
\vskip.2cm\noindent One distinguishes two cases:
\begin{enumerate}
\item if $E_0$ belongs to $I_ \pi$ and if the distance from $E_0$ to
  the edges of $I_\pi$ is of the same order of magnitude as the length
  of $I_\pi$, then $\Sigma(\varepsilon)$ consists of the interval
  $I_\pi$ without a ``gap'' of length roughly $\varepsilon t_h(\bar
  E)/t_{v,\pi}(\bar E)$ and containing $E_0$ (see
  figure~\ref{fig:rho_grand1}). Moreover, the distance from $E_0$ to
  any edge of the gap is also of order $\varepsilon t_h(\bar
  E)/t_{v,\pi}(\bar E)$.
\item if $E_0$ is outside $I_ \pi$ and at a distance from $I_\pi$ at
  least of the same order of magnitude as the length of $I_\pi$, then
  $\Sigma(\varepsilon)$ consists in the union of $I_\pi$ and an
  interval $I_0$ (see figure~\ref{fig:rho_grand0}). The interval $I_0$
  is contained in neighborhood of $E_0$ of size roughly $\varepsilon^2
  t_h(\bar E)/|E_0-E_\pi|$. The length of $I_0$ is of size
  $\varepsilon^2 t_h(\bar E)/|E_0-E_\pi|+ \varepsilon t_{v,0}(\bar E)$.
\end{enumerate}
%
%
When $\rho$ is exponentially large, the Lyapunov exponent may vary
very quickly on the intervals containing spectrum. Consider the case
$|E_\pi-E_0|\ll\varepsilon t_{v,\pi}(\bar E)$. For $E$ close the gap
surrounding $E_0$, $\tau^2\xi_0(E)$ is of order $1$ whereas
$\tau^2\xi_\pi(E)$ is exponentially small. Hence, for $E$ near the gap
surrounding $E_0$, Theorem~\ref{thr:7} implies
$\Theta(E,\varepsilon)=o(1)$. On the other hand, at the external edges
of the intervals containing spectrum, $\tau^2\xi_0(E)$ is of size
roughly $\rho^2$; this factor being exponentially large, at such
energies, the Lyapunov exponent is positive and given by
\begin{equation*}
  \Theta(E,\varepsilon)=\frac{\varepsilon}{\pi}\log\rho+o(1).
\end{equation*}
This phenomenon is similar to that observed for $\tau\gg1$ except
that, now, the Lyapunov exponent sharply drops to a value that is
small and that may even vanish.
In fact, on most of $\Sigma(\varepsilon)$, the Lyapunov exponent stays
positive and, the spectrum is singular (by Corollary~\ref{cor:1}).
Moreover, near the lacuna surrounding $E_0$, neither
Corollary~\ref{cor:1}, nor Theorem~\ref{thr:1} apply. These zones are
similar to the zones where asymptotic Anderson transitions were found
in~\cite{MR2003f:82043}.
\subsection{The model equation}
\label{sec:model-equation}
The study of the spectrum of $H_{z,\varepsilon}$  is reduced
to the study  of  the  finite difference
equation (the monodromy equation, see
section~\ref{sec:monodr-matr-monodr}):
\begin{equation}
  \label{model-eq}
  \Psi_{k+1}=M(kh+z,E) \Psi_k,\quad \Psi_k\in \C^2,\quad k\in\Z,
\end{equation}
where $h=\frac{2\pi}{\varepsilon}\,{\rm mod}\, 1$,  and $(z,E)\to
M(z,E)$    is a matrix function
taking  values in $SL(2,\C)$  (the monodromy matrix, see
section~\ref{sec:Monodromy-matrix}). The asymptotic  of $M$  is
described in section~\ref{sec:mon-mat-as}; here we write down its
leading term. Assume additionally that 
$\tau^2(\bar E)\geq\min_{\nu\in\{0,\pi\}}t_{v,\nu}(\bar E)$. 
Then, $M$ has the most simplest asymptotic
(see Corollary~\ref{cor:M-matrix} and Remark~\ref{principal-terms}),
and,  for  $E\in I_0\cup I_\pi$, one has
\begin{equation}
  \label{model-mat}
  M(z,E)\sim\begin{pmatrix}
    \tau^2g_0(z)g_\pi(z)+\theta_n^{-1} &\tau g_0(z)
    \\{}\\\theta_n \tau g_\pi(z) & \theta_n
  \end{pmatrix},
\end{equation}
where 
\begin{gather*}
  g_\nu=\xi_\nu+\sin(2\pi (z- z_\nu)),\quad
  \nu\in\{0,\pi\},
\end{gather*} 
$\theta_n$ is the solution to $2\Lambda_n=\theta_n+ \theta^{-1}_n$
in $[1,+\infty[$, and $(z_\nu)_{\nu\in\{0,\pi\}}$ are constants.\\
The behavior of the solutions to~\eqref{model-eq} mimics that of those
to $H_{z,\varepsilon}\psi=E\psi$ in the sense of Theorem 2.1
from~\cite{MR2003f:82043}. Equation~\eqref{model-eq} in which the
matrix is replaced with its principal term is a model equation of our
system.  All the effects we have described can be seen when analyzing
this model equation.
\subsection{When $\tau$ is of order 1}
\label{sec:quand-tau-e}
When $\tau$ is of order 1, the principal term of the monodromy matrix
asymptotics is the one described by~\eqref{model-mat}.  If $\tau$ and
$|\xi_0|$ and $|\xi_\pi|$ are of order of $1$, the principal term does
not contain any asymptotic parameter.  This regime is similar to that
of the asymptotic Anderson transitions found in~\cite{MR2003f:82043}.
If at least one of the ``local variables'' $\xi_\nu(E)$ becomes large,
then, the spectrum can again be analyzed with the same precision as in
sections~\ref{sec:quand-tau-g} and~\ref{sec:quand-tau-p}.
\subsection{Numerical computations}
\label{sec:numer-comp}
\begin{figure}[h]
  \centering
  \includegraphics[bbllx=71,bblly=309,bburx=716,bbury=722,width=14cm]{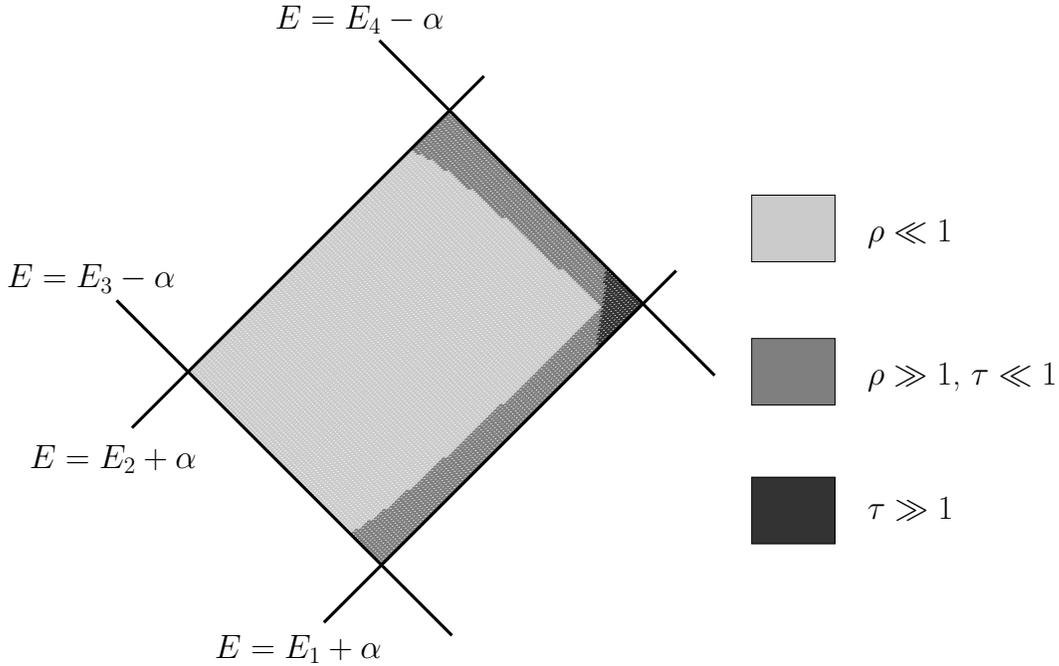}
  \caption{Comparing the coefficients $\tau$ and $\rho$ to $1$}
  \label{fig:tau}
\end{figure}

We now turn to some numerical results showing that the multiple
phenomena described in section~\ref{sec:plus-proche-de} do occur.\\
All these phenomena only depend on the values of the actions $S_h$,
$S_{v,0}$, $S_{v,\pi}$. We pick $V$ to be a two-gap potential; for such
potentials, the Bloch quasi-momentum $k$ (see
section~\ref{sec:le-quasi-moment}) is explicitly given by a
hyper-elliptic integral (\cite{MR52:11181,Ke-Mo:75}). The actions then
become easily computable. As the spectrum of $H_0=-\Delta+V$ only has
two gaps, we write $\sigma(H_0)=[E_1,E_2]\cup[E_3,E_4]\cup
[E_5,+\infty[$. In the computations, we take the values
\begin{equation*}
  E_1=0,\ E_2=3.8571,\ E_3=6.8571,\ E_4=12.1004,\text{
    and }E_5=100.7092.
\end{equation*}
On the figure~\ref{fig:tau}, we represented the part of the
$(\alpha,E)$-plane where the condition (TIBM) is satisfied for $n=1$.
Denote it by $\Delta$.  Its boundary consists of the straight lines
$E=E_1+\alpha$, $E=E_2+\alpha$, $E=E_3-\alpha$ and $E=E_4-\alpha$.\\
The computation shows that~(T) is satisfied in the whole of $\Delta$.
As $n=1$, one has $E_{2n-2}=-\infty$. It suffices to check~(T) for
$\zeta_{2n+3}=\zeta_5$. (T) can then be understood as a consequence of
the fact that $E_5-E_4$ is large.\\
On figure~\ref{fig:tau}, we show the zones where $\tau$ and $\rho$ are
large and small.
%
%
So, for carefully chosen $\alpha$, all the phenomena described in
section~\ref{sec:quand-tau-g},~\ref{sec:quand-tau-p}
and~\ref{sec:les-scen-princ}, that is in
figures~\ref{fig:tau_grand},~\ref{fig:rho_petit}
and~\ref{fig:tau_petit}, do occur.
\subsection{The outline of the paper}
The main idea of our analysis is to reduce the spectral study
of~\eqref{family} to the study of a finite difference equation (the
monodromy equation) the coefficients of which, in adiabatic limit,
take a simple asymptotic form. Therefore, we use the first step of a
renormalization procedure. Such renormalization procedure, called the
monodromization, was first suggested to study spectral properties of
the finite difference equations (on the real line) with periodic
coefficients in~\cite{MR96m:47060}. We have understood that this idea
can be generalized to study quasi periodic systems with two
frequencies and, in~\cite{MR2003f:82043}, we applied it to the
analysis of the differential Schr{\"o}dinger equation~\eqref{family}.\\
In section~\ref{sec:MM}, we recall the definition of the monodromy
matrix and the monodromy equation for the quasi-periodic Schr{\"o}dinger
equation. Then, we get the asymptotic of the monodromy matrix in the
adiabatic limit in the resonant case. Note that most of the technical
work was already done in~\cite{Fe-Kl:04a}, where we have got this
asymptotic in a general case; here, first, we analyze the error terms
in the general formula and show that, in the resonant case, one can
get for them much better estimates, second we show that, in the
resonant case, one can simplify the principal term of the
asymptotic.\\
In section~\ref{sec:big-tau}, we prove our spectral results for 
the case of big $\tau$. The analysis made here is quite standard:
similar calculations  can be found in~\cite{Fe-Kl:03f}
and~\cite{Fe-Kl:04a}.\\
In section~\ref{sec:small-tau}, we prove our results for the case of
small $\tau$. This is the most complicated case. Though that we
roughly follow the analysis preformed in~\cite{MR2003f:82043}
and~\cite{Fe-Kl:04a}, we have to develop some new ideas to be able
carry out rather delicate computations. This is due to two facts:
first, in the case of small $\tau$, one observes a rich set of new
spectral phenomena, and, second, one has to control simultaneously
several objects having different orders of exponential smallness in
$\varepsilon$. \\
In section~\ref{sec:les-scen-princ}, we analyze the results of the
previous section and describe the possible spectral scenarios for small
$\tau$.\\
The main goal of section~\ref{S3} is to study $\Lambda_n(V)$, the
quantity responsible for the gap between the two resonant intervals
containing spectrum. In, particular, we show that, generically, it
satisfies~(\ref{eq:41}).


%
\section{The monodromy matrix}
\label{sec:MM}
In this section, we first recall the definitions of the monodromy
matrix and of the monodromy equation for the quasi-periodic
differential equation
\begin{equation}
  \label{G.2z}
  -\frac{d^2}{dx^2}\psi(x)+(V(x-z)+\alpha\cos(\varepsilon x))\psi(x)=
   E\psi(x),\quad x\in\R,
\end{equation}
and recall how these objects related to the spectral theory of the
operator $H_{z,\varepsilon}$ defined in~\eqref{family}. In the second
part of the section, we describe a monodromy matrix for~\eqref{G.2z}
in the resonant case.
\subsection{The monodromy matrices and the monodromy equation}
\label{sec:monodr-matr-monodr}
We now follow~\cite{MR2002h:81069,MR2003f:82043}, where the reader can
find more details, results and their proofs.
\subsubsection{The definition of the monodromy matrix}
\label{sec:Monodromy-matrix}
For any $z$ fixed, let $(\psi_j(x,\,z))_{j\in\{1,2\}}$ be two linearly
independent solutions of equation~\eqref{G.2z}.  We say that they form
a {\it consistent basis} if their Wronskian is independent of $z$,
and, if for $j\in\{1,2\}$ and all $x$ and $z$,
\begin{equation}
  \label{consistency}
  \psi_j(x,\,z+1)=\psi_j(x,\,z).
\end{equation}
As $(\psi_j(x,\,z))_{j\in\{1,2\}}$ are solutions to
equation~\eqref{G.2z}, so are the functions
$((x,z)\mapsto\psi_j(x+2\pi/\varepsilon,z+2\pi/\varepsilon))_{j\in\{1,2\}}$.
Therefore, one can write
\begin{equation}
  \label{monodromy}
  \Psi\,(x+2\pi/\varepsilon,z+2\pi/\varepsilon)= M\,(z,E)\,\Psi\,(x,z),\quad
  \Psi(x,z)=\begin{pmatrix}\psi_{1}(x,\,z)\\ \psi_{2}(x,z)\end{pmatrix},
\end{equation}
where $M\,(z,E)$ is a $2\times 2$ matrix with coefficients independent of
$x$. The matrix $M(z,E)$ is called {\it the monodromy matrix}
corresponding to the basis $(\psi_j)_{j\in\{1,2\}}$. To simplify the
notations, we often drop the $E$ dependence when not useful.
\smallpagebreak For any consistent basis, the monodromy matrix
satisfies
\begin{equation}
  \label{Mproperties}
  \det M\,(z)\equiv 1,\quad
  M\,(z+1)=M\,(z),\quad \forall z.
\end{equation}
\subsubsection{The monodromy equation and the link with the spectral
  theory of $H_{z,\varepsilon}$}
\label{sec::Mon-eq}
Set
\begin{equation}\label{h}
h=\frac{2\pi}{\varepsilon}\,{\rm mod}\, 1.
\end{equation}
Let $M$ be the monodromy matrix corresponding to the consistent basis
$(\psi_{j})_{j=1,2}$. Consider the {\it monodromy equation}
\begin{equation}
  \label{Mequation}
  F(n+1)=M\,(z+nh)F(n),\quad\quad F(n)\in \C^2,\quad \forall n\in\Z.
\end{equation}
The spectral properties of $H_{z,\varepsilon}$ defined
in~\eqref{family} are tightly related to the behavior of solutions
of~\eqref{Mequation}. This follows from the fact that the behavior of
solutions of the monodromy equation for $n\to\pm \infty$ mimics the
behavior of solutions of equation~\eqref{family} for $x\to\mp\infty$,
see Theorem 3.1 from~\cite{MR2003f:82043}. \\
\subsubsection{Relations between the equations family~\eqref{family}
and the monodromy equation}     
Here,  we describe only two consequences from Theorem 3.1
from~\cite{MR2003f:82043}; more examples will be given in the course
of the paper. One has
\begin{Th}
  \label{Resolvent-set}
  Fix $E\in\R$. Let $h$ be defined by~\eqref{h}.  Let $z\mapsto
  M(z,E)$ be a monodromy matrix for equation~\eqref{G.2z}
  corresponding to a basis of consistent solutions that are locally
  bounded in $(x,z)$ together with their derivatives in $x$.\\
  Suppose
  that the monodromy equation has two linearly independent solutions
  $(\chi_+,\chi_-)$ such that, for some $C>0$, for $n\in \N$, one has
  $\|\chi_+(n)\|+\|\chi_-(-n)\|\le C e^{- n/C}$.
  Then, $E$ belongs to the resolvent set of the operator
  $H_{z,\varepsilon}$.
\end{Th}
\noindent The proof of this theorem mimics the proof of Lemma 4.1
in~\cite{MR2003f:82043}.\\
The second result we now present is the relation between the Lyapunov
exponents of the family of equations~\eqref{family} and of the
monodromy equation.
\smallpagebreak Recall the definition of the Lyapunov exponent for a
matrix cocycle.  Let $z\mapsto M(z)$ be an $SL(\C,\,2)$-valued
$1$-periodic function of the real variable $z$.  Let $h$ be a positive
irrational number. The Lyapunov exponent for the {\it matrix cocycle}
$(M,h)$ is the limit (when it exists)
\begin{equation}
  \label{eq:44}
  \theta(M,h)=
  \lim_{L\to+\infty}\frac{1}{L}\log\Vert M(z+Lh)\cdot M(z+(L-1)h)\cdots
  M(z+h)\cdot M(z)\Vert.
\end{equation}
Actually, if $M$ is sufficiently regular in $z$ (say, belongs to
$L^\infty$), then $\theta(M,h)$ exists for almost every $z$ and does
not depend on $z$, see e.g.~\cite{MR94h:47068}.\\
One has
\begin{Th}[\cite{MR2002h:81069}]
  \label{Lyapunov}
  Let $h$ be defined by~\eqref{h}.  Let $z\mapsto M(z,E)$ be a
  monodromy matrix for equation~\eqref{G.2z} corresponding to a basis
  of consistent solutions that are locally bounded in $(x,z)$
  together with their derivatives in $x$.  \ 
  The Lyapunov exponents $\Theta(E,\varepsilon)$ and
  $\theta(M(\cdot,E),h)$ satisfy the relation
\begin{equation}
  \label{eq:Lyapunov}
  \Theta(E,h)=\frac\varepsilon{2\pi}\theta(M(\cdot,E),h).
\end{equation}
\end{Th}
\subsection{Asymptotics of the monodromy matrix}
\label{sec:mon-mat-as}
Consider the sequences $(E_0^{(l)})_l$ and $(E_\pi^{(l')})_{l'}$
defined by the quantization conditions~\eqref{eq:11}.  Let $E_\pi$ be
one of the points $(E_\pi^{(l')})_{l'}$, and let $E_0$ be the point
from the sequence $(E_0^{(l)})_l$ closest to $E_\pi$. Define
\begin{equation}
  \bar E=\frac{E_0+E_\pi}2\quad\text{and}\quad\Delta=\frac{E_0-E_\pi}2.
\end{equation}
We assume that $E_0$ and $E_\pi$ are resonant, i.e., that they satisfy
\begin{equation}
  \label{Delta}
  |\Delta|\le 2e^{-\frac{\delta_0}{\varepsilon}}.
\end{equation}
We now describe the asymptotic of the monodromy matrix for the family
of equations~\eqref{family} for (complex) energies $E$ such that
\begin{equation}
  \label{neighb}
  |E-\bar E|\le 4e^{-\frac{\delta_0}{\varepsilon}},\quad 
\end{equation}
We shall use the following notations:
\begin{enumerate}
\item the letter $C$ denotes various positive constants independent of
  $z$, $E$, $E_\pi$, $E_0$ and $\varepsilon$;
\item the symbol $O(f_1, f_2, \dots f_n)$ denotes functions satisfying
  the estimate
$  |O(f_1, f_2, \dots f_n)|\le C(|f_1|+|f_2|+\dots |f_n|)$.
\item when writing $f\asymp g$, we mean that there exists $C>1$ such
  that $C^{-1} |g|\le |f|\le C|g|$ for all $\varepsilon$, $\zeta$,
  $E$, $E_\pi$, $E_0$ in consideration.
\item when writing $f=o(g)$, we mean that there exists
  $\varepsilon\mapsto c(\varepsilon)$, a function such that
\begin{itemize}
\item $|f|\le c(\varepsilon) |g|$ for all $\varepsilon$, $\zeta$, $E$,
  $E_\pi$ and $E_0$ in consideration;
\item $c(\varepsilon)\to0$ when $\varepsilon\to 0$.
\end{itemize} 
\end{enumerate}
We let
\begin{equation}
  \label{p}
  p(z)=e^{2\pi |\im z|}.
\end{equation}
Recall that  $\delta_0$ is the constant defined in~\eqref{eq:6}.
One has
\begin{Th}
  \label{th:M-matrix} 
  Pick $E_*\in J$. There exists $V_*$, a neighborhood of $E_*$, such
  that, for sufficiently small $\varepsilon$,
  there exists a consistent basis of solutions of~\eqref{G.2z} for
  which the monodromy matrix $(z,E)\mapsto M^\pi(z,E)$ is analytic in
  the domain $ \left\{z\in \C;\,2\pi |\im z|<\frac{\delta_0}\varepsilon
  \right\}\times V_*^\varepsilon$ where $V_*^\varepsilon=\{E\in V_*; \ |\im
  E|\le \varepsilon\}$.
  Its coefficients take real values for real $E$ and $z$. Fix
  $J_*\subset V_*\cap \R$, a compact interval. If $E_\pi\in J_*$ and
  satisfies~\eqref{Delta}, then, in the domain
  \begin{equation}
    \label{analyticity:dom:a}
    \left\{z\in \C;\,2\pi|\im z|<\frac{\delta_0}\varepsilon\right\}\times 
    \left\{E\in \C;\,  |E-E_\pi|<
      4e^{-\frac{\delta_0}{\varepsilon}}\right\},
  \end{equation}
  one has
  \begin{equation}
    \label{mm-as}
    M^\pi=\sigma \begin{pmatrix} \tau^2 g_0 g_\pi+\bar\theta^{-1} &
      r\tau g_0\\\frac {\bar\theta \tau}r g_\pi & \bar\theta
    \end{pmatrix}+
    e^{-\frac{\delta}\varepsilon}\,
    \begin{pmatrix} O\left(\tau^2|g|_0|g|_\pi, p\right) &  O\left(r\tau\,|g|_0,p\right)
      \\O\left(\frac\tau{r} |g|_\pi p,\,\bar T_h p\right) & O\left(p\right)
    \end{pmatrix}.
  \end{equation}
  where $0<\delta<\delta_0$ is a constant depending only on $J$,
  $\sigma$ is constant in $\{+1,-1\}$, and, for $\nu\in\{0,\pi\}$, one
  has
  \begin{gather}
    \label{g-nu}
    g_{\nu}(z,E)=\xi_\nu(E)+\sin(2\pi(z-z_\nu)), \quad
    |g|_\nu=|\xi_\nu|+p,\\
    \label{xi-nu}
    \xi_{\nu}(E)=\gamma_\nu\cdot(E-E_\nu).
  \end{gather}
  Furthermore, $\tau$, $r$, $\gamma_\nu$, $\bar \theta$ and $z_\nu$
  are real constants (independent of $z$ and $E$), one has
  \begin{gather}
    \label{tau:ext}
    \tau=2\sqrt{\frac{\bar T_{v,0}\,\bar T_{v,\pi}}{\bar T_h}},\quad
    r=\frac{ \tau}{\bar T_{v,\pi}},\\
    \label{gamma:ext}
    \gamma_\nu=\frac{\check\Phi'_\nu(\bar E)}{\varepsilon\cdot \bar
      T_{v,\nu}},\\
    \label{theta-bar}
    \bar\theta=\theta_n(V)(1+o(1)),
  \end{gather}
  where $\theta_n(V)$ is the positive constant depending only on $n$
  and $V$ defined in~\eqref{theta-omega}. and
  \begin{equation}
    \label{T-bars}
    \bar T_h=t_h(\bar E)\,(1+o(1)),\quad \bar T_{v,0}=t_{v,0}(\bar
    E)\,(1+o(1)),\quad
    \bar T_{v,\pi}=t_{v,\pi}(\bar E)\,(1+o(1)),
  \end{equation}
\end{Th} 
\noindent We prove this theorem in section~\ref{sec:Proof:mm-as}.
\begin{Rem}
  \label{rem:1}
  We shall also need a more detailed description of the error terms
  for the coefficient $M_{11}^\pi$. The proof of
  Theorem~\ref{th:M-matrix} (see the part of the proof in
  section~\ref{sec:asympt-monodr-matr} also yields
  \begin{equation}
    \label{M11:details}
    M_{11}^\pi=\sigma\tau^2 g_0 g_\pi+\bar\theta^{-1}+
    e^{-\delta/\varepsilon}\,
    \left(\tau^2\left[ O(\xi_0\xi_\pi) +O(\xi_0  p)+O(\xi_\pi
        p)+O(p^2)\right]+ O(p)\right),
  \end{equation}
  where all the error terms are analytic in $E$ and in $z$.
\end{Rem}
\begin{Rem}
  \label{rem:2}
  Being a consequence of Theorem 2.2 of\cite{Fe-Kl:04a},
  Theorem~\ref{th:M-matrix} stays valid if one swaps the indexes $0$
  and $\pi$, and the quantities $\theta$ and $1/\theta$. One thus
  obtains $M^0$, a different monodromy matrix. Though that all the
  spectral results we have announced can be obtained directly by
  analyzing the monodromy equation with the matrix $M^\pi$ from
  Theorem~\ref{th:M-matrix}, to simplify the analysis, we use from
  time to time the monodromy matrix $M^0$ instead of $M^\pi$.
\end{Rem}
\begin{Rem}
  \label{notations} 
  The notation $\tau$ (resp. $\xi_\nu$) in sections~\ref{sec:intro}
  and~\ref{sec:MainResults} and the notation $\tau$ (resp. $\xi_\nu$)
  in the rest of the paper denote different quantities. They differ by
  a constant factor of the form $1+o(1)$ (when $\varepsilon\to0$)!
\end{Rem}
\noindent Let us mention a case when, preserving good error estimates,
one can get a more symmetric monodromy matrix:
\begin{Cor}
  \label{cor:M-matrix}
  In the case of Theorem~\ref{th:M-matrix}, assume that $\D \tau^2\ge
  \min_{\nu\in \{0,\pi\}}\{\bar T_{v,\nu}\}$.
  Then, there exists a consistent basis of solutions of~\eqref{G.2z}
  for which the monodromy matrix $(z,E)\mapsto \tilde M^\pi(z,E)$ is analytic in
  the domain $ \left\{z\in \C;\,|\im z|<\frac{\delta_0}\varepsilon
  \right\}\times V_*^\varepsilon$;
  its coefficients take real values for real $E$ and $z$. Fix
  $K\subset V_*\cap \R$ compact. If $E_\pi$ belongs to $K$ and
  satisfies~\eqref{Delta}, then, in the
  domain~\eqref{analyticity:dom:a}, one has
  \begin{equation}
    \label{new-Mmatrix:as}
    \tilde M^\pi=\sigma \begin{pmatrix} \tau^2 g_0 g_\pi+\bar\theta^{-1} &
      \tau g_0 \\\bar\theta \tau g_\pi & \bar\theta
    \end{pmatrix}+  e^{-\frac{\delta}\varepsilon}\,
    \begin{pmatrix} O(\tau^2|g|_0|g|_\pi,\,p) & O(\tau |g|_0)
      \\O(p\tau |g|_\pi) & O(p)
   \end{pmatrix}.
 \end{equation}
\end{Cor}
\begin{Rem}\label{principal-terms} The principal term
  in~(\ref{model-mat}) is obtained from~\eqref{new-Mmatrix:as} by
  replacing $\tau$, $\bar\theta$ and $g_\nu$ (more precisely, the
  quantity $\xi_\nu$ from the definition of $g_\nu$) by the principal
  terms of their asymptotics (see also Remark~\ref{notations}).
\end{Rem}
\demo Let $(f_1, f_2)$ be the consistent basis for which the monodromy
matrix is given by Theorem~\ref{th:M-matrix}. Then $r^{-1/2} f_1$ and
$r^{1/2} f_2$ form a consistent basis; the corresponding monodromy
matrix is
\begin{equation*}
  \tilde M^\pi=\begin{pmatrix} r^{-1/2} & 0 \\ 0 &
    r^{1/2} \end{pmatrix} M^\pi \begin{pmatrix} r^{1/2} & 0 \\ 0 &
    r^{-1/2} \end{pmatrix}.
\end{equation*}
The asymptotic representation~\eqref{new-Mmatrix:as} follows
from~\eqref{mm-as} and the observations that
\begin{equation*}
  r^{-1}p(z)=\frac{\bar T_{v,\pi}}\tau\, p(z)\le \tau |g|_0,\quad\quad
  \text{and}\quad\quad  
  r\bar T_{h}p(z)= \frac{4r\bar T_{v,0}\bar T_{v,\pi}}{\tau^2} p(z)=
  \frac{4\bar T_{v,0}}{\tau}  p(z)\le 4 \tau |g|_\pi.
\end{equation*}
Thus, Corollary~\ref{cor:M-matrix} is proved.\qed
\subsection{Proof of Theorem~\ref{th:M-matrix}}
\label{sec:Proof:mm-as}
The matrix $M^\pi$ is introduced in section 3.3 in~\cite{Fe-Kl:04a}
($M^\pi$ was denoted by $M^U$); there, its asymptotics are described
in Theorem 3.1. Under the hypothesis~\eqref{Delta} and~\eqref{neighb},
we improve the estimates of the error terms in the asymptotic of the
coefficient $M^\pi_{12}$ and simplify the leading terms of the
asymptotics of all the coefficients of $M^\pi$.
This will give~\eqref{mm-as}.\\
We shall use the following notations. For $(z_1,\cdots,z_n)\mapsto
g(z_1,\cdots,z_n)$, an analytic function, we define
\begin{equation}
  \label{star}
  g^*(z_1,\cdots,z_n)=\overline{g(\overline{z_1},\cdots,\overline{z_n})}.
\end{equation}
For $Y>0$, we let
\begin{equation}
  \label{TY}
  T_Y=e^{-2\pi Y/\varepsilon}.
\end{equation}
\subsubsection{General asymptotic representation for the monodromy
  matrix $M^\pi$}
\label{sec:gener-asympt-repr}
Here, we give the asymptotics of $M^U$ valid without the
hypotheses~\eqref{Delta} and~\eqref{neighb}.  We check
\begin{Pro} 
  \label{new-M} Pick $E_*\in J$.
  There exists $V_*$, a complex neighborhood of $E_*$, such that, for
  sufficiently small $\varepsilon$, the following holds.  Let
                    in~\eqref{new-Mmatrix:as},             %
  \begin{equation}
    \label{Y's}
    Y_m=\frac1{2\pi}\inf_{E\in J\cap V_*}\min(S_{0,\nu}(E),
    S_{\pi,\nu}(E))
    ,\quad\quad Y_M=\frac1{2\pi}\sup_{E\in J\cap V_*} \max(S_{v,0}(E),
    S_{v,\pi}(E), S_h(E)).
  \end{equation}
  Fix $0<y<Y_m$. There exist $Y>Y_M$ and a consistent basis of
  solutions of~\eqref{G.2z} for which the monodromy matrix
  $(z,E)\mapsto M^U(z,E)$ is analytic in the domain $\left\{z\in
    \C:\,|\im z|<\frac{y}\varepsilon\right\}\times V_*^\varepsilon$. Its
  coefficients are real analytic. One has
  \begin{equation}
    \label{M-new:as}
    M^\pi(z,E)= P(z,E)+Q(z,E)+R(z,E),
  \end{equation}
  where
  \begin{equation}
    \label{P}
    P(z,E)=\frac4{T_h}\,\begin{pmatrix} \tilde C_\pi(z,E)\,C_0(z,E) &
      -\tilde S_\pi(z,E) C_0(z,E)\\0 & 0\end{pmatrix},
  \end{equation}
  \begin{equation}
    \label{Q}
    Q(z,E)=\begin{pmatrix} \frac1\theta\,\cos\frac{\check\Phi_\pi-
        \check\Phi_0}\varepsilon\,+\theta\,\cos\frac{\check\Phi_\pi}
        \varepsilon\,\cos\frac{\check\Phi_0}\varepsilon& 
        -\frac1\theta\,\sin\frac{\check\Phi_\pi-\check\Phi_0}
        \varepsilon- \theta\,\sin\frac{\check\Phi_\pi}\varepsilon
        \,\cos\frac{\check\Phi_0}\varepsilon \\
        -\theta \sin\frac{\check\Phi_0}\varepsilon\, \tilde C_\pi(z) &
        \theta\,\sin\frac{\check\Phi_\pi}\varepsilon\,
        \sin\frac{\check\Phi_0}\varepsilon
    \end{pmatrix},
  \end{equation}
  and
  \begin{gather}\label{R}
    R_{11}(z), \ R_{12}(z), \ R_{22}(z) \ = \ O\left(T_h,\,p(z)
      T_YT_h^{-1} ,\, p(z)T_{v}\right),\\
     \label{R21}
     \begin{split}
       R_{21}(z)
     &
       = \cos\frac{\check\Phi_\pi}\varepsilon\,O\left(T_h,\,
         p(z) T_YT_h^{-1}, p(z)T_{v}\right)+
       \cos\frac{\check\Phi_0}\varepsilon\,O\left(T_h\right)
     \\&\hskip6cm
       +O\left(pT_Y,\, pT_vT_h,\, T_h^2,\,p^2T_vT_{v,\pi},\,
         p^2T_{v,\pi}T_YT_h^{-1}\right).
     \end{split}
  \end{gather}
  In these formulae,
  \begin{gather}
    \label{CS-pi}
    \tilde C_\pi=\frac{\tilde\alpha_\pi
      e^{i\check\Phi_\pi/\varepsilon}+ \tilde\alpha_\pi^*
      e^{-i\check\Phi_\pi/\varepsilon}}2,\quad \tilde
    S_\pi=\frac{\tilde\alpha_\pi e^{i\check\Phi_\pi/\varepsilon}-
      \tilde\alpha_\pi^*e^{-i\check\Phi_\pi/\varepsilon}}{2i}, \\
    \label{C0}
    C_0=\frac{\alpha_0 e^{i\check\Phi_0/\varepsilon}+ \alpha_0^*
      e^{-i\check\Phi_0/\varepsilon}}2.
  \end{gather} 
  The functions $(z,E)\mapsto\alpha_0(z,E)$ and $(z,E)\mapsto\tilde
  \alpha_\pi (z,E)$ are analytic in $\left\{z\in \C:\,|\im
    z|<\frac{y}\varepsilon\right\}\times V_*^\varepsilon$; they are
  $1$-periodic in $z$ and admit the asymptotics
  \begin{gather}
    \label{alpha:as:0}
    \alpha_0(z,\theta) =1+ T_{v,0} e^{2\pi i(z-z_0(E))}+O\left(T_Y\,
      p(z)\right)=1+O\left(T_{v,0}\, p(z)\right)=1+o(1),\\
    \label{alpha:as:pi}
    \begin{split}
      \tilde\alpha_\pi(z,\theta)
      =&1+T_{v,\pi}\,\left[\cos(2\pi(z-z_\pi))+
        i\sin(2\pi(z-h-z_\pi))\right]+O(p^2(z)T_{v,\pi}^2,\,p(z)T_Y)\\
      =& 1+O\left(T_{v,\pi}\, p(z)\right)=1+o(1).
    \end{split}
  \end{gather}  
  The functions $E\mapsto\check\Phi_\nu(E)$, $E\mapsto T_{v,\nu}(E)$,
  $E\mapsto z_{\nu}(E)$ and $E\mapsto T_h(E)$, $E\mapsto\theta(E)$ are
  real analytic in $V_*$; they are independent of $z$. In $V_*$, one
  has
  \begin{gather}
    \label{check-Phi:as}
    \check\Phi_\nu(E)=\Phi_\nu(E)+o(\varepsilon), \\
    \label{T:as}
    T_h(E)=t_h(E)(1+o(1)),\quad T_{v,\nu}(E)=t_{v,\nu}(E)\,(1+o(1)),
  \end{gather}
  where $\Phi_\nu$ and $t_h$, $t_{v,\nu}$ are the phase integrals and
  the tunneling coefficients defined in the introduction;
  \begin{equation}
    \label{theta:as}
    \theta(E)=\theta_n(V)\,(1+o(1)),
  \end{equation}
  where $\theta_n(V)$ is the constant defined in~\eqref{theta-omega}
  (it is positive and depends only on $n$ and $V$);
  \begin{equation} 
    \label{z-nu-prime}
    z_\nu'(E)=O(1/\varepsilon).
  \end{equation}
  In all the above formulae, all the error terms are analytic in $E$
  and $z$.
  Finally, in the error term estimates, $T_v(E)=\max\{T_{v,0}(E),\,
  T_{v,\pi}(E)\}$.
\end{Pro}
\begin{Rem} Proposition~\ref{new-M} stays valid if one swaps the
  indexes $0$, $\pi$, and the quantities $\theta$, $1/\theta$.
\end{Rem}
\noindent Proposition~\ref{new-M} differs from Theorem 3.1
from~\cite{Fe-Kl:04a} by more precise estimate of the coefficient
$R_{21}$, and, an additional information provided by Theorem 2.2,
formula (2.26) and Lemma 3.4 in~\cite{Fe-Kl:04a}. So, to prove
Proposition~\ref{new-M}, we have only to check the estimate for
$R_{12}$.
\smallpagebreak\noindent
{\it Proof of the estimate for $R_{21}$.\/} \  
We now analyze the structure of the term $R_{21}$ in
detail. Therefore, recall the description of the matrix $M^\pi$ given in
section 3.2 of~\cite{Fe-Kl:04a} (it was denoted by $M^U$). It has the form
\begin{equation}
    \label{M-gauge}
    M^\pi(z)=U(z+h)\,\begin{pmatrix} A_\pi(z) & B_\pi(z) \\ 
    B^*_\pi (z) & A^*_\pi(z)\end{pmatrix}\,U(z)^{-1},\quad
    U(z)=U(z)=\frac12\begin{pmatrix}1&1\\ -i & i\end{pmatrix}\,
  \begin{pmatrix}\gamma(z) & 0\\ 0& \gamma^*(z)\end{pmatrix},
\end{equation}
where $\gamma(z+h)=\sqrt{{\dsize\frac{\alpha_\pi^*(z)}
    {\alpha_\pi(z)}}}\, e^{-i\check\Phi_\pi/\varepsilon}$.  Here and
below, we often drop the dependence on $E$.\\
The function $(z,E)\mapsto\alpha_\pi(z,E)$ is analytic in $\left\{z\in
  \C:\,|\im z|<\frac{y}\varepsilon\right\}\times V_*^\varepsilon $; it is
$1$-periodic in $z$ and admits the asymptotic representations (see
Theorem 2.2 and formula (2.26) in~\cite{Fe-Kl:04a})
\begin{equation}
  \label{alpha:as:pi-old}
  \alpha_\pi =1+ T_{v,\pi} e^{2\pi i(z-z_\pi)}+O\left(T_Y\,
    p(z)\right)=1+O(p(z)T_{v,\pi})=1+o(1).
\end{equation}
The branch of the square root in the definition of $\gamma$ is
chosen so that $\gamma(z)=e^{-i\check\Phi_\pi/\varepsilon}(1+o(1))$.\\
The coefficients $A_\pi$ and $B_\pi$ are described by the formulae
\begin{equation}
  \label{A-pi}
  A_\pi=2\,\frac{\alpha_\pi
  e^{i\frac{\check\Phi_\pi}\varepsilon}\,C_0}{T_h}+
  \alpha_\pi\,\alpha_0^*\,e^{i\frac{\check\Phi_\pi-
      \check\Phi_0}\varepsilon}\,\left\{
    \frac{\theta+1/\theta}2+\frac{T_h}4+\frac{O_\pi+O_0^*}{T_h}+
    \frac{O_\pi/\theta+O_0^*\theta}2+\frac{O_\pi O_0^*}{T_h}\right\},
\end{equation}
and
\begin{equation}
  \label{B-pi}
  B_\pi=2\,\frac{\alpha_\pi
  e^{i\frac{\check\Phi_\pi}\varepsilon}\,C_0}{T_h}+
  \alpha_\pi\,e^{i\frac{\check\Phi_\pi}\varepsilon}\,
  \left\{\frac{\alpha_0
      e^{i\frac{\check\Phi_0}\varepsilon}/\theta+\alpha_0^* 
      e^{-i\frac{\check\Phi_0}\varepsilon}\theta}2+
    \frac{\alpha_0e^{i\frac{\check\Phi_0}\varepsilon}O_0+
          \alpha_0^*e^{-i\frac{\check\Phi_0}\varepsilon}
      O_\pi}{T_h}\right\}.
\end{equation}
where, for $\nu\in\{0,\pi\}$, the factor $O_\nu$ satisfies the
estimate (see formulae (5.18) and (1.10) in~\cite{Fe-Kl:04a})
\begin{equation}
 \label{O-nu}
     O_\nu=O(T_{h}^2,\, T_Y\,p(z),\, T_{h} T_{v,\nu} p(z)).
\end{equation}
Note that formulae~\eqref{A-pi} and~\eqref{B-pi} are respectively the
formula (5.52) and, up to a constant factor, the formula (5.53)
in~\cite{Fe-Kl:04a}. The constant factor is omitted as it can be
removed by conjugation (see the explanations in the last lines of
section 5.3.2 in~\cite{Fe-Kl:04a}).\\
Finally, we recall that the functions $\alpha_\pi$ and $\tilde
\alpha_\pi$ are related by the formula
\begin{equation}
  \label{tilde-alpha-alpha}
  \tilde\alpha_\pi(z)={\dsize\frac{\gamma(z+h)}{\gamma(z)}}\,
  \alpha_\pi(z)
\end{equation}
which is formula (3.16) in~\cite{Fe-Kl:04a}.\\
Now, we are ready to prove the estimate~\eqref{R21} for $R_{21}$.
For an analytic function $f$, we let $\Im(f)=\frac{f-f^*}{2i}$.
Representation~\eqref{M-gauge} implies that
\begin{equation*}
  M^\pi_{21}=\Im(D),\quad\text{where}\quad 
  D=\frac{\gamma(z+h)}{\gamma(z)}\,\left[A_\pi(z)-
    e^{\frac{2i\check\Phi_\pi}\varepsilon}
    \frac{\alpha_\pi(z)}{\alpha_\pi^*(z)}B^*_\pi(z)\right].
\end{equation*} 
Substituting representations~\eqref{A-pi} and~\eqref{B-pi} into this
formulae, we get
\begin{equation*}
  M^\pi_{21}=c_0+c_1+c_2+c_3+c_4,
\end{equation*}
where
\begin{gather*}
  c_0=-\theta\,\tilde C_\pi\,\Im\left(\alpha_0
    e^{\frac{i\check\Phi_0}\varepsilon}\right),\quad
  c_1=\frac{T_h}4\,\Im\left(\tilde \alpha_\pi \alpha_0^*
    e^{\frac{i(\check\Phi_\pi-\check\Phi_0)}\varepsilon}\right), \quad
  c_2=-\frac{2\tilde C_\pi}{T_h}\,\Im\left(\alpha_0
    e^{\frac{i\check\Phi_0}{\varepsilon}}O_\pi^*\right),\\
  c_3=\frac12\Im\left( \alpha_0^*\tilde\alpha_\pi
    e^{\frac{i(\check\Phi_\pi-\check\Phi_0)}\varepsilon}
    (O_\pi/\theta+O_0^*\theta)\right), \quad\quad
  c_4=\frac1{T_h}\Im\left(\alpha_0^*\tilde\alpha_\pi
    e^{\frac{i(\check\Phi_\pi-\check\Phi_0)}\varepsilon} O_\pi
    O_0^*\right).
\end{gather*}
Using the representations~\eqref{alpha:as:0} and~\eqref{alpha:as:pi}
for $\alpha_0$ and $\tilde\alpha_\pi$, we get
\begin{gather*}
  c_0=-\theta\sin\frac{\check\Phi_0}\varepsilon\,\tilde C_\pi +
  \cos\frac{\check\Phi_\pi}\varepsilon\,O(p T_{v,0})+O(p^2
  T_{v,0}T_{v,\pi}),\\
  c_1=\sin\frac{\check\Phi_\pi-\check\Phi_0}\varepsilon\,T_h/4+O(pT_v
  T_h,p^2T_v^2 T_h)= \cos\frac{\check\Phi_\pi}\varepsilon\,O(T_h)+
  \cos\frac{\check\Phi_0}\varepsilon\,O(T_h)+ O(pT_v T_h),
\end{gather*}
where, in the last step, we have also used that $O(pT_v)=o(1)$ for
$|\im\zeta|\le y<Y_m$.  In view of the estimate~\eqref{O-nu} for
$O_\pi$, we get also
\begin{equation*}
  c_2=\cos\frac{\check\Phi_\pi}\varepsilon\,O\left(T_h,\, p T_Y/T_h,\,p
    T_{v,\pi} \right)+ O\left(pT_{v,\pi}T_h,\, p^2
    T_{v,\pi}T_Y/T_h,\,p^2 T_{v,\pi}^2 \right).
\end{equation*}
Finally, as $|\alpha_0|+|\tilde\alpha_\pi|=O(1)$, in view
of~\eqref{O-nu} and as $|\im z|\le y<Y_M<Y$, we get
\begin{equation*}
c_3=O\left(T_h^2,\, p T_Y,\,p T_v T_h \right),\quad 
c_4=O\left(T_h^3, p^2T_Y^2/T_h, p^2 T_v^2 T_h\right).  
\end{equation*}
These estimates imply the announced representation for $M^\pi_{21}$, and
complete the proof of the estimate for $R_{21}$.  \qed
\subsubsection{Asymptotics of the monodromy matrix in the resonant case}
\label{sec:asympt-monodr-matr}
We now simplify the asymptotics for the coefficients of the monodromy
matrix given by Proposition~\ref{new-M} in the resonant case.\\
In this section, $J_*$ always denotes a compact interval in $V_*\cap J$.\\
For $\nu\in\{0,\pi\}$, we let
\begin{equation}
  \label{sigma-def}
  \sigma_\nu=-\sin\frac{\check\Phi_\nu(E_\nu)}\varepsilon,\quad
  \text{and}\quad \sigma=\sigma_0\sigma_\pi.
\end{equation}
Note that, by the definition of $E_\nu$, one has
$\sigma_\nu\in\{+1,-1\}$, and $\sigma\in\{+1,-1\}$.\\
Define
\begin{equation}
  \label{bar-T-def}
  \bar T_{v,\nu}=T_{v,\nu}(\bar E), \quad \bar T_{h}=T_{h}(\bar E),
  \quad\text{and}\quad \bar \theta=\theta(\bar E).
\end{equation}
Clearly, these quantities satisfy~\eqref{T-bars} and~\eqref{theta-bar}.\\
 First, for later use, we recall
\begin{Le}[\cite{Fe-Kl:04a}, Lemma 2.1]
  \label{le:Phi'}
  There exists
  a neighborhood of $J_*$, say $\tilde V_*$, and $C>0$ such that, for
  sufficiently small $\varepsilon$, for $E\in\tilde V_*$ and
  $\nu\in\{0,\pi\}$, one has
  \begin{equation*}
    |\check\Phi_\nu'(E)|+|\check\Phi_\nu''(E)|\le C,
  \quad                              %
  {\rm and} 
  \quad
    \frac1C\le |\check\Phi_\nu'(E)|.
  \end{equation*}
\end{Le}
\noindent 
Now, we check two simple lemmas.
\begin{Le}
  \label{cos-sin} Pick $\nu\in \{0,\pi\}$. Fix $C>0$.  For $\varepsilon$
  sufficiently small, if $E_\nu\in J_*$ and 
  $|E-E_\nu|\le Ce^{-\delta_0/\varepsilon}$, one has
  \begin{gather}
    \label{cos-1}
    \cos\frac{\check\Phi_\nu(E)}\varepsilon=\sigma_\nu \frac{\check
      \Phi_\nu'(\bar E)}\varepsilon\cdot (E-E_\nu)\left(1+
    O\left(\varepsilon^{-1}e^{-\delta_0/\varepsilon}\right)\right)
    \\
    \label{sin}
    \sin\frac{\check\Phi_\nu(E)}\varepsilon=-\sigma_\nu+
    O\left(\varepsilon^{-2}e^{-2\delta_0/\varepsilon}\right).
  \end{gather} 
\end{Le}
\demo Both estimates follow from Lemma~\ref{le:Phi'} and the
definition of $E_\nu$. We note only that to get~\eqref{sin} one
uses~\eqref{cos-1}. This completes the proof of
Lemma~\ref{cos-sin}. \qed\\
Estimate~\eqref{cos-1} implies that
\begin{equation}
  \label{cos-1-1}
  \cos\frac{\check\Phi_\nu(E)}\varepsilon= \sigma_\nu \bar T_{v,\nu}
  \xi_\nu\left(1+
    O\left(\varepsilon^{-1}e^{-\delta_0/\varepsilon}\right)\right),
  \quad\text{and}\quad
  \left|\cos\frac{\check\Phi_\nu(E)}\varepsilon\right|\le
  C\varepsilon^{-1}e^{-\delta_0/\varepsilon},
\end{equation}
where  $\xi_\nu$ are the local variables  defined in~(\ref{xi-nu}).\\
We now prove
\begin{Le}
  \label{le:T-bar-T}  Let $\bar E\in J_*$. For $\varepsilon$
  sufficiently small, for $E$ satisfying~(\ref{neighb}), one has
  \begin{gather*}
    T_h(E)=\bar
    T_h\,(1+O(\varepsilon^{-1}e^{-\frac{\delta_0}{\varepsilon}})),\quad
    T_{v,0}(E)=\bar
    T_{v,0}\,(1+O(\varepsilon^{-1}e^{-\frac{\delta_0}{\varepsilon}})),\quad
    T_{v,\pi}(E)=\bar T_{v,\pi}\,(1+O(\varepsilon^{-1}
    e^{-\frac{\delta_0}{\varepsilon}}),\\
    \theta=\bar \theta \,(1+O(e^{-\frac{\delta_0}{\varepsilon}})),
    \quad\quad
    z_\nu(E)=z_\nu(\bar E) +O(\varepsilon^{-1}e^{-\frac{\delta_0}{\varepsilon}})
  \end{gather*}
\end{Le}
\demo Prove the representation for $T_h$.  Recall that, in $V_*$, a
neighborhood of $E_*$ independent of $\varepsilon$, \ $T_{h}$ admits
the asymptotics from~\eqref{check-Phi:as}.  This and the Cauchy
estimates for the derivatives of analytic functions imply that, for
$E$ in any fixed compact of $V_*$, one has $|\frac{d}{dE}\log
T_{h}|\le C\varepsilon^{-1}$. So, for $E$ satisfying~\eqref{neighb},
we get $T_{h}(E)=\bar T_{h}(1+O(\varepsilon^{-1}
e^{-\delta_0/\varepsilon})$. The estimates for $T_{v,0}$ and
$T_{v,\pi}$ are proved similarly.\\
Furthermore, (\ref{theta:as})  implies that, for $E$ in any fixed
compact of $V_*$,  one has $|\theta'(E)|\le C$. This implies the 
asymptotic representation for $\bar\theta$. The asymptotic
representation for $z_\nu$ follows from~(\ref{z-nu-prime}) in the 
same way. This completes the proof of Lemma~\ref{le:T-bar-T}. \qed \\
We now derive simplified asymptotic representations for the functions
$\tilde C_\pi$, $\tilde S_\pi$ and $C_0$ defined in~\eqref{CS-pi}
and~\eqref{C0}. Redefine $z_0:=z_0(\bar E)$, and $z_\pi:=z_\pi(\bar
E)-2\pi h$.  We prove
\begin{Le}
  \label{CCS} Let $E_\pi\in J_*$, let~(\ref{Delta}) hold.
  For $\varepsilon$ sufficiently small, for all $(z,E)$ in 
  the domain~(\ref{analyticity:dom:a}), one has
  \begin{equation}
    \label{eq:10}
    \begin{split}
    C_0&=\sigma_0\bar T_{v,0} g_0+ \varepsilon^{-1}
    e^{-\delta_0/\varepsilon}\bar
    T_{v,0}\,O\left(\xi_0,p\right)+ O\left( T_Y p\right),\\
    \tilde C_\pi&= \sigma_\pi\bar T_{v,\pi} g_\pi+
    \varepsilon^{-1} e^{-\delta_0/\varepsilon}\bar
    T_{v,\pi}O\left(\xi_\pi,p\right)+O\left(T_Y p\right),\quad\quad
    \tilde S_\pi= -\sigma_\pi
    +O\left(e^{-\delta_0/\varepsilon}\right).      
    \end{split}
  \end{equation}
\end{Le}
\demo Prove~\eqref{eq:10}.  From~\eqref{C0} and~\eqref{alpha:as:0}, we
get
\begin{equation*}
  C_0=\cos\frac{\check\Phi_0(E)}\varepsilon+ 
     T_{v,0}\cos\left(2\pi(z-z_0)+
       \frac{\check\Phi_0(E)}\varepsilon\right)+
     O(p T_Y)
\end{equation*}
with the ``old'' $z_0$. By means of
Lemmas~\ref{cos-sin},and~\ref{le:T-bar-T}, this yields
\begin{equation*}
  C_0=\sigma_0\bar T_{v,0} (\xi_0+\sin(2\pi(z-z_0))+
  \varepsilon^{-1}e^{-\delta_0/\varepsilon}\bar  T_{v,0}\,O(\xi_0, p)+O( T_Yp)
\end{equation*}  
already with the ``new'' $z_0$. This result and~\eqref{g-nu}
imply~\eqref{eq:10}.  \\
The asymptotic representations for $\tilde C_\pi$ and $\tilde S_\pi$
are proved similarly; we only note that one uses the estimate $p\bar
T_{v,\nu}\le e^{-\delta_0/\varepsilon}$ which follows from the
definitions of $p$ and $\delta_0$ (as $2\pi |\im z|\le
\delta_0/\varepsilon$). This completes the proof of Lemma~\ref{CCS}.\qed\\
Now, we are ready to derive Theorem~\ref{th:M-matrix} from
Proposition~\ref{new-M}. Begin by computing $M^\pi_{11}$ for $E$ and
$E_\pi\in J_*$ satisfying~\eqref{neighb} and~\eqref{Delta}.
By~\eqref{M-new:as}, we have
\begin{equation}
  \label{M11}
  M^\pi_{11}=P_{11}+Q_{11}+R_{11}.
\end{equation} 
By~\eqref{P} and Lemmas~\ref{CCS} and~\ref{le:T-bar-T}, for sufficiently
small $\varepsilon$, we get
\begin{equation*}  
  \begin{split}
    P_{11}=&\sigma_0\sigma_\pi\,\frac{4}{\bar
      T_h}\,\left(1+O(\varepsilon^{-1}e^{-\delta_0/\varepsilon})
    \right)\cdot\\ &\left( \bar T_{v,0} g_0+ \varepsilon^{-1}
      e^{-\delta_0/\varepsilon}\bar
      T_{v,0}O\left(\xi_0,p\right)+\,\,O\left(T_Y p\right)\right)\,
    \left(\bar T_{v,\pi} g_\pi+ \varepsilon^{-1}
      e^{-\delta_0/\varepsilon}\bar
      T_{v,\pi}O\left(\xi_\pi,p\right)+O\left( T_Y p\right)\right).
\end{split}
\end{equation*}
Now, recalling the definitions of $\tau$ and $g_\nu$,
see~\eqref{tau:ext} and~\eqref{g-nu}, we get
\begin{equation*}
  P_{11}=\sigma_0\sigma_\pi\,\tau^2 g_0 g_\pi+
  \tau^2 \varepsilon^{-1}e^{-\delta_0/\varepsilon} O\left(\xi_0
  \xi_\pi, p\xi_0,p\xi_\pi, p^2\right)+
  O\left(p\frac{T_Y}{\bar T_{h}}\left(
      e^{-\delta_0/\varepsilon}/\varepsilon+(\bar
      T_{v,\pi}+\bar T_{v,0}) p+T_Yp\right)\right)
\end{equation*}
Note that, by~(\ref{eq:6}), for $2\pi|\im z|\leq
\delta_0/\varepsilon$, one has
\begin{equation}
  \label{tp}
  (\bar T_{h}+\bar T_{v,0}+\bar T_{v,\pi}) p\le
  Ce^{-2\delta_0/\varepsilon}p\le Ce^{-\delta_0/\varepsilon}. 
\end{equation}
Recall that $Y>Y_M$ (see Theorem~\ref{new-M}) and note that $T_{h}\ge
e^{-2\pi Y_M/\varepsilon}$, see~(\ref{Y's}).
Let $\delta_1=2\pi (Y-Y_M)$. Then,
\begin{equation}
  \label{TY-Th}
  T_Y/\bar T_h\le Ce^{-\delta_1/\varepsilon}.
\end{equation}
Therefore, we get
\begin{equation}
  \label{P11}
  P_{11}=\sigma_0\sigma_\pi\,\tau^2 g_0 g_\pi+\tau^2
  \varepsilon^{-1}e^{-\delta_0/\varepsilon} O\left(\xi_0
  \xi_\pi, p\xi_0,p\xi_\pi, p^2\right)+
  O\left(pe^{-(\delta_1+\delta_0)/\varepsilon}\right).  
\end{equation}
Furthermore, by~\eqref{Q} and Lemmas~\ref{cos-sin} and~\ref{le:T-bar-T}, we have
\begin{equation*}
  Q_{11}=\theta^{-1}\,\cos\frac{\check\Phi_\pi-
    \check\Phi_0}\varepsilon\,+\theta\,\cos\frac{\check\Phi_\pi}
  \varepsilon\,\cos\frac{\check\Phi_0}\varepsilon=
  \bar\theta^{-1}\sigma_\pi\sigma_0+O(
  e^{-\delta_0/\varepsilon}).
\end{equation*}
Finally, by~\eqref{R} and estimates~\eqref{tp} and~\eqref{TY-Th}, we
get
\begin{equation*}
  R_{11}=O\left(T_h,\,p(z) T_YT_h^{-1},
\,p(z)T_{v}\right)=O(pe^{-\delta_1/\varepsilon}, e^{-\delta_0/\varepsilon}).
\end{equation*}
Fix $0<\delta<\min(\delta_0,\delta_1)$.  Substituting the estimates
obtained for $P_{11}$, $Q_{11}$ and $R_{11}$ into~\eqref{M11}, and
using the notation $\sigma=\sigma_0\sigma_\pi$, we get the
representation for $M_{11}^\pi$ announced in
Theorem~\ref{th:M-matrix}. The representations for the other
coefficients of the matrix $M^\pi$ are proved with the same
technique.\\
We only briefly comment on how to derive the representation for
$M_{21}^U$.  By~\eqref{M-new:as} and~\eqref{P}, one has
\begin{equation*}
  M_{21}^U=Q_{21}+R_{21}.
\end{equation*}
First, one shows that 
\begin{equation*}
  Q_{21}=\sigma\bar \theta\tau
  g_\pi/r+\varepsilon^{-1}e^{-\delta_0/\varepsilon}\,\tau
  r^{-1}\,O(\xi_\pi,p)+O(e^{-\delta_1/\varepsilon} T_h p).
\end{equation*}
The representation for $M_{21}^U$ follows from the one for $Q_{21}$
and the estimate
\begin{equation*}
  R_{21}=e^{-\delta/\varepsilon}\,\tau
  r^{-1}\,O(p\xi_\pi, p^2)+e^{-\delta/\varepsilon}\,O(\bar T_h p)
\end{equation*}
which follows from~\eqref{R21} and the estimates
\begin{gather*}
  \label{R-1}
  \cos\frac{\check\Phi_\pi}\varepsilon\,O\left(T_h,\,
    p(z)T_{v}\right)=e^{-\delta_0/\varepsilon} \tau r^{-1}
  O(\xi_\pi),\quad \cos\frac{\check\Phi_\pi}\varepsilon\,
  O\left(pT_Y/T_h\right)=e^{-\delta_1/\varepsilon}
  \tau r^{-1}O(p\xi_\pi),\\
  \label{R-2}
    \cos\frac{\check\Phi_0}\varepsilon\,O\left(T_h\right)=
    \varepsilon^{-1}e^{-\delta_0/\varepsilon}O(\bar T_h),\quad
  \quad
  O\left(p^2T_vT_{v,\pi}\right)=
  e^{-\delta_0/\varepsilon}\tau r^{-1}O(p),\\
   \label{R-5}
   O\left(p^2\frac{T_{v,\pi}T_Y}{T_h}\right)=
   e^{-\delta_1/\varepsilon}\tau r^{-1}O\left(p^2 \right),\quad\quad
  O\left(pT_Y,\, pT_vT_h,\, T_h^2\right)=e^{-\delta/\varepsilon} O(p
  \bar T_h).
\end{gather*}
We omit further details of the proof of Theorem~\ref{th:M-matrix}.
\qed


%
\section{The case of large $\tau$}
\label{sec:big-tau}
In this section, we prove Theorem~\ref{th:tib-res:sp:1}. We fix
$E_*\in J$, assume that $\varepsilon$ is so small that
Theorem~\ref{thr:2} holds,
and systematically use  notations from this theorem.\\
In this section, we work under the condition $\tau\gg1$;
by~(\ref{eq:28}), this means that, for some $\delta_\tau>0$,
\begin{equation}
  \label{big-tau}
  \tau\ge e^{\delta_\tau/\varepsilon}.
\end{equation} 
Let $\bar V=\{E\in\C: \ |E-\bar E|\le 4
e^{-\delta_0/\varepsilon}\}$. Here, we study the spectrum of
$H_{z,\varepsilon}$ in $R=\bar V\cap\R$.\\
As $\tau\gg1$, we shall use the monodromy matrix $\tilde M$ described
by Corollary~\ref{cor:M-matrix}.
\subsubsection{Intervals containing spectrum}
\label{sec:interv-cont-hte}
Fix $N>0$. For each $\nu\in\{0,\pi\}$, we let 
\begin{equation*}
  I_\nu(N)=\{E\in J;\ |\xi_\nu(E)|\le 1+\varepsilon^N\}. 
\end{equation*}
We prove
\begin{Th}
  \label{place:tau-grand} 
  For sufficiently small $\varepsilon$, the spectrum of
  $H_{z,\varepsilon}$ in $R$ is contained in $I_0(N)\cup I_\pi(N)$.
\end{Th}
\noindent The proof of this theorem is based upon 
\begin{Pro}[\cite{MR96m:47060}, Proposition 3.1]
  \label{resolvent-set:0} 
  Fix $E\in\R$ and define $h$ by~\eqref{h}.   
  Let $z\mapsto
  M(z,E)$ be a monodromy matrix for equation~\eqref{G.2z}
  corresponding to a basis of consistent solutions that are locally
  bounded in $(x,z)$ together with their derivatives in $x$.\\
  Define
  \begin{equation}
    \label{rho,v}
    \rho(z)=M_{12}(z)/M_{12}( z-h),\quad
    v(z)=M_{11}(z)+\rho( z)\,M_{22}( z).
  \end{equation}
  Suppose that 
  \begin{equation}
    \label{resolvent-set:cond}
    \min_{z\in\R}|M_{12}(z)|>0, \quad 
    \max_{z\in\R} |\rho(z)|< 
    \left(\frac12\,\min_{z\in\R} |v(z)|\right)^2,\quad
    \ind\rho=\ind v=0,
  \end{equation}
  where $\ind g$ is the index of a continuous periodic function $g$.
  \smallpagebreak Then, $E$ is in the resolvent set of~\eqref{family}.
\end{Pro}
\noindent Note that  the proof of this proposition is based on
Theorem~\ref{Resolvent-set}.\\
\noindent{\it Proof of Theorem~\ref{place:tau-grand}.\/} \
It suffices to prove that, for $\varepsilon$ small enough, for $E$ in
$R\setminus(I_0(N)\cup I_\pi(N))$, the monodromy matrix $\tilde M$
described in Corollary~\ref{cor:M-matrix} satisfy the assumptions of
Proposition~\ref{resolvent-set:0}.\\
Below, we always assume that $z\in \R$. \\
{\bf 1.} \ In terms of the coefficients of $\tilde M^\pi$ (see
Corollary~\ref{cor:M-matrix}), define the function $\rho$
by~\eqref{rho,v}. We prove that, for sufficiently small $\varepsilon$,
for $E\in R\setminus I_0(N)$ and $z\in\R$, one has
\begin{equation}
  \label{eq:place:tau-grand:1}
   |\tilde M^\pi_{12}|>0,\quad{\rm and}\quad
   |\rho| \le C \varepsilon^{-N}. 
\end{equation}
By Corollary~\ref{cor:M-matrix},  we have
\begin{equation}\label{eq:100}
  \tilde M^\pi_{12}=\sigma\tau g_0(1+e^{-\delta/\varepsilon} O(|g|_0/g_0)).
\end{equation}
By~\eqref{g-nu}, for $E\in R\setminus I_0(N)$, one has
\begin{equation}\label{eq:101}
 |g_0|\ge \varepsilon^N,\quad  1\le \left|\frac{|g|_0}{g_0}\right|\le
  \frac{|\xi_0|+1}{||\xi_0|-1|}\le
  \varepsilon^{-N}(2+\varepsilon^N),\quad
  \left|\frac{g_0(z)}{g_0(z-h)}\right|\le 
   \varepsilon^{-N} (1+\varepsilon^N).
\end{equation}
For $\varepsilon$ sufficiently small,~(\ref{eq:100})
and~(\ref{eq:101}) imply~(\ref{eq:place:tau-grand:1}).\\
{\bf 2.} \ Here, we assume that $E\in R\setminus
(I_0(N)\cup I_\pi(N))$.  In terms of the coefficients of $\tilde M^\pi$, define the
function $v$ by~\eqref{rho,v}. Check that
\begin{equation}
  \label{eq:place:t-g:3}
  v=\sigma\tau^2 g_0 g_\pi(1+o(1))\quad\text{and}\quad 
  |v|\ge C\varepsilon^{2N}e^{2\delta_\tau/\varepsilon}.
\end{equation}
We have
\begin{equation}
  \label{eq:place:t-g:4}
  |g_0 g_\pi|\ge ( |\xi_0|-1)( |\xi_\pi|-1)\ge
  \varepsilon^{2N}\quad\text{and}\quad
  \frac{|g|_0|g|_\pi}{|g_0 g_\pi|}\le
  \frac{(|\xi_0|+1)(|\xi_\pi|+1)}{(|\xi_0|-1)(|\xi_\pi|-1)}\le 
  \varepsilon^{-2N}(2+\varepsilon^N)^2.
\end{equation}
Using the asymptotics of $\tilde M^\pi$ given
by~\eqref{new-Mmatrix:as}, estimates~\eqref{eq:place:t-g:4} 
and~\eqref{eq:place:tau-grand:1}, we get
\begin{equation*}
  v= \sigma\tau^2 g_0 g_\pi+\bar\theta^{-1}+e^{-\frac{\delta}\varepsilon}\,
   O(\tau^2|g|_0|g|_\pi,\,1)+\rho(z) (\bar \theta
   +O(e^{-\frac{\delta}\varepsilon}))=
   \sigma\tau^2 g_0 g_\pi(1+O(\varepsilon^{-2N}
   e^{-\frac{\delta}\varepsilon},\,\varepsilon^{-3N} \tau^{-2})).
\end{equation*}
In view of~\eqref{big-tau}, this implies the first estimate
in~\eqref{eq:place:t-g:3}. The latter,~\eqref{big-tau} and the first
estimate from~\eqref{eq:place:t-g:4} imply the second estimate
in~\eqref{eq:place:t-g:3}. \\
{\bf 3.} \ Steps 1 and 2 imply that, for sufficiently small
$\varepsilon$ and $E\in R\setminus(I_0(N)\cup I_\pi(N))$, the matrix
$\tilde M$ satisfies the conditions of
Proposition~\ref{resolvent-set:0} (note that the equalities $\ind
v=\ind\rho=0$ follow from the inequalities $|v|, \, |\tilde
M^{\pi}_{12}|>0$ and the fact that the coefficients of the monodromy
matrix $\tilde M^\pi$ are real valued). This implies the statement of
Theorem~\ref{place:tau-grand}.\qed\\
One has
\begin{Cor}
  \label{cor:place:grand-tau}
  Fix $N>0$. For sufficiently small $\varepsilon$, the
  interval $I_0(N)$ (resp. $I_\pi(N)$)
  is contained in the $(C\varepsilon t_{v,0}(\bar E))$-neighborhood
  (resp.  $(C\varepsilon \bar t_{v,\pi}(\bar E))$-neighborhood) of the
  point $E_0$ (resp $E_\pi$).
\end{Cor}
\demo The result follows from Theorem~\ref{place:tau-grand}, the definitions of
$I_\nu(N)$, the definitions of $\xi_\nu$, see~(\ref{xi-nu})
and~(\ref{gamma:ext}),  Lemma~\ref{le:Phi'} and~(\ref{T-bars}).\qed \\ 
\noindent We shall also use a rougher result
\begin{Cor}
  \label{corcor:place:grand-tau}
  For sufficiently small $\varepsilon$, in the case of
  Theorem~\ref{place:tau-grand}, $I_0(N)$ (resp. $I_\pi(N)$) is
  contained in the $(C\varepsilon
  e^{-2\delta_0/\varepsilon})$-neighborhood of the point $E_0$ (resp.
  $E_\pi$).
\end{Cor}
\demo This follows from the previous statement as, for each
$\nu\in\{0,\pi\}$, one has $t_{v,\nu}(\bar E)\le
e^{-2\delta_0/\varepsilon}$. \qed
\subsection{Computation of the integrated density of states}
\label{sec:comp-integr-dens}
We now compute the increment of the integrated density of states on
the intervals $I_0$ and $I_\pi$ described in the previous subsection.
We prove
\begin{Th}\label{ids:tau-grand} 
  Fix $N>0$.  For sufficiently small $\varepsilon$, 
  \begin{equation}
    \label{eq:ids:t-g:1}
    \int_{I_0(N)\cup I_\pi(N)} n_{\varepsilon}(dE)=
    \frac\varepsilon{\pi},\quad{\rm and}\quad
    \int_{I_0(N)} n_{\varepsilon}(d E)=\int_{\check
      I_\pi} n_{\varepsilon}(dE) =\frac\varepsilon{2\pi}\
    \text{if}\ I_0(N)\cap I_\pi(N)=\emptyset,
  \end{equation}
  where
  $n_{\varepsilon}(dE)$ denotes the density of states measure of
  $H_{z,\varepsilon}$.
\end{Th}
\noindent{\it Proof.} The proof of this theorem is based upon 
\begin{Pro}[\cite{Fe-Kl:04a}, Proposition 4.2]
  \label{pro:ids:1} 
  Pick two points $a<b$ of the real axis. Let $\gamma$ be a
  continuous curve in $\C_+$ connecting $a$ and $b$.\\
  Assume that, for all $E\in\gamma$, there is a consistent basis such
  that the following holds.
  \begin{itemize}
  \item The basis solutions are locally bounded in $(x,z)$ together
    with their first derivatives in $x$.
  \item There exists $V(\gamma)$, a neighborhood of $\gamma$ such that
    the monodromy matrix is continuous in $(z,E)\in \R\times V(\gamma)$ and
    analytic in $E\in V(\gamma)$.
  \item On $\gamma$, the coefficients of $M$, the monodromy matrix, satisfy
    the conditions~\eqref{resolvent-set:cond} in which
    $\rho$ and $v$ are defined by~\eqref{rho,v} with $h$
    from~\eqref{h}.
  \item the coefficients of $M$ are real for real $E$ and $z$.
  \end{itemize}
  Then, the increment of the integrated density of states on the
  interval $[a,b]$ is given by
  \begin{equation}
    \label{DeltaN}
    \int_a^b n_\varepsilon(dE)=-\left.\frac\varepsilon{2\pi^{2}}
      \int_0^1\arg v(z,E)\,dz\right|_{\gamma},
  \end{equation}
  where $f|_{\gamma}$ denotes the increment of $f$ when going from $a$
  to $b$ along $\gamma$.
\end{Pro}
\noindent The proof of Theorem~\ref{ids:tau-grand} consists of the
following steps.\\
{\bf 1.} First, we assume that $I_0(N)\cap I_\pi(N)=\emptyset$ and
prove that $\int_{I_0(N)} n_{\varepsilon}(d E)=\varepsilon/2\pi$.\\
By~\eqref{xi}, $\xi_0$ is a non constant affine function of $E$. To
apply Proposition~\ref{pro:ids:1}, as $\gamma$ we choose the half
circle
\begin{equation*}
  \gamma=\{E\in\C:\ |\xi_0(E)|=1+\varepsilon^{N+1},\,\im E\geq 0\}.
\end{equation*}
Recall that $a<b$ denote the ends of $\gamma$. Now, 
$[a,b]=I_0(N)$. So, $I_0(N+2)\subset]a,b[$, and $[a,b]\cap
I_\pi(N)=\emptyset$.  So, by Theorem~\ref{place:tau-grand}, 
$a$ and $b$ are in the resolvent set of $H_{z,\varepsilon}$.\\
Note also that, for sufficiently small $\varepsilon$, for $E\in
\gamma$, one has
\begin{equation}
  \label{eq:dist-bar}
  |E-\bar E|\le 2e^{-\delta_0/\varepsilon}.
\end{equation}
Indeed, by~\eqref{Delta}, one has $|E_0-\bar E|\le
e^{-\delta_0/\varepsilon}$, and, for sufficiently small $\varepsilon$,
for $E\in\gamma$, we get
\begin{equation*}
  |E-E_0|= 
  \varepsilon \bar T_{v,0} |\check\Phi_0'(\bar
  E)|^{-1}\,\,|\xi_0(E)|\le C\varepsilon t_{v,0}(\bar E)\le C\varepsilon
  e^{-2\delta_0/\varepsilon},
\end{equation*} 
where we have used the definitions of $\xi_0$, of $\gamma$, 
Lemma~\ref{le:Phi'},~\eqref{T-bars} and the definitions of $\delta_0$
and $t_{v,0}$.\\
By~(\ref{eq:dist-bar}),  $(z,E)\in\R\times\gamma$ is in the 
domain~(\ref{analyticity:dom:a}), so, we can use the matrix $\tilde
M^\pi$ from Corollary~\ref{cor:M-matrix} and its 
asymptotics~(\ref{new-Mmatrix:as}).  Define $\rho$ and $v$ in terms of 
$\tilde M^\pi$ by~\eqref{rho,v}.  To apply the
Proposition~\ref{pro:ids:1}, we need only to check
that, along $\gamma$, condition~\eqref{resolvent-set:cond} is
satisfied.   This follows from the
\begin{Le}
  \label{le:ids:t-g:1}
  For sufficiently small $\varepsilon$, for
  $E\in\gamma$ and $z\in \R$, one has
  \begin{equation}
    \label{eq:ids:t-g:2} 
    \begin{split}
      \tilde M^\pi_{12}&\ne 0,\quad\quad \ind
      \rho=0,\quad\quad|\rho|\le C\varepsilon^{-N- 1}, \\
      v&=\sigma\tau^2 g_0 g_\pi (1+o(1)), \quad\quad \ind
      v=0,\quad\quad |v|\ge
      C\varepsilon^{2(N+1)}e^{2\delta_\tau/\varepsilon}.
    \end{split}
  \end{equation}
\end{Le}
\demo When proving this lemma, one uses almost the same arguments as
in the proof of Theorem~\ref{place:tau-grand} with $N$ replaced by
$N+1$. The only difference is that now one deduces the equalities
$\ind \rho=\ind v=0$ from the asymptotic $\tilde M^\pi_{12}=\sigma\tau
g_0(1+o(1)$ and the asymptotics of $v$ from~\eqref{eq:ids:t-g:2}.
We omit  further details.\qed\\
As the integrated density
of states is constant outside the spectrum, formula~\eqref{DeltaN} and
the representation for $v$ in~\eqref{eq:ids:t-g:2} give
\begin{equation*}
  \frac{2\pi^2}\varepsilon\int_{I_0(N)}
  n_{\varepsilon}(dE)= 
  -\left.\int_0^1\arg v(z,E)\,dz\right|_{\gamma}
  =-\left.\int_0^1\arg (g_0 g_\pi) dz\right|_{\gamma}.
\end{equation*}
where, in the last step, we have used the fact that, for $z$ and $E$
real, the functions $v$, $g_0$ and $g_\pi$ take real values, and,
therefore, the last two integrals coinciding up to $o(1)$, they are
equal. As, for $E\in\gamma$, one has $|g_\nu/\xi_\nu-1|<1$,
$\nu\in\{0,\pi\}$, and as $\xi_\nu$ and $g_\nu$ are real for real $z$
and  $E$, we get finally
\begin{equation*}
  \int_{I_0(N)} n_{\varepsilon}(dE)
  =-\frac\varepsilon{2\pi^2}\left.\int_0^1\arg (\xi_0 \xi_\pi)
  dz\right|_{\gamma}=
  -\left.\frac\varepsilon{2\pi^2}\cdot\arg((E-E_0)(E-E_\pi))
  \right|_{\gamma}=\frac\varepsilon{2\pi},
\end{equation*}
where we have used the fact that only $E_0$ is located between the
ends of $\gamma$ and $E_\pi$ is not.\\
{\bf 2.} \ Now, to complete the proof of Theorem~\ref{ids:tau-grand},
it suffices to check the first equality in~\eqref{eq:ids:t-g:1}.
Therefore, we use the same techniques as in the previous step.
So, we only outline the proof.\\
Now, we take  
\begin{equation*}
  \gamma=\{E\in\C:\ |E-\bar E|=2e^{-\delta_0/\varepsilon},\,\im E\geq0\}.
\end{equation*}
Now, both $E_0$ and $E_\pi$ are between the ends of $\gamma$.
Moreover, by Corollary~\ref{corcor:place:grand-tau},
both $E_0$ and $E_\pi$ are between the ends of $\gamma$.
Define $\rho$ and $v$ in terms of $\tilde M^\pi$ by~\eqref{rho,v}.  One
proves
\begin{Le}
  \label{le:2}
  For sufficiently small $\varepsilon$, for $E\in\gamma$ and $z\in\R$,
  one has
  \begin{equation*}
    \tilde M^\pi_{12}\ne 0,\quad
    \rho=1+o(1),\quad \ind \rho=0,\quad \quad v=\sigma\tau^2 \xi_0
    \xi_\pi (1+o(1)),\quad \ind v=0,\quad |\tau^2\xi_0\xi_\pi|\ge C/\varepsilon^2.
  \end{equation*}
\end{Le}
\demo One uses essentially the same analysis
as when proving Theorem~\ref{place:tau-grand} and
Lemma~\ref{le:ids:t-g:1}. We omit the details, noting only that, now, 
for sufficiently small $\varepsilon$, for each 
$\nu\in\{0,\pi\}$ and $E\in\gamma$, 
\begin{equation*}
  |\xi_\nu(E)|\ge C\varepsilon^{-1} t_{v,\nu}^{-1}(\bar E)
  |E-E_\nu|\ge C\varepsilon^{-1}e^{2\delta_0/\varepsilon}(
  |E-\bar E|-|\bar E-E_\nu|)\ge C\varepsilon^{-1}
  e^{\delta_0/\varepsilon}.\quad\quad \text{ \qed}
\end{equation*}
\noindent This lemma immediately implies that the conditions of
Proposition~\ref{pro:ids:1} are satisfied. The points $E_0$ and
$E_\pi$ being between the ends of $\gamma$, one obtains
\begin{equation*}
  \int_{I_0(N)\cup I_\pi(N)} n_{\varepsilon}(dE)=
  -\left.\frac\varepsilon{2\pi^2}
    \cdot\arg((E-E_0)(E-E_\pi))\right|_{\gamma}=
  \frac\varepsilon{\pi}.
\end{equation*}
This completes the proof of~\eqref{eq:2}.
\subsection{Computation of the Lyapunov exponent}
\label{sec:comp-lyap-expon}
Here, we compute the asymptotics of the Lyapunov exponent
$\Theta(E,\varepsilon)$ on the intervals $I_0(N)$ and $I_\pi(N)$ and
prove
\begin{Th}
  \label{th:Le:t-g}
  Fix $N>0$. For sufficiently small $\varepsilon$, for $E\in
  I_0(N)\cup I_\pi(N)$, one has~\eqref{Theta-res}.
\end{Th}                                %
\noindent To compute $\Theta(E,\varepsilon)$, we use
Theorem~\ref{Lyapunov} and the matrix cocycle $(\tilde M^\pi,h)$.  In
the next two subsections, we get an upper and a lower bound for
$\theta(\tilde M^\pi,h)$. They will coincide up to smaller order
terms, and, thus, lead to the asymptotic formula for
$\Theta(E,\varepsilon)$.\\
In
sections~\ref{sec:upper-bound-lyapunov},~\ref{sec:lower-bound-lyapunov}
and~\ref{sec:completing-analysis}, we always assume that $E\in
I_0(N)\cup I_\pi(N)$.
\subsubsection{The upper bound}
\label{sec:upper-bound-lyapunov}
We now prove that
\begin{equation}
  \label{Le:14}
  \theta(\tilde M^\pi,h)\le 2\log\left(\tau\sqrt{1+|\xi_0(E)|+
  |\xi_\pi(E)|}\right)+C.
\end{equation}
Therefore, we first note that~\eqref{new-Mmatrix:as} implies that 
\begin{equation*}
   \|M^\pi(z,E)\|\le  C\tau^2(|\xi_0|+1)(|\xi_\pi|+1),\quad z\in\R.
\end{equation*}
Note that to get this estimate, we have used that $\tau>1$.\\
As $E\in (I_0(N)\cup I_\pi(N))$, then, by
Theorem~\ref{place:tau-grand}, for sufficiently small $\varepsilon$,
at least one of the inequalities
\begin{equation*}
   |\xi_0|\le 2,\quad\text{and}\quad |\xi_\pi|\le 2.
\end{equation*}
is satisfied. Therefore, we get 
\begin{equation*}
   \|M^\pi(z,E)\|\le  C\tau^2(|\xi_0|+|\xi_\pi|+1),\quad z\in\R.
\end{equation*}
Now, this estimate and the definition of Lyapunov exponent for matrix
cocycles~\eqref{eq:44} imply~\eqref{Le:14}.
\subsubsection{The lower bound for the Lyapunov exponent}
\label{sec:lower-bound-lyapunov}
Here, we prove that
\begin{equation}
  \label{Le:7}
  \theta(\tilde M^\pi,h)\geq 2\log\left(\tau\sqrt{1+|\xi_0(E)|+
      |\xi_\pi(E)|}\right)+O(1).
\end{equation}
Therefore, we use the following construction.\\
Assume that a matrix function $M:\C\to SL(2,\C)$ is $1$-periodic and
depends on a parameter $\varepsilon>0$. One has
\begin{Pro}
  \label{le:Le:2}
  Pick $\varepsilon_0>0$. Assume that there exist $y_{0}$ and $y_{1}$
  satisfying the inequalities $0<y_0<y_1<\infty$ and such that, for
  any $\varepsilon\in(0,\varepsilon_0)$ one has
  \begin{itemize}
  \item the function $z\to M(z,\varepsilon)$ is analytic in the strip
    $S=\{z\in\C;\ 0\le\im z\le y_1/\varepsilon\}$;
  \item in the strip $S=\{z\in\C;\ y_0/\varepsilon\le\im z\le
    y_1/\varepsilon \}\subset S$, $M(z,\varepsilon)$ admits the
    following uniform in $S$ representation
    \begin{equation}
      \label{Le:form}
      M(z,\varepsilon)=\lambda(\varepsilon)e^{2\pi i m z}\cdot
      \left(\begin{pmatrix}1& 0\\0 & 0 \end{pmatrix}+o(1)\right),\quad
      \varepsilon\to 0,
    \end{equation}
    where $\lambda(\varepsilon)$ and $m$ are independent of $z$, and
    $m$ is an integer independent of $\varepsilon$.
  \end{itemize}
  Then, there exists $\varepsilon_1>0$ such that, if
  $0<\varepsilon<\varepsilon_1$, one has
  \begin{equation}
    \label{Le:8}
    \theta(M,h)>\log|\lambda(\varepsilon)|+o(1);
  \end{equation}
  the number $\varepsilon_1$ and the error estimate in~\eqref{Le:8}
  depend only on $\varepsilon_0$, $y_0$, $y_1$ and the norm of the
  term $o(1)$ in~\eqref{Le:form}.
\end{Pro}
\smallpagebreak This proposition immediately follows from
Proposition~10.1 from~\cite{MR2003f:82043}. Note that the proof of the
latter is based on the ideas of~\cite{MR93b:81058} generalizing
Herman's argument~\cite{MR85g:58057}.
\smallpagebreak Consider the case where $E\in I_0(N)$. Then, one has  $|\xi_0|< 2$
(for $\varepsilon< 1$).
Fix $0<y_0<y_1<\frac{\delta}{2\pi}$, where $\delta$ is the constant
from~(\ref{new-Mmatrix:as}). We shall describe the precise choice
of $y_0$ and $y_1$ later. For sufficiently small $\varepsilon$ and
$\frac{y_0}\varepsilon\le |\im z|\le \frac{y_1}\varepsilon$,
representation~\eqref{new-Mmatrix:as} implies that
\begin{equation}
  \label{new-MMmatrix:as}
  \tilde M^\pi=\frac{\sigma\tau^2}{2i}\,e^{-2\pi i (z-z_0)}\cdot 
  \left[\begin{pmatrix} 
      \xi_\pi (1+o(1))+\frac1{2i}e^{-2\pi i (z-z_\pi)}(1+o(1)) &
      0  \\ \xi_\pi\cdot  o(1)  & 0
    \end{pmatrix}+o(1)\right].
\end{equation}
We have used~\eqref{big-tau}.  Now, let
$y=\frac{\varepsilon}{2\pi}\log(1+|\xi_\pi(E)|)$.
Fix $0<A<\delta/2\pi$. \\
To compute the Lyapunov exponent for $E\in I_0(N)$ such that $y\le A$,
we choose $y_0>A$. Then, for sufficiently small $\varepsilon$, for
such $E$, we get
\begin{equation*}
      \tilde M^\pi=-\frac{\sigma\tau^2}{4}\,e^{-2\pi i (2z-z_0-z_\pi)}\cdot 
   \left[\begin{pmatrix} 1 & 0  \\0  & 0\end{pmatrix}+o(1)\right],
\end{equation*}
and so, Proposition~\ref{le:Le:2} implies that $\theta(\tilde
M,h)\ge\log(\tau^2/4)+o(1)$.\\
On the other hand, to compute the Lyapunov exponent for all $E\in
I_0(N)$ such that $y\ge A$, we choose $y_1<A$. Then, for sufficiently
small $\varepsilon$, for all such $E$, we get
\begin{equation*}
      \tilde M^\pi=\frac{\sigma\tau^2}{2i}\,\xi_\pi\, e^{-2\pi i
   (z-z_\nu)}\cdot \left[\begin{pmatrix} 1 & 0  \\0  & 0
     \end{pmatrix}+o(1)\right],
\end{equation*}
and so, Proposition~\ref{le:Le:2} implies that $\theta(\tilde
M,h)\geq\log(\tau^2|\xi_\pi|/2)+o(1)$. \\
For sufficiently small $\varepsilon$ and for $E\in I_0(N)$, the
obtained lower bounds for $\theta(\tilde M^\pi,h)$ imply~\eqref{Le:7}.\\
For $E\in I_\pi(N)$, one proves~\eqref{Le:7} similarly.
\subsubsection{Completing the analysis}
\label{sec:completing-analysis}
Theorem~\ref{Lyapunov} and estimates~\eqref{Le:14} and~\eqref{Le:7}
imply that, for sufficiently small $\varepsilon$, on $I_0\cup I_\pi$,
one has~\eqref{Theta-res}.  This completes the proof of
Theorem~\ref{th:Le:t-g}.


%
\section{The case of small $\tau$}
\label{sec:small-tau}
We now turn to the case $\tau\ll1$; by~(\ref{eq:14}), this means that,
for some $\delta_\tau>0$,
\begin{equation}
  \label{small-tau}
  \tau\le e^{-\delta_\tau/\varepsilon}.
\end{equation}
We shall assume that~(\ref{eq:41}) holds. As before, we fix $E_*\in
J$, assume that $\varepsilon$ is so small that Theorem~\ref{thr:2}
holds, and we systematically use its notations.\\
Let $\bar V=\{E\in\C: \ |E-\bar E|\le 4e^{-\delta_0/\varepsilon}\}$
and study the spectrum in $R=\bar V\cap\R$.  Now, we use the monodromy
matrix $M^\pi$ described by Theorem~\ref{th:M-matrix}.
\subsection{The location of the spectrum}
\label{sec:location-spectrum}
In this section, we prove Theorem~\ref{thr:5}. The central point of
its proof is the a priori estimate provided by
\begin{Le}
  \label{le:apriory-est}
  Under the above conditions, for sufficiently small $\varepsilon$,
  if $E\in R\cap\sigma(H_{z,\varepsilon})$, then
  \begin{equation}
    \label{apriory-est}
    \tau^2(|\xi_0(E)|+1)(|\xi_\pi(E)|+1)\ge C(1+o(1)),
  \end{equation} 
  where $o(1)$ depends only on $\varepsilon$, and $C$ depends only on
  $\theta_n(V)$.
\end{Le}
\noindent First, in sections~\ref{sec:proof-priori-estim}
and~\ref{sec:proof-prop-refl}, we prove Lemma~\ref{le:apriory-est}.
Then,  in
section~\ref{sec:compl-proof-theor}, by means of this lemma, we get a 
description of the resolvent
set of $H_{z,\varepsilon}$ inside $\tilde R=\{E\in R$
satisfying~\eqref{apriory-est}$\}$. This will yield
Theorem~\ref{thr:5}.
\subsubsection{Proof of the a priori estimate}
\label{sec:proof-priori-estim}
The proof of Lemma~\ref{le:apriory-est} is based on the following
construction.\\
Consider the finite difference equation
\begin{equation}
  \label{f-d-e-on-R}
  \psi(z+h)=M(z)\psi(z),\quad z\in\R,
\end{equation}
where $h$ is a fixed positive number, and $M$, a fixed matrix function
in $L^\infty(R,SL(2,\C))$. One has
\begin{Pro}
  \label{le:resolvent-set:1}
  Suppose that, for $z\in\R$, $M(z)$ can be represented as
  \begin{equation}
    \label{f-d-e-on-R:Mform}
    M(z)=\begin{pmatrix} \theta^{-1} & 0 \\ 0 &
    \theta\end{pmatrix}+\tilde M(z),
  \end{equation}
  where $\theta$ is a real number, and  this number and  the matrix valued function
  $z\mapsto\tilde M(z)$ satisfy
  \begin{equation}
    \label{f-d-e-on-R1}
    |\theta+\theta^{-1}| >2\quad\text{and}\quad |\theta-\theta^{-1}|\geq
    4\sup_{1\leq i,j\leq 2}\sup_{z\in\R}|\tilde M_{ij}(z)|.
  \end{equation}
  Then, there exists $\psi_+$ and $\psi_-$, two vector solutions
  to~\eqref{f-d-e-on-R} in $L^{\infty}_{\rm loc} (\R,\C^2)$, such
  that, for $z\in\R$,
  \begin{equation}
    \label{psi-prop}
    \det(\psi_+(z),\psi_-(z))>0\quad\text{and}\quad
    \|\psi_+(z)\|_{\C^2}+\|\psi_-(-z)\|_{\C^2}\le
    Ce^{-z\log\frac{|\theta+\theta^{-1}|}2},
    \quad\text{if}\quad z>0.
  \end{equation}
\end{Pro}
\noindent Let us first derive Lemma~\ref{le:apriory-est} from
Proposition~\ref{le:resolvent-set:1} applied to $M=M^\pi$ and $h$
defined by~\eqref{h}. We represent $M^\pi$ in the
form~\eqref{f-d-e-on-R:Mform} with $\theta=\bar\theta$. Then, for
$z\in \R$, by Theorem~\ref{th:M-matrix}, one has
\begin{gather} 
  \nonumber |\tilde M_{11}|\le \tau^2(|\xi_0|+1)(|\xi_\pi|+1)
  (1+o(1))+O(e^{-\delta/\varepsilon}),\quad
  |\tilde M_{22}|\le O(e^{-\delta/\varepsilon}),\\
  \label{eq:12}
  |\tilde M_{12}\tilde M_{21}|\le \bar
  \theta\tau^2(|\xi_0|+1)(|\xi_\pi|+1)
  (1+o(1))+O(e^{-\delta/\varepsilon}).
\end{gather} 
Only the last estimate requires to be checked. From~\eqref{mm-as}, for
$z\in\R$, we
get
\begin{align*}
  |\tilde M_{12}\tilde M_{21}|&\le \left(r\tau(|\xi_0|+1)(1+o(1))+ C
    e^{-\delta/\varepsilon}\right)\, \left(\bar \theta
    r^{-1}\tau(|\xi_\pi|+1)(1+o(1))+ C \bar
    T_he^{-\delta/\varepsilon}\right)\\
  &\le \bar\theta\tau^2(|\xi_0|+1)(|\xi_\pi|+1) (1+o(1))+ C
  e^{-\delta/\varepsilon}\left(\bar T_{v,0}((|\xi_0|+1)+\bar
    T_{v,\pi}((|\xi_\pi|+1)+
    e^{-\delta/\varepsilon}\bar T_{h}\right)\\
  &\le \bar\theta\tau^2((|\xi_0|+1)(|\xi_\pi|+1)
  (1+o(1))\\&\hspace{3cm} + C
  e^{-\delta/\varepsilon}\left(\varepsilon^{-1}\,|E-E_0|+\bar T_{v,0}+
    \varepsilon^{-1}\,|E-E_\pi|+\bar T_{v,\pi}+
    e^{-\delta/\varepsilon}\bar T_{h}\right)
\end{align*}
which implies~\eqref{eq:12}.\\
The estimates for the coefficients of the matrix $\tilde M$ show that
$M^\pi$ is similar to a matrix of the
form~\eqref{f-d-e-on-R:Mform} for which
\begin{equation*}
  \sup_{1\leq i,j\leq 2}\sup_{z\in\R}|\tilde M_{ij}(z)|\le
  \max\{a,\sqrt{\bar\theta a}\}\quad\text{where}\quad 
  a=\tau^2((|\xi_0|+1)(|\xi_\pi|+1) (1+o(1))+Ce^{-\delta/\varepsilon}.
\end{equation*}
This and Proposition~\ref{le:resolvent-set:1} imply that, if
\begin{equation}
  \label{C(theta)-1}
  \bar \theta+{\bar\theta}^{-1}>2\quad\text{and}\quad a<
  C(\bar\theta)=\min\left\{(\bar\theta-\bar\theta^{-1})/4, 
    (\bar\theta-\bar\theta^{-1})^2/(16\theta)\right\},
\end{equation} 
then, there exist $(\psi_+,\psi_-)$, the two solutions
to~\eqref{f-d-e-on-R} for $M=M^\pi$, that have all the properties
described in Proposition~\ref{le:resolvent-set:1}. Define functions
$\chi_\pm:\ \Z\mapsto \C^2$ by $\chi_\pm(n)=\psi_\pm(nh+z)$. The
functions $\chi_+$ and $\chi_-$ are solutions to the monodromy
equation~\eqref{Mequation} satisfying the conditions of
Theorem~\ref{Resolvent-set}. So, $E$ is in the resolvent set of
$H_{z,\varepsilon}$.\\
Finally, discuss the conditions~\eqref{C(theta)-1}.  Fix $0<q<1$.
Recall that $\bar\theta$ admits the asymptotics~\eqref{theta:as}, and
that $\theta_n(V)+\theta^{-1}_n(V)=2\Lambda_n(V)>2$.  Therefore, if
$\tau^2((|\xi_0|+1)(|\xi_\pi|+1)<q C(\bar\theta)$, then, for
sufficiently small $\varepsilon$, the conditions~\eqref{C(theta)-1}
are satisfied. This implies Lemma~\ref{le:apriory-est}. So, to
complete the proof of this result, we only have to check
Proposition~\ref{le:resolvent-set:1}.
\subsubsection{Proof of Proposition~\ref{le:resolvent-set:1}}
\label{sec:proof-prop-refl}
Set
\begin{equation}
  \label{m-def}
    m=\sup_{1\leq i,j\leq2}\sup_{z\in\R}|\tilde M_{ij}(z)|.
\end{equation}
Note that, if $\psi$ is a solution to~\eqref{f-d-e-on-R:Mform}, then
$e^{\pi i z/h}\psi(z)$ satisfies the same equation with $M$ replaced
by $-M$, and for $\sigma=\begin{pmatrix} 0 & 1\\ 1 & 0
\end{pmatrix}$, $\sigma\psi$ satisfies
equation~\eqref{f-d-e-on-R:Mform} with $M$ replaced by $\sigma
M\sigma$. Therefore, it suffices to consider the case
$\theta>1$. The proof then consists of five steps.\\
{\bf 1.} \ We begin by the following elementary observation.
Let $G$ be a solution of the equation
\begin{equation}
  \label{homothety}
  G(z+h)=\frac{M_{11}(z) G(z)+M_{12}(z)}{M_{21}(z) G(z)+M_{22}(z)},
  \quad z\in\R,
\end{equation}
and let $\psi_2(z)$ be a solution of the equation
\begin{equation}
  \label{homothety:1}
  \psi_2(z+h)=(M_{21}(z) G(z)+M_{22}(z))\psi_2(z),\quad z\in\R.
\end{equation}
Then, the vector function defined by $\psi(z)=\psi_2(z)
\begin{pmatrix} G(z) \\ 1\end{pmatrix}$ is a solution to~\eqref{f-d-e-on-R}.
The proof of this observation being elementary, we omit it.\\
{\bf 2.} \ Let 
\begin{equation*}
  q=\frac{\theta-\theta^{-1}}{2m}-1-\sqrt{
    \left(\frac{\theta-\theta^{-1}}{2m}-1\right)^2-1}. 
\end{equation*}
Recall that $\theta>1$ and satisfies conditions~(\ref{f-d-e-on-R1}).
One has
\begin{gather} 
  \label{q-eq}
  q^2-2\left(\frac{\theta-\theta^{-1}}{2m}-1\right) q+1=0;\\
  \nonumber 0<q<1;\\ \label{p-ineq}
  p=\theta-m(q+1)>\frac{\theta^{-1}+\theta}2>1.
\end{gather}
The first relation is obvious; the lower bound follows from the second
condition in~\eqref{f-d-e-on-R1}; the upper bound follows from the
facts that the second solution to~\eqref{q-eq} is greater than $q$ and
that the product of the solutions is equal to one;~\eqref{p-ineq}
follows from the equality 
\begin{equation*}
  \theta-m(q+1)=\frac{\theta^{-1}+\theta}2+
  m\sqrt{\left(\frac{\theta-\theta^{-1}}{2m}-1\right)^2-1} 
\end{equation*}
and the first condition in~\eqref{f-d-e-on-R1}.\\
{\bf 3.} Let us construct a bounded solution to~\eqref{homothety}.
For $z\in\R$ and $k\geq1$, let
\begin{equation} 
  \label{eq:3}
  G_{k+1}(z+h)=\frac{M_{11}(z) G_k(z)+M_{12}(z)}{M_{21}(z)
    G_k(z)+M_{22}(z)},\quad\text{and}\quad G_0(z)=0.
\end{equation}
One has 
\begin{equation}
  \label{G-bound} 
  \sup_{z\in\R}|G_{k+1}(z)|\le q,\quad k\in\N^*.
\end{equation}
Indeed, this estimate is valid for $G_0$. Assume that it has been
proved for a positive integer $k$; using~\eqref{eq:3}
and~\eqref{q-eq}, we get
\begin{equation*} 
  \sup_{z\in\R}|G_{k+1}(z)|\le\frac{(\theta^{-1}+m) q+m}{\theta- m-m
    q}=q.
\end{equation*}
Now, check prove that $\{G_k(\cdot)\}_{k=0}^\infty$ converges in
$L^\infty$.  It suffices to prove that, for $k\in\N^*$,
\begin{equation}
  \label{DG} 
  \sup_{z\in\R}|G_{k+1}(z)-G_k(z)|\le \frac{m}{p^{2k+1}}. 
\end{equation} 
Here, $p$ has been defined in~\eqref{p-ineq}. In view
of~\eqref{G-bound} and~\eqref{m-def}, we get
\begin{gather*}
  \sup_{z\in\R}|G_{1}(z+h)-G_0(z+h)|\le m/(\theta-m)<m/p;\\
  \intertext{and for $k\geq1$ and $z\in\R$}
  \begin{split}
  |G_{k+1}(z+h)-G_k(z+h)|&=\left|\frac{\det M(z-h)\cdot (
      G_k(z-h)-G_{k-1}(z-h))} {(M_{21}(z) G_k(z)+M_{22}(z))(M_{21}(z)
      G_{k-1}(z)+M_{22}(z))}\right|\\&\le
  \frac1{p^2}|G_k(z-h)-G_{k-1}(z-h)|.    
  \end{split}
\end{gather*}
Here, we have used $\det M=1$; this is the only place in the proof of
Proposition~\ref{le:resolvent-set:1} where we use this property. The
above estimates then  imply~\eqref{DG}.\\
Denote by $G$ the limit of $\{G_k(\cdot)\}_{k=0}^\infty$. Clearly, $G$
is a solution to~\eqref{homothety} and satisfies $|G(z)|\le q$.\\
{\bf 4.} Consider equation~\eqref{homothety:1} with $G$ constructed in
the previous step. To construct a solution to this equation, it
suffices to define it on the interval $[0,h)$ and continue it outside
this interval by induction using equation~\eqref{f-d-e-on-R}; that is,
for $n\geq1$ and $z\in[0,h)$, one sets
\begin{gather*}
  \psi_2(z+nh)=(M_{21}(z+(n-1)h)
  G(z+(n-1)h)+M_{22}(z+(n-1)h))\psi_2(z+(n-1)h),\\
  \psi_2(z-nh)=(M_{21}(z-nh)
  G(z-nh)+M_{22}(z-nh))^{-1}\psi_2(z-(n-1)h).
\end{gather*}
Note that, for all $z\in\R$, one has $|M_{21}(z) G(z)+M_{22}(z)|\ge
\theta-m(q+1)=p>0$, and, so, for $n\geq1$ and $z\in[0,h)$, the second
formula correctly defines $\psi_2$ for negative $z$.\\
Let $\psi_2(z)=1$ for $z\in[0,h)$. Then, by construction, for $z\in
[-nh, -(n-1)h)$ and $n\in\N$, one has $|\psi_2(z)|\le p^{-n}$. In
terms of $\psi_2$, we construct a vector solution
to~\eqref{f-d-e-on-R} as described in step 1. We denote the thus
constructed solution by $\psi_-$; it satisfies
\begin{equation*}
  \psi_-(z)=\begin{pmatrix} G(z) \\ 1\end{pmatrix}\text{ for }\ 
  z\in[0,h)\quad\text{and}\quad\|\psi_-(-z)\|\le\sqrt{1+q^2}e^{-\log
  p\cdot z/h}\le \sqrt{1+q^2}e^{-\gamma z/h}\text{ for } z\ge 0, 
\end{equation*}
where $\gamma=\log\frac{\theta+\theta^{-1}}2$.\\
{\bf 5.} \ Construct the solution $\psi_+$. Therefore, consider the
equation~\eqref{f-d-e-on-R} with the matrix $\sigma
M^{-1}(-z-h)\sigma$ replacing $M(z)$. The matrix $\sigma
M^{-1}(-z-h)\sigma$ can be written in the
form~\eqref{f-d-e-on-R:Mform} with the old $\theta$ and the matrix
$\sigma \tilde M^{-1}(-z-h)\sigma$ instead of $\tilde M(z)$. Clearly,
with the matrix $M$, it satisfies all the hypotheses of
Proposition~\ref{le:resolvent-set:1}. So, as when constructing
$\psi_-$, we can construct $\tilde\psi_-$, a solution to the
equation~\eqref{f-d-e-on-R} for the matrix $\sigma
M^{-1}(-z-h)\sigma$.  When constructing this solution, in the last
step, we normalize it by setting $\tilde\psi_2=1$ on the interval
$(-h,0]$. Then, we get
\begin{equation*}
  \tilde \psi_-(z)=\begin{pmatrix}\tilde G(z) \\ 1\end{pmatrix}\text{
  for }\ z\in[-h,0)\quad\text{and}\quad \|\tilde \psi_-(-z)\|\le
  \sqrt{1+q^2}e^{-\gamma (z-h)/h}\text{ for } z\ge 0.
\end{equation*}
Here, $\tilde G$ is a function satisfying the estimate $|\tilde
G(z)|\le q$ for all $z\in\R$.\\
Having constructed $\tilde \psi_-$, we define $\psi_+:\R\mapsto\C^2$
by the formula $\psi_+(z)=\sigma\tilde\psi_-(-z)$.  The
function $\psi_+$ satisfies~\eqref{f-d-e-on-R} for the matrix $M$, and
one has
\begin{equation*}
  \psi_+(z)=\begin{pmatrix}1\\ \tilde G(-z) \end{pmatrix}\text{
  for }\ z\in[0,h)\quad\text{and}\quad\|\psi_+(z)\|\le
  e^{\gamma}\sqrt{1+q^2}e^{-\gamma z/h}\text{ for } z\ge 0.
\end{equation*}
{\bf 6.} To complete the proof of
Proposition~\ref{le:resolvent-set:1}, we need only to check that
$\det(\psi_+,\psi_-)\ne 0$. From equation~\eqref{f-d-e-on-R}, it
follows that this determinant is $h$-periodic.  So, it suffices to
consider $z\in[0,h)$. Then, one has $\det(\psi_+,\psi_-)=1-G(z)\tilde
G(-z)$, and so, $|\det(\psi_+,\psi_-)-1|= |G(z)\tilde G(-z)|\le
q^2<1$. This completes the proof of
Proposition~\ref{le:resolvent-set:1}.\qed
\subsubsection{Completing the proof of Theorem~\ref{thr:5}}
\label{sec:compl-proof-theor}
The proof of Theorem~\ref{thr:5} consists of three steps.  In the
first two steps, we apply Proposition~\ref{resolvent-set:0} to the
monodromy equations with the matrix $M^\pi$ and $M^0$ (see
Remark~\ref{rem:2}).  When we can do it, $E$ is outside the spectrum
of $H_{z,\varepsilon}$. In this case, we see that, if $E$ is in the
spectrum of $H_{z,\varepsilon}$, then, it satisfies~\eqref{eq:37}. In
the third step, we analyze the case when one can not apply
Proposition~\ref{resolvent-set:0}; using the a priori estimate from
Lemma~\ref{le:apriory-est}, we see that then $E$ is
outside the spectrum of $H_{z,\varepsilon}$.\\
Below, we consider only $z\in\R$.
Fix $\delta_\xi>0$. The precise choice of this constant will be
described later using the a priori estimate~\eqref{apriory-est}.\\
{\bf 1.} \ By Theorem~\ref{th:M-matrix}, one has
\begin{equation*}
  M_{12}^\pi=\sigma r\tau (\xi_0+\sin(2\pi(z-z_0))
  +e^{-\frac{\delta}\varepsilon}\cdot O(r\tau (|\xi_0|+1),1).
\end{equation*}
Assume that $E$ satisfies
\begin{equation}
  \label{xi0-big}
  \xi_0(E)\ge e^{\delta_\xi/\varepsilon}\cdot
  \max\{1,\,(r\tau)^{-1}e^{-\frac{\delta}\varepsilon}\}.
\end{equation}
Then, we get
\begin{equation}\label{M12:small-tau}
M_{12}^\pi=\sigma\,r\,\tau\,\xi_0(1+o(1)).
\end{equation}
So, we have $M_{12}^\pi\ne 0$. \\
In terms of the monodromy matrix $M^\pi$,
define the functions $\rho$ and $v$ by~\eqref{rho,v}. By means
of~\eqref{M12:small-tau} and Theorem~\ref{th:M-matrix}, for
sufficiently small $\varepsilon$, we get
\begin{gather}
  \label{rho:small-tau}
  \rho=1+o(1),\\
  \label{v:small-tau}
  v=\sigma \left(\tau^2 \left(\xi_0+\sin(2\pi(z-z_0))\right)
    \left(\xi_\pi+\sin(2\pi(z-z_\pi))+
      o(|\xi_\pi|+1)\right)+2\Lambda_n(V)\right)+o(1);
\end{gather}
when deriving the representation for $v$, we have also
used~\eqref{theta:as} and~\eqref{theta-omega}. \\
As the coefficients of $M^\pi$ are real when $E$ and $z$ are real, one
has $\ind \rho=\ind v=0$ as soon as $|\rho|$ is bounded away from
zero, and $v$ satisfy the second condition
in~\eqref{resolvent-set:cond}.
By~\eqref{small-tau},~\eqref{rho:small-tau} and~\eqref{v:small-tau},
there exists a function $\varepsilon\mapsto f_0(\varepsilon)$ that is
$o(1)$ when $\varepsilon$ is small such that the first bound
in~\eqref{resolvent-set:cond} is satisfied when
\begin{equation}
  \label{0-condition}
  \left|\tau^2 \xi_0\xi_\pi+2\Lambda_n(V)\right|\ge 
  \left(2+\tau^2|\xi_0|+\tau^2|\xi_\pi|\right)\,(1+f_0).
\end{equation}
So, for sufficiently small $\varepsilon$, if $E$
satisfies~\eqref{0-condition}, it is outside the
spectrum of $H_{z,\varepsilon}$.\\
{\bf 2.} \ Now, assume that $E$ satisfies
\begin{equation}
  \label{xipi-big}
  |\xi_\pi(E)|\ge e^{\delta_\xi/\varepsilon}\cdot
  \max\{1,\,r\tau^{-1}e^{-\frac{\delta}\varepsilon}\bar T_h\}.
\end{equation}
In this case, for sufficiently small $\varepsilon$, there
exists a function $\varepsilon\mapsto f_\pi(\varepsilon)$ that is
$o(1)$ for $\varepsilon$ small such that, if $E$ satisfies
\begin{equation}
  \label{pi-condition}
  \left|\tau^2 \xi_0\xi_\pi+2\Lambda_n(V)\right|\ge 
  \left(2+\tau^2|\xi_0|+\tau^2|\xi_\pi|\right)\,(1+f_\pi).
\end{equation}
then $E$ is outside the spectrum of $H_{z,\varepsilon}$.\\
Though this result can be obtained by using directly the matrix
$M_\pi$, the proof becomes immediate if, instead of $M^\pi$, one uses
the matrix $M^0$, see Remark~\ref{rem:2}.  Note that the
conditions~\eqref{xi0-big} and~\eqref{xipi-big} are equivalent
respectively to
\begin{equation*}  
\xi_0(E)\ge e^{\delta_\xi/\varepsilon}\cdot
  \max\{1,\, 2e^{-\frac{\delta}\varepsilon} \bar T_{h}/\bar
  T_{v,0}\}\quad {\rm and } \quad 
\xi_\pi(E)\ge e^{\delta_\xi/\varepsilon}\cdot
  \max\{1,\, 2e^{-\frac{\delta}\varepsilon} \bar T_{h}/\bar
  T_{v,\pi}\},
\end{equation*}
and  when one swaps the indices $0$ and $\pi$, one swaps these two
conditions. So, the proof of~\eqref{pi-condition} is obtained from
the one of~\eqref{0-condition} just by swapping the indices. 
This completes the second step.\\
{\bf 3.} We prove
\begin{Le}
  \label{xi-on-spectrum}
  Fix $\delta_\xi$ so that
  \begin{equation}
    \label{delta-xi-def}
    0<\delta_\xi< \max\{\delta_\tau,\delta_0\}.
  \end{equation}
  For sufficiently small $\varepsilon$, in the case of
  Theorem~\ref{thr:5}, all the energies $E$ that satisfy
  neither~\eqref{xi0-big} nor~\eqref{xipi-big} are outside the
  spectrum of $H_{z,\varepsilon}$.
\end{Le}
\demo Pick $E$ that satisfies neither~\eqref{xi0-big}
nor~\eqref{xipi-big}. Then, one has
\begin{align*}
  \tau^2(|\xi_\pi(E)|+1)(|\xi_0(E)|&+1)\le C \tau^2
  e^{2\delta_\xi/\varepsilon}\cdot 
  \left(1+r\tau^{-1}e^{-\frac{\delta}\varepsilon}\bar T_h\right)\,
  \left(1+(r\tau)^{-1}e^{-\frac{\delta}\varepsilon}\right)\\
  &=C e^{2\delta_\xi/\varepsilon}\cdot  \left(\tau^2+
    e^{-\frac{\delta}\varepsilon}\left(\bar T_{v,\pi}+ \bar
      T_{v,0}\right)+
    e^{-2\frac{\delta}\varepsilon}\bar T_h\right)\le C
  e^{2\delta_\xi/\varepsilon}\cdot\left(e^{-\frac{2\delta_\tau}\varepsilon}+
    e^{-\frac{\delta+2\delta_0}\varepsilon}\right),
\end{align*}
where, in the second step, we have used~\eqref{tau:ext}, and, in the
last step, we have used~\eqref{T-bars} and~\eqref{eq:6}. In view of
the last computation and~\eqref{delta-xi-def}, we get that
$\tau^2(|\xi_\pi(E)|+1)(|\xi_0(E)|+1)=o(1)$, and, in view of
Lemma~\ref{le:apriory-est}, for sufficiently small $\varepsilon$, this
implies, that $E$ is outside the spectrum of $H_{z,\varepsilon}$.
This completes the proof of Lemma~\ref{xi-on-spectrum}.\qed\\
Now, let $f(\varepsilon)=\max\{f_0(\varepsilon),
f_\pi(\varepsilon)\}$.  Clearly, for sufficiently small $\varepsilon$,
the function $f$ is well defined and satisfies $f=o(1)$ near $0$. By
Lemma~\ref{xi-on-spectrum} and the first two steps, we see that, if
$E$ in $R$ is in the spectrum of $H_{z,\varepsilon}$, then, it
satisfies $\left|\tau^2 \xi_0\xi_\pi+2\Lambda_n(V)\right|\le
\left(2+\tau^2|\xi_0|+\tau^2|\xi_\pi|\right)\, (1+f)$. This completes
the proof of Theorem~\ref{thr:5}.\qed
\subsection{Properties of the set defined by~\eqref{eq:37}}
\label{sec:G-F:37}
We now analyze~(\ref{eq:37}) and prove Proposition~\ref{pro:1}. \\
Let
\begin{equation}
  \label{G-F}
  G(E)=\tau^2\xi_0(E)\xi_\pi(E)+2\Lambda_n(V),\quad\text{and}\quad 
  F(E)=\left(2+\tau^2|\xi_0(E)|+\tau^2|\xi_\pi(E)|\right)(1+
  f(\varepsilon)),
\end{equation}
where the function $\varepsilon\mapsto f(\varepsilon)$ is defined in
Theorem~\ref{thr:5}. So, the condition~\eqref{eq:37} takes the form:
\begin{equation}
  \label{G-F:37}
  |G(E)|\le F(E).
\end{equation}
\subsubsection{Individual properties of $F$ and $G$}
\label{sec:indiv-prop-f}
We first discuss various properties of the functions $G$ and $F$
without comparing their graphs, i.e., without
analyzing~(\ref{G-F:37}).
\begin{Le}
  \label{le:G,F}
  One has
  \begin{enumerate}
  \item $G$ is a quadratic polynomial in $E$; 
  \item it reaches its maximum at $\bar E$;
  \item $G(E_0)=G(E_\pi)=2\Lambda_n(V)>2$;
  \item $F$ is affine except at the points $E_0$ and $E_\pi$;
  \item $F$ is strictly increasing for $E>\max\{E_0,E_\pi\}$, strictly
    decreasing for $E<\min\{E_0,E_\pi\}$;
  \item between $E_0$ and $E_\pi$, the absolute value of the
    derivative of $F$ is smaller than it is for $E>\max\{E_0,E_\pi\}$
    or $E<\min\{E_0,E_\pi\}$;
  \item $F(E_\pi)=(2+\tau^2|\xi_0(E_\pi)|)(1+f)$ and
    $F(E_0)=(2+\tau^2|\xi_\pi(E_0)|)(1+f)$.
  \end{enumerate}
\end{Le}
\demo Lemma follow from~\eqref{xi-nu} and~(\ref{G-F}). To prove point
(2) one also uses~(\ref{gamma:ext}) and~(\ref{eq:21}). \qed
\subsubsection{The intervals $I_{\rm in}$ and $I_{\rm out}$}
\label{sec:intervals-i_rm-in}
Now, we begin the analysis of condition~\eqref{G-F:37}.  First,
we describe the set $\{E\in\R;\ G(E)\ge F(E)\}$.
\begin{Cor}
  \label{I-in}
  For sufficiently small $\varepsilon$, the set $I_{\rm in}=\{E\in\R;\ 
  G(E)\ge F(E)\}$ is a compact interval of positive length. It is
  located strictly between the zeros of the polynomial $G$.
\end{Cor}
\demo Lemma~\ref{le:G,F} implies that $G$ is concave (points (1) and
(2)). As $F$ is positive, $I_{\rm in}$ (if not empty) is located
between the zeros of $G$.  By the points (4) -- (6) of
Lemma~\ref{le:G,F}, $F$ is convex.  So, now, it suffices to prove,
that for sufficiently small $\varepsilon$, there exists a point
$\tilde E$, where $G(\tilde E)>F(\tilde E)$.  Therefore, note that
$0=|\xi_0(E_0)|\le |\xi_\pi(E_0)|$, and that $0=|\xi_\pi(E_\pi)|\le
|\xi_0(E_\pi)|$.  This implies that, between $E_0$ and $E_\pi$, there
is $\tilde E$, a point where $\xi_0(\tilde E)=\xi_\pi(\tilde E)$.
Denote this common value by $\xi$.  At $\tilde E$, one has
\begin{equation*}
  \begin{split}
  G(\tilde E)-F(\tilde
  E)&=\tau^2|\xi|^2+2\Lambda_n(V)-2(1+\tau^2|\xi|)(1+f)\\&=
  \tau^2(|\xi|-1-f)^2+2(\Lambda_n(V)-1-f)-(1+f)^2 \\
  &\ge 2(\Lambda_n(V)-1-f) -\tau^2(1+f)^2.    
  \end{split}
\end{equation*}
Therefore, for sufficiently small $\varepsilon$, one has $G(\tilde
E)-F(\tilde E)>\Lambda_n(V)-1>0$. This completes
the proof of Corollary~\ref{I-in}.\qed\\
We now prove
\begin{Cor}
  \label{I-out} 
  The set $I_{\rm out}=\{E\in\R;\ G(E)\ge -F(E)\}$ is a compact interval of
  positive length. Moreover, $I_{\rm in}$ is
  contained in $\dot I_{\rm out}$, the interior of $I_{\rm out}$.
\end{Cor}
\demo Let $\varepsilon$ be so small that $I_{\rm in}$ exists and that
$F>0$ for all  $E\in\R$. On the compact interval
bounded by $E_\pi$ and $E_0$, the function $G$ is positive (see points
(2) and (3) of Lemma~\ref{le:G,F}), whereas $-F$ is negative. This,
the facts that $G$ is a concave quadratic polynomial  and $-F$ is
piecewise affine  and concave (by  Lemma~\ref{le:G,F}) 
imply that $I_{\rm out}$ is a compact  interval of positive length.  
The inclusion $I_{\rm in}\subset\dot I_{\rm out}$  follows from the 
inequality  $G(E)\ge F(E)>-F(E)$ valid for $E\in I_{\rm in}$.\qed
\subsubsection{The intervals $I_{\rm l}$ and $I_{\rm r}$}\label{I-lr}
\label{sec:intervals-i_rm-l}
Let  $\Sigma(\varepsilon)$ be the set where condition~\eqref{G-F:37}
is  satisfied. One has $\Sigma(\varepsilon)=I_{\rm out}\setminus\dot 
I_{\rm  in}$. Corollaries~\ref{I-in} and~\ref{I-out} imply 
\begin{Cor}
  \label{two-int-cor}  
  For sufficiently small $\varepsilon$, the set $\Sigma(\varepsilon)$
  consists of two disjoint compact intervals of positive length.
\end{Cor}
\noindent We denote these intervals by $I_{\rm l}$ and $I_{\rm r}$ so
that $I_{\rm l}$ be to the left of $I_{\rm r}$. We finally check
\begin{Le}
  \label{le:I0Ipi:neighb}
  For $\varepsilon$ sufficiently small, both $I_l$ and $I_r$ are
  inside the $(2e^{-\delta_0/\varepsilon})$-neighborhood of
  $\bar E$.
\end{Le}
\demo It suffices to check that outside the
$(2e^{-\delta_0/\varepsilon})$-neighborhood of $\bar E$, one has
$|G(E)|\ge F(E)$. By~\eqref{eq:6} and Lemma~\ref{le:Phi'}, for $E$ such that $|E-\bar E|\geq
2e^{-\delta_0/\varepsilon}$ and $\nu\in\{0,\pi\}$, one has
\begin{equation*}
  |\xi_\nu(E)|=\frac{|\check\Phi_\nu'(\bar E)(E-E_\nu)|}
  {\varepsilon\bar T_{v,\nu}}\,\geq\frac{e^{\frac{\delta_0}\varepsilon}}{C\varepsilon},
  \quad
  \tau^2|\xi_0(E)\xi_\pi(E)|=
  \frac{4|\check\Phi_0'(\bar E)\check\Phi_\pi'(\bar E)(E-E_0)(E-E_\pi)|}
  {\varepsilon^2\bar T_h}\geq\frac1{C\varepsilon^2}.
\end{equation*}
For sufficiently small $\varepsilon$, these estimates imply that,
for $|E-\bar E|\geq 2e^{-\delta_0/\varepsilon}$ and $\nu\in\{0,\pi\}$,
\begin{equation*}
  |G(E)|-F(E)\ge
  \tau^2(|\xi_0(E)|-1-f)(|\xi_\pi(E)|-1-f)-2(\Lambda_n(V)-1-f)-
  \tau^2(1+f)^2\ge \frac1{C\varepsilon^2}.
\end{equation*}
This completes the proof of Lemma~\ref{le:I0Ipi:neighb}.\qed\\ \\
Corollary~\ref{two-int-cor} and Lemma~\ref{le:I0Ipi:neighb} prove
Proposition~\ref{pro:1}.
\subsection{Computing the density of states}
\label{sec:comp-dens-stat}
We now compute the increments of the integrated density of states on
each of the intervals defined in Proposition~\ref{pro:1} and, thus,
prove Theorem~\ref{thr:6}.\\
We assume that
\begin{equation}
  \label{small-tau:ids:T-cases}
  \bar T_{v,\pi}\ge \bar T_{v,0}.
\end{equation}
The complementary case is treated similarly, but instead of working
with $M^\pi$, one uses the monodromy matrix $M^0$ mentioned in
Remark~\ref{rem:2}.  For sake of definiteness, we assume that
$\sigma=1$ in~\eqref{mm-as} and that
\begin{equation}\label{E0:greater}
  E_\pi\le E_0.
\end{equation}
The cases where $E_\pi\ge E_0$ or $\sigma=-1$ are analyzed similarly.\\
Our main tool is Proposition~\ref{pro:ids:1}.  First, we compute the
increment of the integrated density of states on the whole set
$\Sigma_\varepsilon$ and, then, we compute it on one of the intervals of
this set.
\subsubsection{The increment of the IDS on the set
  $\Sigma_\varepsilon$}
\label{sec:increment-ids-whole}
As the curve used to apply Theorem~\ref{pro:ids:1}, we choose
\begin{equation}\label{small-tau:ids:curve:1}
\gamma=\{E\in\C; \ |E-\bar E|=2\,e^{-\delta_0/\varepsilon}, \ \im E\ge0\}.
\end{equation} 
Lemma~\ref{le:I0Ipi:neighb} implies that the set
$\Sigma_\varepsilon$ is strictly between the ends of $\gamma$.\\
The analysis of the increment of the integrated density of states
between the ends of $\gamma$ is standard, see
section~\ref{sec:comp-integr-dens}.  We omit details and note only
that, first, in terms of $M^\pi$, one defines $v$ and $\rho$
by~(\ref{rho,v}), and, then, one checks that for $\gamma$ defined
in~\eqref{small-tau:ids:curve:1}, the statements of Lemma~\ref{le:2}
hold.  As in part 2 of section~\ref{sec:comp-integr-dens}, this
implies that the increment of the integrated density of states between
the ends of $\gamma$ is equal to $\varepsilon/\pi$.
\subsubsection{The  increment of the IDS on one of the intervals of
  $\Sigma_\varepsilon$} 
\label{sec:increment-ids-one}
To complete the proof of Theorem~\ref{thr:6}, we pick $E_*$, a point
in between the connected the components of the set
$\Sigma_\varepsilon$, and prove that, between $E_*$ and a point
located on $\R$ outside the
$(2e^{-\delta_0/\varepsilon})$-neighborhood of $\bar E$, the increment
of the integrated density of is equal to $\varepsilon/(2\pi)$. \\
We define $E_*=E_0-\bar T_h-\bar T_{v,0}$ and
\begin{equation*} 
  \gamma_{\perp}=\{E\in \C; \ \re E=E_*,\ E\text{ is between }\R
  \text{ and }\gamma \}.
\end{equation*}
Now, let $\gamma_*$ be the curve going from $E_*$ along $\gamma_\perp$
to $\gamma$ and, then, along $\gamma$ to $\R$ in the clockwise
direction. This is the curve we use to apply Theorem~\ref{pro:ids:1}.\\
Let us study $v$ and $\rho$ on the curve $\gamma_\perp$.  We shall use
\begin{Le} 
  \label{E0-neighb} 
  For sufficiently small $\varepsilon$, for $z\in \R$ and $E=E_*$, one
  has
  \begin{gather}
    \label{xi0-1}
    |\xi_0|\asymp \varepsilon^{-1}(1+\bar T_h/\bar T_{v,0}),\\
    \label{xi0-2}  
    |\xi_0|\ge C\varepsilon^{-1},\quad |r\tau \xi_0|\ge
    C\varepsilon^{-1},\\
    \label{xi0-3}
    \tau^2|\xi_0|=o(1).
  \end{gather}
\end{Le}
\demo Estimate~(\ref{xi0-1}) follows from the definitions of $E_*$ and
$\xi_0$ and the bounds $\check\Phi'\asymp 1$, see Lemma~\ref{le:Phi'}.  
The other estimates follow from~(\ref{xi0-1})
and the definitions of $r$ and $\tau$. \QED
Now, we can easily check 
\begin{Le} 
  \label{small-tau:ids:rho}
  For sufficiently small $\varepsilon$, for $z\in \R$ and
  $E\in\gamma_\perp$, one has
  \begin{equation}
    M_{12}^\pi\ne 0,\quad \rho=1+o(1),\quad \ind \rho=0.
  \end{equation}
\end{Le}
\demo By~(\ref{mm-as}), for $z\in \R$, we have $M_{12}^\pi= r\tau
\xi_0(1+o(1))+ O(r\tau)+ o(1)$.  Note that estimates~(\ref{xi0-2}) are
valid as at $E=E_*$ so on the whole curve $\gamma_\perp$. Therefore,
under the conditions of Lemma~\ref{small-tau:ids:rho}, $M_{12}^\pi=
r\tau \xi_0(1+o(1))$. This implies all the statements of
Lemma~\ref{small-tau:ids:rho}.\QED
Now, we turn to the function $v$. We prove
\begin{Pro}
  \label{small-tau:ids:pro}
  For sufficiently small $\varepsilon$, for $z\in\R$, one has
  \begin{itemize}
  \item if $E\in\gamma_\perp$, then $\re v\ge 2\Lambda_n+o(1)$;
  \item if $E=E_*$, then $v=(\tau^2 \xi_0\,\xi_\pi
    +2\Lambda_n)(1+o(1))$.
  \end{itemize}
\end{Pro}
\begin{Rem}
  \label{rem:4}
  When proving Proposition~\ref{small-tau:ids:pro}, we shall see that,
  for sufficiently small $\varepsilon$ and $E=E_*$,
  \begin{equation*}
    \tau^2  \xi_0=o(1),\quad 
    \tau^2  \xi_\pi=o(\tau^2 \xi_\pi\, \xi_0), \quad \text{and}\quad
    \tau^2\xi_0\xi_\pi\ge o(1),
  \end{equation*}
  see~(\ref{small-tau:ids:errors:1}) and~(\ref{4.38}). 
  This implies that, for sufficiently small $\varepsilon$, the point
  $E_*$ is inside the interval $I_{\rm in}$, i.e., between the
  intervals of the set $\Sigma_\varepsilon$ described
  by~\eqref{eq:37}.
\end{Rem}
\noindent Proposition~\ref{small-tau:ids:pro} immediately follows from
the next two lemmas.
\begin{Le}
  \label{small-tau:ids:le:1} 
  For sufficiently small $\varepsilon$, for $z\in\R$ and $E\in
  \gamma_{\perp}$, one has
  \begin{equation}
    \label{small-tau:ids:ReV}
    \re v= \tau^2 \re\xi_0\,\re\xi_\pi (1+o(1))
    -\tau^2 \im\xi_0\,\im\xi_\pi (1+o(1))+ 2\Lambda_n+o(1).
  \end{equation}
\end{Le}
\demo Recall that $\tau$ is small.  Using~(\ref{M11:details}), the representation
$M_{22}^\pi= \bar \theta+o(1)$ following from~(\ref{mm-as}), and the
asymptotics $\rho=1+o(1)$, see Lemma~\ref{E0-neighb}, we get for $z\in
\R$ and $E\in\gamma_\perp$
\begin{equation}
  \label{small-tau:ids:ReV:a}
  \begin{split}
    \re v=\tau^2&\left(\re\xi_0\,\re\xi_\pi
      -\im\xi_0\,\im\xi_\pi\right)+2\Lambda_n\\   
      &+\tau^2\re(\xi_0)\sin(2\pi(z-z_0))+
      \tau^2\re (\xi_\pi)\sin(2\pi(z-z_\pi))\\
      &\hspace{1cm} + e^{-\delta/\varepsilon}\, \tau^2\,\re
      \left(O(\xi_0\xi_\pi) +O(\xi_0)+O(\xi_\pi)\right)+o(1),
    \end{split}
\end{equation}
where the terms $O(\xi_0\xi_\pi)$, $O( \xi_0)$ and $O(\xi_\pi)$ are
analytic in $E$. Let us study the terms in the second and the third
lines of this formula.\\
First, we prove that
\begin{equation}
  \label{small-tau:ids:errors:1}   
  \tau^2 \re (\xi_0)=o(1),\quad 
  \tau^2 \re (\xi_\pi)=o(\tau^2\re \xi_\pi\,\re \xi_0).
\end{equation}
Clearly,
\begin{equation*}   
  \re (\xi_0)=\xi_0(E_*),\quad 
  \re (\xi_\pi)=\xi_\pi(E_*).
\end{equation*}
As $|\xi_0(E_*)|\gg1$ and $\tau^2|\xi_0(E_*)|\ll1$, see~\eqref{xi0-2}
and~\eqref{xi0-3}, this implies~\eqref{small-tau:ids:errors:1}.\\
Now, prove the estimate
\begin{equation}
  \label{small-tau:ids:errors:2}   
  \re O(\xi_0\xi_\pi)= O(\re\xi_0\,\re\xi_\pi)
  +O(\im\xi_0\,\im\xi_\pi).
\end{equation}
Note that as in~(\ref{M11:details}), the term $O(\xi_0\xi_\pi)$ is a
real analytic function of $E$ bounded by $C|\xi_0\xi_\pi|$ uniformly
in the $(4e^{-\frac{\delta_0}\varepsilon})$-neighborhood of $\bar E$.
Therefore, it can be represented in the form
\begin{equation*}
  O(\xi_0\xi_\pi)=\xi_0\xi_\pi\,f,
\end{equation*}
where $g$ is a real analytic function of $E$
satisfying the estimates 
\begin{gather*}
  |g(E)|\le C\text{ for } |E-\bar E|\le  4e^{-\frac{\delta_0}\varepsilon};\\
  |g'(E)|\le C e^{\frac{\delta_0}\varepsilon},\text{ for } |E-\bar
  E|\le 3e^{-\frac{\delta_0}\varepsilon};\\
  |\im g(E)|\le C\,|\im E|\, e^{\frac{\delta_0}\varepsilon},\text{ for
  } |E-\bar E|\le 3e^{-\frac{\delta_0}\varepsilon}.
\end{gather*}
As
\begin{equation*}   
  |\re O(\xi_0\xi_\pi)|\le|\re (\xi_0\xi_\pi) \re
  g|+|\im(\xi_0\xi_\pi)\im g|,  
\end{equation*}
the estimates for $\im g$ and $|g|$  imply that
\begin{equation*}
  \re O(\xi_0\xi_\pi)=
  O(\re\xi_0\,\re\xi_\pi)+O(\im\xi_0\,\im\xi_\pi)+
  e^{\delta_0/\varepsilon}\left(O(\im\xi_0\,\re \xi_\pi\, \im E)+
    O(\im \xi_\pi\,\re \xi_0\,\im E)\right).
\end{equation*}
Using the definitions of $\xi_0$ and $\xi_\pi$, on $\gamma_\perp$, we
get
\begin{equation*}
  \im\xi_0\,\re \xi_\pi\, \im E=\im\xi_0\im\xi_\pi
  (E_*-E_\pi)\text{ and }\im \xi_\pi\,\re \xi_0\,\im
  E=\im\xi_0\im\xi_\pi (E_*-E_0).
\end{equation*}
As both $|E_*-E_0|$ and $|E_*-E_\pi|$ are bounded by
$Ce^{-\delta_0/\varepsilon}$, we finally
get~(\ref{small-tau:ids:errors:2}).\\
Using the same techniques, for each $\nu\in\{0,\pi\}$, one also proves that
\begin{equation*}
  \begin{split}
  \re O(\xi_\nu)&=O(\re\xi_\nu)+e^{\delta_0/\varepsilon} O(\im
  \xi_\nu\,\im E)\\&=O(\re \xi_\nu)+ \bar T_{v,\mu}
  e^{\delta_0/\varepsilon}o(\im\xi_0\,\im\xi_\pi)=O(\re\xi_\nu)
  +o(\im\xi_0\,\im\xi_\pi),     
  \end{split}
\end{equation*}
where $\mu$ is the index complementary to $\nu$ in $\{0,\pi\}$.  In
view of~(\ref{small-tau:ids:errors:1}), this implies that
\begin{equation}
  \label{small-tau:ids:errors:3}   
  \tau^2\re ( O( \xi_0)+O(\xi_\pi))=o(1)+
  o(\tau^2\re\xi_0\re\xi_\pi)+o(\tau^2\im\xi_0\,\im\xi_\pi).
\end{equation}
Substituting estimates~\eqref{small-tau:ids:errors:1}
--~\eqref{small-tau:ids:errors:3} into~\eqref{small-tau:ids:ReV:a}, we
come to~\eqref{small-tau:ids:ReV}. This completes the proof of
Lemma~\ref{small-tau:ids:le:1}.\QED
\begin{Le}
  \label{small-tau:ids:poly} 
  For sufficiently small $\varepsilon$ and $E\in \gamma_{\perp}$, one
  has
  \begin{equation}\label{4.38}
    \im\xi_0\,\im\xi_\pi\le 0,\quad \tau^2 \re\xi_0\re\xi_\pi\ge o(1).
  \end{equation}
\end{Le}
\demo It follows from the definitions of $\xi_0$ and $\xi_\pi$ that
\begin{equation}\label{small-tau:ids:xi0xipi}
\im\xi_0\im\xi_\pi= \frac{\check \Phi_0'(\bar E)\check \Phi_\pi'(\bar
  E)}{\varepsilon^2 \bar T_{v,0}\bar T_{v,\pi}}\,(\im E)^2,\quad 
\tau^2\re\xi_0\re\xi_\pi= \frac{4\check \Phi_0'(\bar E)\check \Phi_\pi'(\bar
  E)}{\varepsilon^2 \bar T_h}\,(E_*-E_0)(E_*-E_\pi).
\end{equation}
For each $\nu\in\{0,\pi\}$, one has $\check \Phi_\nu'(\bar
E)=\Phi_\nu'(\bar E)+o(1)$ (which follows from~(\ref{eq:17}) and the
Cauchy estimates). So, in view of~(\ref{eq:21}), we get $\check
\Phi_0'(\bar E)\check \Phi_\pi'(\bar E)<0$.
Therefore,~\eqref{small-tau:ids:xi0xipi} implies that
$\im\xi_0\,\im\xi_\pi\le 0$, and that, if $E_\pi\le E_*<E_0$, \ 
$\tau^2\re\xi_0\re\xi_\pi\ge 0$. To complete the proof of
Lemma~\ref{small-tau:ids:poly}, we need only to check that, if $E_*\le
E_\pi\le E_0$, one has
$\tau^2\re\xi_0\re\xi_\pi=o(1)$. But, for such values of $E_\pi$, the
second formula in~\eqref{small-tau:ids:xi0xipi} and the definition of
$E_*$  imply that
\begin{equation*}
|\re\xi_0\re\xi_\pi|\le \frac{C}{\varepsilon^2 \bar T_h}\,(\bar
 T_{v,0}+\bar T_h)^2=\frac{C}{\varepsilon^2}\,(\bar
 T_{v,0}^2/\bar T_h+2\bar T_{v,0}+ \bar T_h)\le
 \frac{C}{\varepsilon^2}\,(\tau^2+2\bar T_{v,0}+ \bar T_h)
\end{equation*}
which implies the needed estimate. This completes the proof of
Lemma~\ref{small-tau:ids:poly}.\QED
Now, we are ready to prove
\begin{Le}
  \label{le:small-tau:ids:arg} 
  For sufficiently small $\varepsilon$, the matrix $M^\pi$ and
  the curve $\gamma_*$ satisfy the assumptions of
  Theorem~\ref{pro:ids:1}.  One has
  \begin{equation}
    \label{small-tau:ids:arg}
    \left.\left.\int_0^1 \arg v(x,E) dx \right|_{\gamma_*}=
      \arg G(E) \right|_{\gamma_*}\quad\text{where} \quad
      G(E)=\tau^2\xi_0\xi_\pi+2\Lambda_n.
  \end{equation}
\end{Le}
\demo The assumptions of Proposition~\ref{pro:ids:1} are satisfied as
\begin{itemize}
\item as we have already mentioned, on $\gamma_*\cap \gamma$, the
  statement of Lemma~\ref{le:2} holds;
\item on $\gamma_*\cap\gamma_\perp$, the function $\rho$ is described
  by Lemma~\ref{small-tau:ids:rho} and, in view of the first point of
  Proposition~\ref{small-tau:ids:pro}, one has $\ind v=0$, and
  $|v|/2\ge \Lambda_n+o(1)$ \ (recall that $\Lambda_n\geq1$ is a
  constant depending only on $V$ and $n$, and that we consider the
  case where $\Lambda_n>1$).
\end{itemize}
Now, let us prove~(\ref{small-tau:ids:arg}).  As both $v$ and $G$ are
real analytic, the left and  the right hand sides
of~\eqref{small-tau:ids:arg} coincide modulo $\pi$. So, it suffices to
prove this equality up to $o(1)$. This follows from the observations:
\begin{itemize}
\item in view of Lemma~\ref{le:2},  on $\gamma_*\cap \gamma$, one has
  $v(x,E)=G(E) (1+o(1))$;
\item in view the previous point and the second point of
  Proposition~\ref{small-tau:ids:pro}, at the ends of $\gamma_*\cap
  \gamma_\perp=\gamma_\perp$, one also has  $v(x,E)=G(E) (1+o(1))$;
\item in view of the first point of
  Proposition~\ref{small-tau:ids:pro} and as $v$ is real analytic, one
  has
  \begin{equation*}
    -\pi/2<\left.\int_0^1 \arg v(x,E) dx\right|_{\gamma_\perp}<\pi/2.   
  \end{equation*}
\item in view of Lemma~\ref{small-tau:ids:poly}, on $\gamma_\perp$,
  $\re G(E)\ge 2\Lambda_n>0$, and so, as $G$ is real analytic,
  one has
  \begin{equation*}
    -\pi/2<\left. \arg G(E) \right|_{\gamma_\perp}<\pi/2.   
  \end{equation*}
\end{itemize}
This completes the proof of Lemma~\ref{le:small-tau:ids:arg}.\QED
Now, we note that $ \left.\arg G(E) \right|_{\gamma_*}=-\pi$.  This
follows from Remark~\ref{rem:4} and the statement of Lem\-ma~\ref{I-lr}
saying that $I_{\rm in}$ is located strictly between the zeros of the
polynomial $G$. So, Theorem~\ref{thr:6} follows from~(\ref{DeltaN})
and~(\ref{small-tau:ids:arg}).\QED
\subsection{Computing the Lyapunov exponent}
\label{sec:comp-lyap-expon-1}
We now prove Theorem~\ref{thr:7}.
The computations are essentially the same as in
section~\ref{sec:comp-lyap-expon}, but, instead of working with the
matrix cocycle $(M^\pi,h)$, we pass to an auxiliary one.\\
Below, we always assume that $E\in\Sigma_\varepsilon$, i.e., that it
satisfies~(\ref{eq:37}). Note that this implies that
\begin{equation}
  \label{small-tau:Le:xis:1}
  \tau^2|\xi_0||\xi_\pi|\le
  C(\tau^2|\xi_0|+\tau^2|\xi_\pi|+1),
\end{equation}
Fix a constant $\delta$ satisfying
\begin{equation}
  \label{delta-xi-def-1}
  0<\delta_\xi<\min(\delta, \delta_\tau,\delta_0).
\end{equation}
Then, by Lemma~\ref{xi-on-spectrum}, for any $E$
satisfying~(\ref{eq:37}), one has either~(\ref{xi0-big})
or~(\ref{xipi-big}). We prove Theorem~\ref{thr:7} assuming that $E$
satisfies~\eqref{xi0-big}. The other case is treated similarly, but,
instead of the matrix $M^\pi$, one uses the monodromy matrix $M^0$
introduced in Remark~\ref{rem:2}.
\subsubsection{Auxiliary matrix cocycle} 
\label{sec:auxil-matr-cocycle}
We use the following 
\begin{Le}[\cite{Fe-Kl:04a}, Lemma 4.7]
  \label{le:Le:1} 
   Let $M\in L^{\infty}(\R, SL(2,\C))$ be $1$-periodic, and let $h$ be an
irrational number. 
  Assume that there exists $A>1$ such that
  \begin{equation}
    \label{Le:1}
    \forall x\in\R,\quad A^{-1} \le M_{12}(x)\le A.
  \end{equation}
  In terms of $M$ and $h$, construct $v$ and $\rho$ by
  formulae~\eqref{rho,v}. Set
  \begin{equation}
    \label{Le:2}
    N(x)=\begin{pmatrix} v(x)/\sqrt{\rho(x)} & -\sqrt{\rho(x)} \\
      1/\sqrt{\rho(x)} & 0
    \end{pmatrix}.
  \end{equation}
  Then, the Lyapunov exponents for the matrix cocycles $(M,h)$ and
  $(N,h)$ are related by the formula
  \begin{equation}\label{Le:3}
    \theta(M,h)=\theta(N,h).
  \end{equation}
\end{Le}
\noindent Now, for $M=M^\pi$ and $h$ defined by~\eqref{h}, we
construct $N$ by formula~\eqref{Le:2}. Under the
condition~(\ref{xi0-big}), Theorem~\ref{th:M-matrix} implies that
\begin{equation}
  \label{small-tau:Le:M12}
  M^\pi_{12}=\sigma r \tau \xi_0 (1+O(pe^{-\delta_\xi/\varepsilon})).
\end{equation} 
So, we are in the case of Lemma~\ref{le:Le:1}.
\subsubsection{The coefficients of $N$}
\label{sec:coefficients-n}
Here, we check
\begin{Le}
  \label{le:small-tau:Le:N} 
  For sufficiently small $\varepsilon$, if $E$
  satisfies~\eqref{xi0-big} and if $2\pi|\im z|\leq
  \delta_\xi/\varepsilon$, one has
  \begin{equation}
    \label{small-tau:Le:N}
    N=\begin{pmatrix}
      \sigma\tau^2(\xi_0\sin(2\pi(z-z_\pi))+
      \xi_\pi\sin(2\pi(z-z_0))) & 0\\ 0 & 0 \end{pmatrix}
    +O\left(\tau^2(|\xi_0|+|\xi_\pi|)+1\right).
  \end{equation}
\end{Le}
\demo For sufficiently small $\varepsilon$, for $z$ and $E$ as in
Lemma~\ref{le:small-tau:Le:N}, representation~\eqref{small-tau:Le:M12}
implies that $\rho=1+o(1)$. 
This,~(\ref{mm-as}),~(\ref{delta-xi-def-1})
and~\eqref{small-tau:Le:xis:1} imply that
(for sufficiently small $\varepsilon$, for $z$ and $E$ we consider)
\begin{equation*}
  v=\sigma\tau^2(\xi_0\sin(2\pi(z-z_\pi))
  +\xi_\pi\sin(2\pi(z-z_0)))+O(\tau^2(|\xi_0|+|\xi_\pi|)+1).  
\end{equation*}
This representation and the representation $\rho=1+o(1)$
imply~\eqref{small-tau:Le:N}.\QED
\subsubsection{An upper  bound for $\Theta(N,h)$}
\label{sec:an-upper-bound}
Lemma~\ref{le:small-tau:Le:N} implies that, for sufficiently
$\varepsilon$, for $z\in \R$ and for $E$ satisfying~\eqref{xi0-big},
one has
\begin{equation*}
  \|N\|\le C(\tau^2(|\xi_0|+|\xi_\pi|)+1).
\end{equation*}
This and the definition of the Lyapunov exponent  for a matrix cocycle
imply that 
\begin{equation}
  \label{small-tau:Le:upper}
  \Theta(N,h)\le \log(\tau^2(|\xi_0|+|\xi_\pi|)+1)+C.
\end{equation}
\subsubsection{A lower  bound for $\Theta(N,h)$}
\label{sec:lower-bound-thetan}
Let us now show that, for sufficiently small $\varepsilon$, for all
$E$ satisfying~\eqref{xi0-big} and such that
\begin{equation}
  \label{small-tau:Le:xis:2}
  \tau^2(|\xi_0|+|\xi_\pi|)\ge 1,
\end{equation}
one has 
\begin{equation}
  \label{small-tau:Le:lower}
  \Theta(N,h)\ge \log(\tau^2(|\xi_0|+|\xi_\pi|))+C.
\end{equation}
The proof consists of two steps. \\
{\bf 1.} \ For sufficiently small $\varepsilon$, for all $E$
satisfying~\eqref{xi0-big} and~(\ref{small-tau:Le:xis:2}), one has either
\begin{gather}
  \label{small-tau:Le:a1}
  |\xi_0|\ge C/\tau^2 \text{ \ and \ } |\xi_\pi|\le C,\\
  \intertext{or}
  \label{small-tau:Le:a2}
  |\xi_\pi|\ge C/\tau^2 \text{ \ and \ } |\xi_0|\le C.     
\end{gather}
Indeed,~(\ref{small-tau:Le:xis:1}) and~(\ref{small-tau:Le:xis:2})
imply that $|\xi_0||\xi_\pi|\le C(|\xi_0|+|\xi_\pi|)$, and, for
sufficiently small $\tau$, this inequality
and~(\ref{small-tau:Le:xis:2}) imply the above alternative.\\
Below, we consider only the case of~(\ref{small-tau:Le:a1}). The
second case is treated similarly.\\
{\bf 2.} \ Fix $y$ so that $0<y<\delta_\xi$.
Lemma~\ref{le:small-tau:Le:N} implies that, for sufficiently
$\varepsilon$, for $y<2\pi\varepsilon |\im z|\le \delta_\xi$ and for
all $E$ satisfying~\eqref{xi0-big} and~\eqref{small-tau:Le:a1}, one
has
\begin{equation}
  \label{small-tau:Le:N:complex-z}
  N= \frac{\sigma\tau^2}{2i} \xi_0 e^{-2\pi i(z-z_\pi)}\left[
  \begin{pmatrix}
    1& 0\\  0 & 0 
  \end{pmatrix}
   +o(1)\right].
\end{equation}
This, Lemma~\ref{le:Le:2} and~\eqref{small-tau:Le:a1}
imply~\eqref{small-tau:Le:lower}.
\subsubsection{Completing the proof}
\label{sec:completing-proof}
From~\eqref{small-tau:Le:upper} and~\eqref{small-tau:Le:lower}, we
conclude that $\theta(N,h)= \log(\tau^2(|\xi_0|+|\xi_\pi|)+1)+O(1)$.
This and Theorem~\ref{Lyapunov} imply~(\ref{eq:40}).  This completes
the proof of Theorem~\ref{thr:7}.  \qed
\subsection{Absolutely continuous spectrum}
\label{sec:absol-cont-spectr}
Here, we prove Theorem~\ref{thr:1}. The proof consists of two main
steps. As when computing the density of states, we work under the
assumptions~(\ref{small-tau:ids:T-cases}) and~(\ref{E0:greater}).
\subsubsection{Properties of  the set $I_c^-$}
\label{sec:properties-set-i_c}
Let us discuss properties of the sets $I_c^-$ and $\tilde \Sigma^{\rm
  ac}(\varepsilon)$ defined in~(\ref{eq:34}) and~(\ref{eq:37:a}).
Recall that $\Delta=|E_\pi-E_0|/2$. Let
\begin{equation}
  \label{eq:M(Delta)}
  M(\Delta)=\min\left\{ \varepsilon \sqrt{\bar T_h},\,
  \frac{\varepsilon^2 \bar T_h}{\Delta}\right\}. 
\end{equation}
First, we check
\begin{Le}
  \label{M(Delta)}
  The set $\tilde\Sigma_{\rm ac}(\varepsilon)$ consists of two
  disjoint intervals $\tilde I_0$ and $\tilde I_\pi$ such that $\tilde
  I_0$ is to the right of $E_0$, $\tilde I_\pi$ is to the left of
  $E_\pi$; for each $\nu\in \{0,\pi\}$, one has
\begin{equation}
  \label{eq:Ipm:size}
  |\tilde I_\nu|\asymp M(\Delta)\quad\text{and}\quad \dist(\tilde
   I_\nu,\, E_\nu)\asymp M(\Delta),
\end{equation}
where $|I|$ denotes the length of an interval $I$. 
\end{Le}
\demo Using definitions of $\xi_0$ and $\xi_\pi$, see~(\ref{xi-nu}),
and the ones of $\tilde \Sigma_{\rm ac}$, see~(\ref{eq:37:a}), one
obtains $\tilde\Sigma_{\rm ac}(\varepsilon)=\tilde I_0\cup \tilde
I_\pi$ where
\begin{gather*}
\tilde I_\pi= [\bar E -b,\bar E-a],\quad \tilde I_0= [\bar E +a,\bar E+b],\\
a=\sqrt{\Delta^2+\varepsilon^2 \bar T_h \gamma (\Lambda_n-1-g)},
\quad b=\sqrt{\Delta^2+\varepsilon^2 \bar T_h \gamma
  (\Lambda_n+1+g)},\\
\gamma=-\frac12\,\left(\check\Phi_0'(\bar E)\,\check\Phi_\pi'(\bar
  E)\right)^{-1}.
\end{gather*}
Note that, by~(\ref{eq:21}) and~(\ref{eq:17}), one  obtains
$\gamma\asymp 1$; furthermore, recall that $\Lambda_n>1$.\\
The above formulae already imply that $I_0$ and $I_\pi$ are disjoint
and their length satisfy~(\ref{eq:Ipm:size}). Moreover,
as $E_0=\bar E+\Delta$ and $E_\pi=\bar E-\Delta$,
see~(\ref{E0:greater}), they also imply the
statements on the positions of  the intervals $\tilde I_0$
and $I_\pi$.  This completes the proof of Lemma~\ref{M(Delta)}. \QED
We shall use 
\begin{Cor}
  \label{cor:xi-on-tildeI}
  Let $d_0=\frac12\,\dist(E_0,\tilde I_0)$. On the $d_0$-neighborhood
  of $\tilde I_0$, 
\begin{equation*}
  |\xi_0(E)|\asymp \frac{M(\Delta)}{\varepsilon\bar
    T_{v,0}}\quad \text{and}\quad|\xi_\pi(E)|\asymp
  \frac{\Delta+M(\Delta)}{\varepsilon\bar 
    T_{v,\pi}}.
\end{equation*}
Let $d_\pi=\frac12\,\dist(E_\pi,\tilde I_\pi)$. On the
$d_\pi$-neighborhood of $\tilde I_\pi$, 
\begin{equation*}
  |\xi_0(E)|\asymp \frac{\Delta+M(\Delta)}{\varepsilon\bar
    T_{v,0}}\quad \text{and}\quad |\xi_\pi(E)|\asymp
  \frac{M(\Delta)}{\varepsilon\bar T_{v,\pi}}.
\end{equation*}
\end{Cor}
\demo Corollary~\ref{cor:xi-on-tildeI} follows from
Lemma~\ref{M(Delta)} and the definitions of $\xi_0$ and $\xi_\pi$.\QED
The following property of  $I_c^-$ plays important  role:
\begin{Le}
  \label{Ic-and-Ipm}
  Fix $c>0$ and $\nu\in \{0,\pi\}$. There is a positive constant $C$
  such that, for sufficiently small $\varepsilon$ the following holds.
  If $I_c^-\cap \tilde I_\nu\ne\emptyset$, then, for all $E\in\tilde
  I_\nu$, one has
  \begin{equation} \label{eq:lambda-eff:I-sigma}
    \tau^2(|\xi_0(E)|+|\xi_\pi(E)|)\le C e^{-c/\varepsilon}.
  \end{equation}
\end{Le}
\demo Consider the case when $I_c^-\cap \tilde I_0\ne\emptyset$.
The complementary case is analyzed similarly.\\
The statement of Lemma follows from the observations:
\begin{itemize}
\item at a point of $\tilde I_0$, one has
  $\tau^2(|\xi_0(E)|+|\xi_\pi(E)|)\le e^{-c/\varepsilon}$ (as
  $I_c^-\cap \tilde I_0\ne\emptyset$);
\item simultaneously for all $E\in \tilde I_0$, one has
  $|\xi_0(E)|+|\xi_\pi(E)|\asymp q$ where $q$ is positive and
  independent of $E$ (Corollary~\ref{cor:xi-on-tildeI} gives $q=
  \frac{M(\Delta)}{\varepsilon\bar
    T_{v,0}}+\frac{\Delta+M(\Delta)}{\varepsilon\bar T_{v,\pi}}$).
\end{itemize}
The proof of Lemma~\ref{Ic-and-Ipm} is completed.\QED 
\subsubsection{The monodromy matrix for $E\in I_c^-$}
When proving our results on the absolutely continuous spectrum, we use
Ishii-Pastur-Kotani Theorem, i.e., we control the Lyapunov exponent
for the family of equations~\eqref{family} using the matrix cocycle
$N$ defined in Lemma~\ref{le:Le:1}. We now study the matrix $N$
constructed by formula~(\ref{Le:2}) in terms of the matrix $M^\pi$.\\
Fix  $0<\delta_{\rm ac}<\min\{c,\delta,\delta_\tau\}$. 
We prove
\begin{Pro}
  \label{le:ac:N} Fix $y>0$.  Pick $\nu\in\{0,\pi\}$. For sufficiently
  small $\varepsilon$, the following holds. If $I_c^-\cap \tilde
  I_\nu\ne\emptyset$, then, for $E$ in the $d_\nu$-neighborhood of
  $\tilde I_\nu$ and for $z$ in the strip $|\im z|\le y$, one has
\begin{equation}
  \label{eq:N-ac}
  N=
  \begin{pmatrix}
    \sigma(\tau^2\xi_0\xi_\pi+\theta_n+1/\theta_n) & -1 \\ 1 & 0
  \end{pmatrix}
   + O(e^{-\delta_{\rm ac}/\varepsilon}).
\end{equation}
\end{Pro}
\demo The cases where $\nu=0$ and $\nu=\pi$ are treated in one and the
same way. We only consider the case of $\nu=0$. Below, we always
assume that $E$ and $z$ are as described in Proposition~\ref{le:ac:N}.
Construct the functions $\rho$ and $v$ in terms of the matrix $M^\pi$.
It suffices to prove that
\begin{equation}
  \label{v,rho:ac}
  \rho=1+O(e^{-\delta_{\rm ac}/\varepsilon}),\quad 
  v=\sigma(\tau^2\xi_0\xi_\pi+\theta_n+1/\theta_n)+O(e^{-\delta_{\rm ac}/\varepsilon}).
\end{equation}
The asymptotic representation for $\rho$ follows from~(\ref{mm-as})
and the estimates
\begin{equation}
  \label{xi-ac}
  |\xi_0|\ge C e^{\delta_{\rm ac}/\varepsilon}\quad \text{and}\quad |r\tau\xi_0|>1.  
\end{equation}
Let us prove these estimates.  By Corollary~\ref{cor:xi-on-tildeI}, we
get 
\begin{equation}
  \label{xi-ac:1}
  |\xi_0|\ge C\frac{M(\Delta)}{\varepsilon \bar T_{v,0}}\quad
  \text{and}\quad \tau r
  |\xi_0|\ge C\frac{M(\Delta)}{\varepsilon \bar T_h}.
\end{equation}
We consider two cases. First, we assume that $\Delta\le
\varepsilon\sqrt{\bar T_h}$. Then, $M(\Delta)=\varepsilon\sqrt{\bar
  T_h}$, and we get the estimates (using
also~(\ref{small-tau:ids:T-cases})),
\begin{equation*}
  |\xi_0|\ge C\frac{\sqrt{\bar T_h}}{\bar T_{v,0}}\ge C\tau^{-1}\ge
  Ce^{\delta_\tau/\varepsilon}\quad{\rm and}\quad
  \tau r|\xi_0|\ge \frac{C}{\sqrt{ \bar T_h}}
\end{equation*}
which imply~\eqref{xi-ac}.  Now, assume that $\Delta\ge
\varepsilon\sqrt{\bar T_h}$. Then, $M(\Delta)=\varepsilon^2\bar
T_h/\Delta$, and we get the estimates
\begin{equation} \label{xi-ac:2}
  |\xi_0|\ge C\frac{\varepsilon\bar T_h}{\bar T_{v,0}\Delta}\quad{\rm and}\quad
  \tau r|\xi_0|\ge \frac{C\varepsilon}{\Delta}.
\end{equation}
In view of~(\ref{Delta}), the second of these estimates implies the second of the
estimates in~(\ref{xi-ac}).  Now, note that Lemma~\ref{Ic-and-Ipm} and
Corollary~\ref{cor:xi-on-tildeI} imply that
\begin{equation*}
  e^{c/\varepsilon}\le C(\tau^2|\xi_\pi|)^{-1}\le C\frac{\varepsilon
  \bar T_h}{\bar T_{v,0}\Delta}.
\end{equation*}
Therefore, the first of the estimates in~(\ref{xi-ac:2}) implies the
first estimate in~(\ref{xi-ac}). This completes the proof of the
asymptotic representation for $\rho$.\\
Prove the asymptotics for $v$. The representation~(\ref{mm-as}) and
the already proved representation for $\rho$ imply that
\begin{equation}
  \label{v-ac-1}
  v=\sigma \tau^2 \xi_0\xi_\pi
  +\bar\theta_n+1/\bar\theta_n+O\left(e^{-\delta_{\rm ac}/\varepsilon}
(\tau^2|\xi_0\xi_\pi|+1)+\tau^2(|\xi_0|+|\xi_\pi|)\right).
\end{equation}
Now, note that, for all  $E\in\tilde I_0$, one
has~(\ref{eq:lambda-eff:I-sigma}) and $\tau^2|\xi_0\xi_\pi|\le C$ (as
$\tilde I_0\subset \tilde \Sigma_{\rm ac}$). Therefore,~(\ref{v-ac-1})
imply the representation for $v$ from (\ref{v,rho:ac}). This completes
the proof of Proposition~\ref{le:ac:N}. 
\QED
\subsubsection{Completing the proof}
Using the representation~(\ref{eq:N-ac}), one completes the proof of
Theorem~\ref{thr:1} as in section 4.5 of~\cite{Fe-Kl:04a}: using
standard KAM techniques, see~\cite{MR2003f:82043}, section 11, one
proves that, under the conditions of Theorem~\ref{thr:1} the Lyapunov
exponent for the matrix cocycle $(N,h)$ vanishes on the most of the
interval $\tilde I_\nu$, i.e., the interval in the neighborhood of
which the representation~(\ref{eq:N-ac}) holds. By Lemma~\ref{le:Le:1}
and Theorem~\ref{Lyapunov}, this implies that the Lyapunov for the
equation family~\eqref{family} vanishes on most of this interval.
Then, by the Ishii-Pastur-Kotani Theorem, this implies that most
of this interval is covered by absolutely continuous spectrum.\\
To complete the proof, we first transform the problem to a form suited
to apply the standard KAM techniques, then, we complete the analysis as
in~\cite{Fe-Kl:04a}. The last part being standard, we only outline it.\\
{\it Step 1: New parametrization.} \ One introduces $\varphi\mapsto
E(\varphi)$, the multivalued analytic function defined by the relation
\begin{equation*}
2\cos \varphi={\rm Tr} ({\rm \ leading \ term \ of \ }
N)=\sigma\tau^2\xi_0(E)\xi_\pi(E)+\bar \theta_n+1/\bar\theta_n. 
\end{equation*}
One picks four positive constants $q_1$, $q_2$, $c_1<1$ and
$c_2<1-c_1$ and considers the function $\varphi\mapsto E(\varphi)$ on
$V(c_1,c_2)$, the complex $(c_2)$-neighborhood of the interval
$[-1+c_1, 1-c_1]$.  Using~\eqref{xi-nu}, one proves that, if $c_1$ and
$c_2$ are sufficiently small, then, for sufficiently small
$\varepsilon$,
\begin{itemize}
\item there exists a real analytic branch of the function
  $\varphi\mapsto E(\varphi)$ that maps the interval $[-1+c_1, 1-c_1]$
  into $\tilde I_\nu$ so that $|\tilde
  I_\nu|-|\varphi((-1+c_1,1-c_1))|\le q_1 M(\Delta)$;
\item it conformally maps $V(c_1,c_2)$ into the $q_2
  M(\Delta)$-neighborhood of $\tilde I_\nu$;
\end{itemize}
From now on, one
considers $N$ as a function of $(z,\varphi)\in
\{z\in\C:  \ |\im  z|\le y\}\times V(c_1,c_2)$.\\
{\it Step 2. Diagonalization of the leading term of $N$.} \ 
Let 
\begin{equation}
  \tilde N=S^{-1} N S\quad\text{ where }\quad S=
\begin{pmatrix}
  e^{i\varphi}  &  e^{-i\varphi}  \\ 
  1             &  1
\end{pmatrix}.
\end{equation}
Clearly, together with $N$, the function $\tilde N$ is unimodular,
$1$-periodic in $z$ and analytic in $(z,\varphi)\in \{z\in\C: \ |\im
z|\le y\}\times V(c_1,c_2)$ (by the previous step, provided that $c_1$ and
$c_2$ are chosen small enough). Furthermore, as $N$ is real analytic,
$\tilde N$ has the form $\begin{pmatrix} a & b \\ b^* & a^*
\end{pmatrix}$ (see~\eqref{star}). Finally, for $(z,\varphi)\in
\{z\in\C: \ |\im z|\le y\}\times V(c_1,c_2)$, one has
\begin{equation}
  \label{tildeN}
  \tilde N =
  \begin{pmatrix}
    e^{i\varphi}  &  0 \\ 0 & e^{-i\varphi}
  \end{pmatrix}
  +O(\lambda)\quad\text{ where }\quad \lambda=e^{-\delta_{\rm
      ac}/\varepsilon}
\end{equation}
(by Proposition~\ref{le:ac:N} and the previous step,  provided that 
$c_1$ and $c_2$ are chosen small enough).\\
{\it 3. Standard KAM theory result.}  \
Clearly, $\theta(N,h)=\theta(\tilde N,h)$. Applying a standard KAM
theory construction from section 4.5.2 of~\cite{Fe-Kl:04a} to the
cocycle $(\tilde N,h)$, one proves that, if the number $h$ satisfies a
Diophantine condition (see Proposition 4.4 in~\cite{Fe-Kl:04a}), and
$\lambda$ is sufficiently small, then $\theta(\tilde N, h)$ vanishes
on $[-1+c_1,1-c_1]$ outside $\Phi_\infty$, a set of measure of order
$\lambda$.\\
Finally, arguing as in the end of section 4.5.3 in~\cite{Fe-Kl:04a},
one proves that the number $h$ satisfies the required Diophantine
condition if $\varepsilon$ belongs to the set $D$ described in
Theorem~\ref{thr:1}.\\
{\it 4. Coming back to $E$.}  The above results imply that, for
$\varepsilon\in D$ sufficiently small, the Lyapunov exponent for the
equation family~\eqref{family} vanishes on the interval
$E([-1+c_1,1-c_1])$ outside a set of the measure $\D
m=\int_{\Phi_\infty} \left|\frac{dE}{d\varphi}\right| d\varphi$. By
the Cauchy estimates for analytic functions and the first step, on
$[-1+c_1,1-c_1]$, we get $\frac{dE}{d\varphi}=O(M(\Delta))$. So,
$m=o(M(\Delta))=o(|\tilde I_\nu|)$ \ (which is obtained
using~\eqref{eq:Ipm:size}).  As $c_1$ can be taken arbitrary small,
this completes the proof of Theorem~\ref{thr:1}.\QED


%
\section{Possible spectral scenarii for small $\tau$}
\label{sec:scenarios}
Consider $\rho$ defined by~(\ref{eq:1}). Being expressed in terms of
the tunneling coefficients, $\rho$ is typically either exponentially
large or exponentially small. In both cases, the results on the
spectral properties of~\eqref{family} for small $\tau$ take a very
explicit form. In this section, we assume that, for some positive
$\delta_\rho$, one has either
\begin{equation}
  \label{small-rho} \rho\le e^{-\delta_\rho/\varepsilon},
  \quad\text{i.e.,}\quad S_h(\bar E)\le \min_\nu\{ S_{v,\nu}(\bar
  E)\}-\delta_\rho,
\end{equation}
or 
\begin{equation}
  \label{big-rho} \rho\ge e^{\delta_\rho/\varepsilon},
 \quad\text{i.e.,}\quad S_h(\bar E)\ge \min_\nu\{ S_{v,\nu}(\bar
 E)\}+\delta_\rho,
\end{equation}
and describe the spectral results for these two cases. These results
easily follow from Theorems~\ref{thr:5}--~\ref{thr:1}. So, we only
outline the analysis omitting most of the elementary calculations.   \\
\subsection{Preliminaries}
\label{sec:preliminaries}
We assume that $\varepsilon$ is sufficiently small so as to be in the
case of Corollary~\ref{two-int-cor}.  The functions $\xi_0$ and
$\xi_\pi$ being linear in $E$, one can explicitly describe the
intervals defined by~(\ref{eq:37}). For sake of definiteness, we
assume~(\ref{E0:greater}). We denote then the leftmost interval
defined in Proposition~\ref{pro:1} by $I_\pi$ and the rightmost by
$I_0$.
We  use the notations   
\begin{equation*}
I_\pi=[E_\pi^{\rm out}, E_\pi^{\rm in}],\quad
\text{and}\quad  I_0=[E_0^{\rm in}, E_0^{\rm out}].
\end{equation*}
Define  also
\begin{equation}
  \label{scenario:seq:t}
  t_v^\pm= \frac12\left(\frac1{|\gamma_0|} \pm
  \frac1{|\gamma_\pi|}\right)\,(1+f),\quad 
  t_h^\pm=\frac{2(\Lambda_n\pm 1\pm f))}{\tau^2|\gamma_0\gamma_\pi|},
\end{equation}
$\gamma_0$ and $\gamma_\pi$ being defined in~(\ref{gamma:ext}), and
$\varepsilon\mapsto f(\varepsilon)$ is the function introduced
in~(\ref{eq:37}). Note that,
\begin{equation*}
  t_v^++t_v^-\asymp \varepsilon \bar T_{v,0},\quad
  t_v^+-t_v^-\asymp\varepsilon \bar T_{v,\pi}. 
\end{equation*}
Furthermore, as $f$ tends to zero when $\varepsilon\to 0$, for
sufficiently small $\varepsilon$, one has
\begin{equation}
  \label{scenarios:eq:order-of-t}
  t_v^\pm\le  C \varepsilon \bar \max_\nu
  T_{v,\nu}\quad\text{and}\quad t_h^\pm \asymp 
  \varepsilon^2 \bar T_h.
\end{equation}
Then, we get
\begin{Le}
  \label{Le:scenario:preliminaries}
  Let $t_h^-\ge 2(t_v^+ + t_v^-)\Delta$ and $t_h^-\ge 2(t_v^+ -
  t_v^-)\Delta$.  Then, one has
  \begin{gather*}
    E_\pi^{\rm out}= \bar E-t_v^+-\sqrt{(t_v^+)^2-2\Delta
      t_v^-+\Delta^2+t_h^+},\quad \quad
    E_\pi^{\rm in}=\bar E+t_v^+-\sqrt{(t_v^+)^2+2\Delta  t_v^-+\Delta^2+t_h^-},\\
%
    E_0^{\rm in}=\bar E-t_v^++\sqrt{(t_v^+)^2-2\Delta
      t_v^-+\Delta^2+t_h^-},\quad\quad
    E_0^{\rm out}=\bar E+t_v^++\sqrt{(t_v^+)^2+2\Delta
      t_v^-+\Delta^2+t_h^+}.
  \end{gather*}
  If $t_h^-\le 2(t_v^+ - t_v^-)\Delta$, then
  \begin{equation*}
    E_\pi^{\rm in}=\bar E-t_v^--\sqrt{(t_v^-)^2-2\Delta\cdot  t_v^++\Delta^2+t_h^-};
  \end{equation*}
  if $t_h^-\le 2(t_v^+ + t_v^-)\Delta$, then
  \begin{equation*}
    E_0^{\rm in}=\bar E-t_v^-+\sqrt{(t_v^-)^2-2\Delta\cdot  t_v^++\Delta^2+t_h^-}.
  \end{equation*}
\end{Le}
\noindent This lemma easily follows from~(\ref{eq:37})
and~(\ref{xi-nu}).
\subsection{Small $\rho$}
\label{sec:small-rho}
Now, let us discuss the intervals $I_0$ and $I_\pi$ and the spectrum
in them for small $\rho$.
Let 
\begin{equation}
  M(\Delta)=\min\{\varepsilon \sqrt{\bar T_h},\, \varepsilon^2 \bar
  T_h/\Delta\},\quad D(\Delta)=\max\{\Delta,\varepsilon \sqrt{\bar T_h}\}. 
\end{equation}
{\it Level repulsion.\/} \ In view of~(\ref{scenarios:eq:order-of-t}),
Lemma~\ref{Le:scenario:preliminaries} implies
\begin{Cor}
  \label{Cor:small-rho:length}
For sufficiently small $\varepsilon$, under the
condition~(\ref{small-rho}), one has 
\begin{equation}
  |I_0|\asymp M(\Delta),\quad |I_\pi|\asymp M(\Delta),\quad{\rm and}\quad
\dist(I_0,I_\pi)\asymp D(\Delta);
\end{equation}
the interval $I_0$ is to the right of $E_0$,  the interval $I_\pi$ is
to the left of $E_\pi$, and for $\nu,\mu\in \{0,\pi\}$, one has
\begin{equation}
  \label{eq:0-dist}
  \dist(E_\nu,I_\mu)\asymp 
  \left\{\begin{array} {ll}
      M(\Delta), & {\rm \ if \ } \mu=\nu,\\
      D(\Delta), & {\rm \ if \ } \mu\ne \nu.
  \end{array}\right.
\end{equation}
\end{Cor}
\noindent This corollary in particular shows that there is always a
repulsion between the intervals $I_0$ and $I_\pi$.\\
{\it The nature of the spectrum.} \ Corollary~\ref{cor:1} and
Theorem~\ref{thr:1} show that the nature of the spectrum depends on
the value of the quantity
\begin{equation}
  \label{eq:lambda}
  \lambda=\varepsilon\log(\tau^2|\xi_0|+\tau^2|\xi_\pi|).
\end{equation}
Corollary~\ref{Cor:small-rho:length} and the definitions of $\xi_0$
and $\xi_\pi$ imply
\begin{Cor}\label{Cor:lambda:small-rho} Pick $\nu\in\{0,\pi\}$. Fix $c_1>0$.  Define
  $\Delta_\nu=\bar T_h/\bar T_{v,\nu}$.  One has:
  \begin{itemize}
  \item There exists $c_2>0$ such that, for sufficiently small
    $\varepsilon$, if $\Delta\le e^{-c_1/\varepsilon}\Delta_\nu$, then
    $\lambda\le -c_2$ for all $E\in I_\nu$.
  \item For sufficiently small
  $\varepsilon$, if $\Delta\ge e^{c_1/\varepsilon}\Delta_\nu$, then
  $\lambda= \varepsilon \log \left(\Delta/\Delta_\nu\right)+o(1)$ uniformly in
  $E\in I_\nu$.
  \end{itemize}
\end{Cor}
\begin{Rem}
  Let $\Delta_*=\varepsilon\sqrt{\bar T_h}$. By
  Corollary~\ref{Cor:lambda:small-rho}, for $\Delta\le \Delta_*$, on
  the intervals $I_0$ and $I_\pi$, one has $\lambda\le -C<0$. Indeed,
  for $\nu\in\{0,\pi\}$, one has $\Delta_*\le\varepsilon\rho
  \Delta_\nu$, and $\rho\le e^{-\delta_\rho/\varepsilon}$ for some
  positive $\delta_\rho$.
\end{Rem}
\subsection{Large $\rho$}
\label{sec:big-rho}
Now, turn to the case of large $\rho$. For the sake of definiteness, we
assume that $\bar T_{v,0}\le \bar T_{v,\pi}$. Note that
\begin{equation}
  \label{eq:tun-rat-big-rho}
  \frac{\bar T_{v,0}}{ \bar T_{v,\pi}}\asymp\frac{\tau^2}{\rho^2}\le 
  e^{-(\delta_\rho+\delta_\tau)/\varepsilon}\ll 1
\end{equation}
so that $t_v^\pm \asymp \varepsilon \bar T_{v,\pi}$ and
$0<t_v^++t_v^-\asymp \bar T_{v,0}\ll\bar T_{v,\pi}$.\\
We consider two cases when the formulae of
Lemma~\ref{Le:scenario:preliminaries} take the simplest form.
Simple computations imply 
\begin{Cor}
  \label{Cor:big-rho:length}
  Fix $0<c<1$. Let the
  conditions~(\ref{big-rho}),~(\ref{small-tau:ids:T-cases})
  and~(\ref{E0:greater}) be satisfied. For sufficiently small
  $\varepsilon$, one has
  \begin{itemize}
  \item if $\Delta<ct_v^+$, then
    \begin{equation*}
      |I_\pi|=  2( t_v^++\Delta)+o(\varepsilon \bar T_{v,\pi}),\quad
      |I_0|=    2(t_v^+-\Delta)+o(\varepsilon \bar T_{v,\pi}),\quad 
      \dist(I_0,I_\pi)\asymp \varepsilon\frac{\bar T_h}{T_{v,\pi}}.
    \end{equation*}
    Moreover, $E_0$ is between $E_\pi^{\rm in}$ and $E_0^{\rm in}$, \
    $E_\pi^{\rm out}$ is to the left of $E_\pi$, and
    \begin{equation*}
      E_0 -E_\pi^{\rm in},\quad\quad E_0^{\rm in}-E_0\asymp
      \varepsilon\frac{\bar T_h}{T_{v,\pi}},\quad\quad 
      E_\pi-E_\pi^{\rm out}=2t_v^++o(\varepsilon \bar T_{v,\pi}).
    \end{equation*}
  \item if $\Delta>c^{-1}t_v^+$, then
    \begin{equation*}
      |I_\pi|=  4 t_v^++o(\varepsilon \bar T_{v,\pi}),\quad\quad
      |I_0|\asymp \varepsilon^2\frac{\bar T_h}{\Delta}+\varepsilon\bar
      T_{v,0},\quad\quad
      \dist(I_0,I_\pi)=2(\Delta-t_{v}^+)+o(\varepsilon \bar T_{v,\pi}).
    \end{equation*}
    Moreover, $E_0$ is to the right of $E_\pi^{\rm in}$, \ $E_\pi^{\rm
      out}$ is to the left of $E_\pi$, and
    \begin{gather}\nonumber
      E_0 -E_\pi^{\rm in}=2(\Delta-t_{v}^+)+o(\varepsilon \bar
      T_{v,\pi}),\quad\quad
      E_\pi-E_\pi^{\rm out}=2t_v^++o(\varepsilon \bar T_{v,\pi}),\\
      \label{big-rho-dist}
      E_0-E_0^{\rm in}=\left(1- 2\frac{\Delta
          (t_v^++t_v^-)}{t_h^-}\right)\,g\quad\text{where}\quad
      g\asymp \varepsilon\frac{\bar T_h}{T_{v,\pi}},\quad\quad g>0.
    \end{gather}
  \end{itemize}
\end{Cor}
\noindent As $c$ in Corollary~\ref{Cor:big-rho:length} can be fixed
arbitrarily close to one, it gives quite a complete picture of what
happens in the case of large $\rho$.
The parameter $\lambda$ governing the nature of the spectrum
(see~\eqref{eq:lambda}) is now described by
\begin{Le} In the case of Corollary~\ref{Cor:big-rho:length}, one has
\begin{itemize}
\item if $\Delta<ct_v^+$, then
\begin{equation*}
  \lambda|_{I_\nu}= \varepsilon\log(1+s_\nu
  \rho^2)+O(\varepsilon)\quad\text{where}\quad  
  s_\nu=\frac{|E-E_\nu^{\rm in}|}{|I_\nu|}\quad\text{for}\quad
  \nu\in\{0,\pi\};
\end{equation*}
\item if $\Delta>c^{-1}t_v^+$, then
\begin{equation*}
  \lambda|_{I_\pi}=\varepsilon\log \left(\frac{\bar T_{v,\pi}}{\bar
      T_h}\Delta\right)+O(\varepsilon),\quad\text{and}\quad 
  \lambda|_{I_0}=\varepsilon\log \left(\frac{\bar T_{v,0}}{\bar
      T_h}\Delta+ \frac{\bar T_{v,\pi}}{\Delta}\right)+O(\varepsilon)
\end{equation*}
\end{itemize}
\end{Le}

%

%
\section{Periodic Schr{\"o}dinger operators}
\label{S3}
\noindent The aim of this section is to give the precise definition of
$\theta_n$ and to study its properties. Recall that
$\Lambda_n=\frac12(\theta_n+\theta_n^{-1})$ is responsible for the gap
between the ``interacting'' intervals $I_0$ and $I_\pi$ in the case of
small $\rho$.\\
The constant $\theta_n$ is defined by the periodic Schr{\"o}dinger
operator~\eqref{Ho}. First, assuming that $V$ is a $1$-periodic, real
valued, $L^2_{loc}$-function, we recall well known results on this
operator (see~\cite{MR2002f:81151,Eas:73,Ma-Os:75,MR80b:30039,Ti:58}).
Then, we define $\theta_n$ and analyze its properties.
\subsection{Analytic theory of Bloch solutions}
\label{sec:analyt-theory-bloch}
\subsubsection{Bloch solutions}
\label{sec:bloch-solutions}
Let $\psi$ be a nontrivial  solution of the equation
\begin{equation}\label{PSE}
  -\frac{d^2}{dx^2}\psi\,(x)+ V\,(x)\psi\,(x)=\mathcal{E}\psi\,(x),
   \quad x\in\R,
\end{equation}
satisfying the relation $\psi\,(x+1)=\lambda\,\psi\,(x)$ for all
$x\in\R$ with $\lambda\in\C$ independent of $x$. Such a solution is
called a {\it Bloch solution}, and the number $\lambda$ is called the
{\it Floquet multiplier}. Let us discuss properties of
Bloch solutions (see~\cite{MR2002f:81151}).
\smallpagebreak As in section~\ref{sec:periodic-operator}, we denote
the spectral bands of the periodic Schr{\"o}dinger equation by
$[E_1,\,E_2]$, $[E_3,\,E_4]$, $\dots$, $[E_{2n+1},\,E_{2n+2}]$,
$\dots$. Consider $\mathcal{S}_\pm $, two copies of the complex plane
cut along the spectral bands. Paste them together to get a Riemann
surface with square root branch points.  We denote this Riemann
surface by $\mathcal{S}$. In the sequel, $\pi_c:\ {\mathcal
  S}\mapsto\C$ is the canonical projection, and $\mathcal{E}$ denotes
a point in $\mathcal{S}$ and $E$ a point in $\C$.
\smallpagebreak For any $E\in\C$, one can construct a Bloch solution
$\psi(x,\mathcal{E})$ of equation~\eqref{PSE}. Normalized by the
condition $\psi(0,E)=1$, the function
$\mathcal{E}\mapsto\psi(\cdot,\mathcal{E})$ is meromorphic on
$\mathcal S$. All its poles are projected by $\pi_c$ either in the
open spectral gaps or at their ends. More precisely, there is exactly
one simple pole per open gap.  The position of the pole is independent
of $x$ (see~\cite{MR2002f:81151}).
\smallpagebreak Let $\hat{\cdot}:\ {\mathcal S}\mapsto {\mathcal S}$
be the canonical transposition mapping; for any point $\mathcal{E}$ in
$mathcal{S}$ different from the branch points, the point
$\hat{\mathcal{E}}$ is the unique solution to the equation
$\pi_c(\mathcal{E})=E$ different from $\mathcal E$.\\
The function $x\mapsto \psi(x,{\hat{\mathcal E}})$ is another Bloch
solution of~\eqref{PSE}. Except at the edges of the spectrum (i.e. the
branch points of $\mathcal{S}$), \ $\psi(\cdot,{\mathcal E}) $ and
$\psi(\cdot,\hat {\mathcal E})$ are linearly independent solutions
of~\eqref{PSE}.
\subsubsection{The Bloch quasi-momentum}
\label{SS3.2}
Consider the Bloch solution $\psi(x,\mathcal{E})$. The corresponding
Floquet multiplier $\mathcal{E}\mapsto\lambda\,(\mathcal{E})$ is
analytic on $\mathcal{S}$. Represent it in the form
$\lambda(\mathcal{E})=\exp(ik(\mathcal{E}))$. The function
$\mathcal{E}\mapsto k(\mathcal{E})$ is the {\it Bloch quasi-momentum}.
\\
The quasi-momentum is defined modulo $2\pi$.  It can be regarded as
single valued analytic function from ${\mathcal S}$ to the cylinder
$\C/(2\pi \Z)$.  Then, the quasi-momenta of $\psi(x,{\mathcal E})$ and
$\psi(x,\hat {\mathcal E})$ differ only by the sign. The imaginary
part of the complex momentum vanishes only if
$\pi_c({\mathcal E})$ belongs to the spectrum of the periodic operator.\\
Below, we denote the derivative with respect to the local coordinate
$E=\pi_c({\mathcal E})$ by the dot $\dot{}$\,. The derivative of the
quasi-momentum is a single valued analytic function on $\mathcal S$.
One has
\begin{equation}
  \label{k-hat-k}
  \dot k(\hat{\mathcal E})=-\dot k({\mathcal E}).
\end{equation}
We note that $\dot k$  vanishes only inside intervals projecting onto
the spectral gaps.\\
\subsection{Constants $\theta_n$ and $\Lambda_n$}
\label{sec:Omega}
These constants are defined in terms of a meromorphic differential on
$\mathcal S$.
\subsubsection{Meromorphic differential $\Omega$}
\label{sec:defin-analyt-prop}
On the Riemann surface $\mathcal S$, consider the function
\begin{equation}\label{omega}
\omega({\mathcal E})=
-\frac{\int_0^1 \psi(x,\hat {\mathcal E})\,\left(\dot\psi(x,{\mathcal
      E})-i\dot k({\mathcal E}) x\,\psi(x,{\mathcal E})\right)\,dx} 
{\int_0^1\psi(x,{\mathcal E})\psi(x,\hat {\mathcal E}) dx}.
\end{equation}
where $k$ is the Bloch quasi-momentum of $\psi$. This function was
introduced in~\cite{Fe-Kl:03f} (the definition given in that paper is
equivalent to~\eqref{omega}). In section 1.4 of~\cite{Fe-Kl:03e}, we
have proved:
\begin{enumerate}
\item the differential $\Omega=\omega\,d{\mathcal E}$ is meromorphic
  on $\mathcal S$; its poles are located at the points of $P\cup Q$,
  where $P$ is the set of the poles of $\psi(x,{\mathcal E})$, and $Q$
  is the set of points where $\dot k({\mathcal E})=0$;
\item all the poles of $\Omega$ are simple;
\item $\res_p \Omega=1$, \ $\forall p\in P\setminus Q$, \ \ 
  $\res_q\Omega=-1/2$, \ $\forall q\in Q\setminus P$, \ \ 
  $\res_r\Omega=1/2$, \ $\forall r\in P\cap Q$.
\end{enumerate}
\subsubsection{Constants $\theta_n$ and $\Lambda_n$}
\label{sec:periods-omega}
We let
\begin{equation}
  \label{theta-omega} 
\Lambda_n(V)=\frac12\left(\theta_n(V)+\frac1{\theta_n(V)}\right)
  \quad\text{ and }\quad
  \theta_n(V)=\exp(l_n(V)),
\end{equation}
where
\begin{equation}
  \label{Lambda-omega}
  l_n(V)=\int_{g_n}\Omega({\mathcal E}),
\end{equation}
and  $g_n\subset\mathcal S$ is a simple closed curve on $\mathcal{S}$
such that
\begin{itemize}
\item $g_n$ is located on $\C\setminus \sigma(H_0)$, the sheet of the
  Riemann surface $\mathcal{S}$ where, for $\im\pi_c({\mathcal E})>0$,
  the Bloch quasi-momentum of $\psi(x,{\mathcal E})$ has positive
  imaginary part;
\item $\pi_c(g_n)$ is positively oriented and going around the $n$-th
  spectral gap of the periodic operator $H_0$.
\end{itemize}
The functional $l_n: V \mapsto l_n(V)$ is defined on $L^2(\mathbb T)$,
the vector space of locally square integrable, real valued
$1$-periodic functions. One has
\begin{Th}
  \label{Lambda:prop}  
  The functional $l_n$ has the following properties:
  \begin{enumerate}
  \item it is real valued;
  \item it vanishes at all even periodic potentials $V$;
  \item for any fixed real $s$, it is invariant under the
    transformations $V(\cdot)\mapsto V(\cdot+s)$.
  \item $l_n(\cdot)$ is non-zero on a dense open set in $L^2(\mathbb T)$;
  \end{enumerate}
\end{Th}
\noindent 
Point 1 in Theorem~\ref{Lambda:prop} is Lemma 6.1 from~\cite{Fe-Kl:04a}.
The remaining part of section~\ref{S3} is devoted to the proof of the
other points.\\
We shall use the representation
\begin{equation}
  \label{new-l-n}
  l_n(V)=\frac12\oint_{g_n} \Omega({\mathcal E})-\Omega(\hat{\mathcal E}).
\end{equation}
This representation follows from the equality $\overline{\oint_{g_n}
  \Omega({\mathcal E})}=-\oint_{g_n}\Omega(\hat{\mathcal E})$ which
follows from formula (6.5) in~\cite{Fe-Kl:04a}, and the fact that
$l_n(V)\in\R$.
\subsubsection{The functional $l_n$ on even functions}
\label{sec:funct-lambda-even}
We now assume that $V$ is even and prove that $l_n(V)=0$.
In view of~\eqref{new-l-n}, it suffices to prove that
\begin{equation}\label{V-pair}
\Omega({\mathcal E})=\Omega(\hat {\mathcal
  E}),  
\end{equation}
which follows from the relation
\begin{equation}
  \label{pm}
  \psi(-x,{\mathcal E})=\psi(x,\hat {\mathcal E}).
\end{equation}
Indeed, making in~\eqref{omega} the change of the variable
$x\,\mapsto\,-x$ and using~\eqref{pm} and the relations
$\psi(x-1,{\mathcal E})=e^{-ik({\mathcal E})}\psi(x,{\mathcal E})$ and
$\psi(x-1,\hat
{\mathcal E})=e^{ik({\mathcal E})}\psi(x,\hat {\mathcal E})$, we come
to~\eqref{V-pair}.\\ 
Finally, check~(\ref{pm}). Both functions in~\eqref{pm} are Bloch
solutions of~\eqref{PSE}. Their quasi-momenta coincide modulo $2\pi$.
Therefore, these solutions are linearly dependent, and, as
$\psi(0,{\mathcal E})=1=\psi(0,\hat {\mathcal E})$, they coincide.
This completes the proof of the point 2 of Theorem~\ref{Lambda:prop}.
\subsubsection{The translation invariance of $l_n$}
\label{sec:transl-invar-lambda}
Let $s$ be a real number. And let $\psi(x,{\mathcal E},s)$ and $k({\mathcal E},s)$ be
the Bloch solution and the Bloch quasi-momentum for the potential
$V_s(x)=V(v+s)$. Then, for $ {\mathcal E}\in \mathcal{S}$, one has
\begin{equation}
  \label{s}
  \psi(x,{\mathcal E},s)=\frac{\psi(x+s,{\mathcal
      E},0)}{\psi(s,{\mathcal E},0)},\quad \dot k({\mathcal E},s)=\dot
  k({\mathcal E},0).
\end{equation}
The proof of~\eqref{s} is similar to that of~\eqref{pm}, hence, we
omit it. Substituting~\eqref{s} into~\eqref{omega}, one sees that
\begin{equation*}
\Omega({\mathcal E},s)=\Omega({\mathcal E},0)+d\log\psi(s,{\mathcal E},0),
\end{equation*}
where $d$ is the external derivative with respect to ${\mathcal E}$.
This, the definition of $l_n$ and the first point in
Theorem~\ref{Lambda:prop} imply the point 3 in
Theorem~\ref{Lambda:prop}.
\subsubsection{For generic $V$, $l_n(V)$ does not vanish}
\label{sec:non-vanishing-lambda}
To prove the last point of Theorem~\ref{Lambda:prop}, we use
\begin{Le} 
  \label{le:soft} 
  One has
  \begin{itemize}
  \item the functional $l_n$ is continuous on $L^2(\mathbb T)$ endowed
    with the natural topology;
  \item for any $(V_0,\,V_1)\in L^2(\mathbb T)\times L^2(\mathbb T)$,
    the function $t\mapsto l_n(V_0 +tV_1)$ is real analytic.
  \end{itemize}
\end{Le}  
\demo Let $V_0\in L^2(\mathbb T)$. Let $K$ be a compact set containing
no branch point and no pole of the Bloch solution ${\mathcal
  E}\mapsto\psi(x,{\mathcal E},V_0)$. It suffices to check that
\begin{enumerate}
\item there exists $\mathcal V$, a neighborhood of $V_0$, such that
  the functions $({\mathcal E},V)\mapsto \dot k({\mathcal E},V)$,
  $(x,{\mathcal E},V)\mapsto \psi(x,{\mathcal E},V)$ and $(x,{\mathcal
    E},V)\mapsto\dot\psi(x,{\mathcal E},V)$ are continuous in
  $(x,{\mathcal E}, V)\in [0,1]\times K\times {\mathcal V}$;
\item for any ${\mathcal E}\in K$ and $x\in [0,1]$, the functions
  $t\mapsto \psi(x,{\mathcal E},V_0+tV_1)$, $t\mapsto
  \dot\psi(x,{\mathcal E},V_0+tV_1)$ and $t\mapsto \dot k({\mathcal
    E},V_0+tV_1)$ are analytic in a neighborhood of $0$.
\end{enumerate}
This easily follows from the classical construction of the Bloch
solutions and the Bloch quasi-momentum presented, for example,
in~\cite{Ti:58}. We omit the elementary details.\qed
\smallpagebreak Now, to prove point 4 of Theorem~\ref{Lambda:prop}, we
need only to prove that $V\mapsto l_n(V)$ is not identically zero.
The facts that $l_n$ vanishes on even potentials and that it is
translation invariant implies that $l_n$ vanishes at one gap periodic
potentials (as well as on many other ``simple'' potentials). In
section~\ref{sec2gap}, we prove
\begin{Pro} 
  \label{two-gap}
  The functional $l_1$ does not vanish identically on the set of two
  gap potentials.
\end{Pro}  
\noindent This proposition implies that $l_n$ does not vanish
identically for any integer $n$. Indeed, let $V$ be a two gap
potential such that $l_1(V)\ne 0$. The potential $V_n(x)=n^2V(nx)$ is
also one periodic.  Denote the corresponding Bloch solution and its
Bloch quasi-momentum by $\psi(x,{\mathcal E},V_n)$ and $k({\mathcal
  E},V_n)$. Clearly, one has
\begin{equation}
  \label{n-gap}
  \psi(x,{\mathcal E},V_n)=\psi(nx,n^{-2}{\mathcal
    E},V_1)\quad\text{and}\quad k({\mathcal E},V_n)=n 
  k({\mathcal E}n^{-2},V_1).
\end{equation}
The second relation from~\eqref{n-gap} implies that, up to scaling by
$n^{-2}$, the $n$-th gap for $V_n(x)$ coincides with the first gap for
$V$; thus, it is open. The definition of $\omega$ (see~\eqref{omega})
and~\eqref{n-gap}) imply also that $l_n(V_n)=l_1(V)$; hence, $l_n(V_n)$
does not vanish. This completes the proof of
Theorem~\ref{two-gap}.\qed
\subsection{The functional $l_n$ on two gap potentials}
\label{sec2gap}
We now prove Proposition~\ref{two-gap}.  Let $V$ be a two gap
potential.  Then, the spectrum of the periodic operator is the
disjoint union of two compact intervals and a closed half-axis going
up to $+\infty$. To the Riemann surface $\mathcal{S}$ constructed for
this spectrum according to the prescriptions in
section~\ref{sec:bloch-solutions}, we add the point $\infty$. Near to
$\infty$, we introduce the standard local coordinate $\tau=E^{-1/2}$.
Then, $\mathcal{S}$ becomes compact of genus $2$. The idea of the
proof of Proposition~\ref{two-gap} is to express the differential
$\Omega({\mathcal E})$ in terms of standard Abelian meromorphic
differentials associated to $\mathcal{S}$; for those, the integrals
along the closed curve $g_1$ can be
controlled. The proof is naturally divided into several steps.\\
{\it The behavior $\Omega$ at infinity.\/} We prove
\begin{Le}
  \label{le:omega-as}
  The function $\omega$ is holomorphic in a neighborhood of $\infty$ and 
  admits the Taylor formula:
  \begin{equation}
    \label{omega-as}
    \omega=-\frac{i\tau^{3}}{4}\,\int_0^1dx\int_0^x dx'\{V\}+
    \frac{\tau^4}{4}\int_0^1(V(x)-V(0))dx+
    \frac{3i\tau^{5}}{16}\,\int_0^1 dx\int_0^x dx'\{V''-V^2\}+
    o\left(\tau^{5}\right)
  \end{equation}
  where,  for $f(x)$, a $1$-periodic function of $x\in\R$, we write
  $\{f\}(x)=f(x)-\int_0^1 f(x)\,dx$.
\end{Le}
\noindent We prove Lemma~\ref{le:omega-as} in
section~\ref{proof:lem:omega-as}. It implies that
$\Omega$ is holomorphic near infinity. \\
{\it Decomposition of $\Omega$ in terms of standard Abelian
  differentials.} First, following~\cite{MR86k:35142}, we recall some
information on Abelian differentials. The surface $\mathcal{S}$ being
hyperelliptic of genus two, there are only two linearly independent
holomorphic differentials on $\mathcal{S}$. Away from the branch
points, in terms of the local coordinate $E=\pi_c({\mathcal E})$, they
have the form
\begin{equation}
  \label{omega-hol}
  \Omega_2(E)=\frac{E\,d E}{\sqrt{R(E)}},\quad
  \Omega_1(E)=\frac{dE}{\sqrt{R(E)}}\quad\text{where}\quad
  R(E)=\prod_{j=1}^5(E-E_j).
\end{equation}
Fix ${\mathcal P}\in {\mathcal  S}$. Let $P=\pi_c({\mathcal P})$.
Define a meromorphic differential by the formula  
\begin{equation}
  \label{omega(E,P)}
  \Omega({\mathcal E},{\mathcal P})= \frac{\sqrt{R(P)}}{\sqrt{R(E)}}\,\frac{dE}{E-P}.
\end{equation}
The differential $\Omega(E,{\mathcal P})$ has on $\mathcal{S}$ only
two simple poles located at the points $\mathcal P$ and $\hat{\mathcal
  P}$. For a suitable choice of the branch of $\sqrt{R}$, the residues
at these points are equal to $+1$ and $-1$ respectively.\\
Consider the differential $\Omega({\mathcal E})-\Omega(\hat{\mathcal
  E})$. As the periodic operator has only two open gaps, the Bloch
solution $\psi(x,{\mathcal E})$ has only two (simple) poles ${\mathcal
  P}_1$ and ${\mathcal P}_2$ located respectively in the closure of
the first and the second gaps. The description of the poles of
$\Omega$ and the analyticity of $\Omega$ at $\infty$ imply that the
differential $\Omega({\mathcal E})-\Omega(\hat{\mathcal
  E})-(\Omega({\mathcal E},{\mathcal P}_1)+\Omega({\mathcal E},
{\mathcal P}_2))$ is holomorphic on $\mathcal{S}$. Therefore, there
exists $(C_1,C_2)\in\C^2$ such that
\begin{equation}
  \label{omega:decomposition}
  \Omega({\mathcal E})-\Omega(\hat{\mathcal E})=\Omega({\mathcal
  E},{\mathcal P}_1)+\Omega({\mathcal E},{\mathcal
  P}_2)+C_1\Omega_1({\mathcal E})+C_2\Omega_2({\mathcal E}).
\end{equation}
Comparing~\eqref{omega:decomposition} with~\eqref{omega-as}, one
easily obtains
\begin{equation}\label{C0C1}
  C_2=-\frac{i}{2}\int_0^1dx\int_0^x dx'\{V\}(x'),\quad 
  C_1=\frac{3i}8\int_0^1 dx\int_0^x dx'
  \{V''-V^2\}(x')-\frac{\sigma}2\,C_2,\quad  \sigma=\sum_{j=1}^5 E_j.
\end{equation}
{\it Dependence on ${\mathcal P} _1$ and ${\mathcal P} _2$.\/} The two gap potentials with a
given spectrum depend only on two parameters, namely, ${\mathcal P} _1$ and ${\mathcal P} _2$,
the poles of the Bloch solution (see~\cite{MR86k:35142}). To
prove Proposition~\ref{two-gap}, we fix the spectrum
and study the dependence of $l_1$ on ${\mathcal P} _1$ and ${\mathcal P} _2$.
The representation~\eqref{omega:decomposition} imply that
\begin{equation}
  \label{L-C} 
  l_1=\oint_{g_1}(\Omega({\mathcal E},{\mathcal P}
  _1,E)+\Omega({\mathcal E},{\mathcal P} _2,E))+
  C_1({\mathcal P}_1,{\mathcal P}_2)\oint_{g_1} \Omega_1+C_2({\mathcal P} _1,{\mathcal P} _2)\oint_{g_1}\Omega_2.
\end{equation}
Note that the number $\oint_{g_1}\Omega_1$ is not zero. So, the
fact that $l_1$ does not vanish identically will follow from
\begin{Le}
  \label{derivatives} Let $c_1\ne0$ and $c_2$  be two  constants. The expression
  $c_1\frac{\partial^2 C_1}{\partial {\mathcal P} _1\partial {\mathcal P} _2}+
  c_2\frac{\partial^2 C_2}{\partial {\mathcal P} _1\partial
  {\mathcal P} _2}$ does not vanish identically.
\end{Le}
\noindent Lemma~\ref{derivatives} is proved in section~\ref{proof:le:der}.
This completes the proof of Proposition~\ref{two-gap}. \qed
\subsubsection{The proof of Lemma~\ref{le:omega-as}}
\label{proof:lem:omega-as}
First, following~\cite{MR86k:35142}, we explain how to describe the
behavior of $\psi$ near $\infty$ on ${\mathcal S}$. Consider the
logarithmic derivative, $\chi(x,{\mathcal E})=-i\frac{\partial}
{\partial x}\log\psi(x,{\mathcal E})$.  It is $1$-periodic in $x$; for
finite gap potentials, $\mathcal{E}\mapsto\xi(\cdot,\mathcal{E})$ is a
meromorphic function on $\mathcal{S}$; in a neighborhood of $\infty$,
its Laurent series has the form
\begin{equation}
  \label{chi-as}
  \chi=\tau^{-1}-{\dsize\frac{V}2}\,\tau-{\dsize\frac{iV'}{4}}\,\tau^2+
  {\dsize\frac{V''-V^2}8}\,\tau^3+\dots.
\end{equation}
The Laurent series for $\chi(x,\hat E)$ is obtained from that of
$\chi(x,E)$ by changing the sign of $\tau$ in~\eqref{chi-as}.\\
In the neighborhood of $\infty$, the representations of $\psi$ and $k$
can be obtained by substituting~\eqref{chi-as} in the formulae
\begin{equation}
  \label{chi}
  \psi(x,{\mathcal E})=\exp\left(i\int_0^x\chi(x,{\mathcal E})\,dx\right),\quad 
  k({\mathcal E})=\int_0^1\chi(x,{\mathcal E})\,dx.
\end{equation}
We now derive~(\ref{omega-as}). Compute the numerator and the
denominator in~\eqref{omega}. One has
\begin{equation}
  \label{omega-as:1}
  \int_0^1\psi(x,{\mathcal E})\psi(x,\hat {\mathcal E}) dx=
  \int_0^1 e^{i\int_0^x(\chi(x',{\mathcal E})+\chi(x',\hat {\mathcal E}))\,dx'}\,dx=
  1+\frac{\tau^2}2\,\int_0^1(V(x)-V(0))dx+O(\tau^4).
\end{equation}
Furthermore, using~\eqref{chi}, we also get
\begin{align*}
   &\int_0^1\psi(x,\hat {\mathcal E})\left(\dot\psi(x,{\mathcal E})-i\dot
    k({\mathcal E})x\,\psi(x,{\mathcal E})\right)\, dx= i\int_0^1 e^{
    i\int_0^x(\chi(x',{\mathcal E})+\chi(x',\hat {\mathcal E}))\,dx'}
    \frac{\partial}{\partial E}\left(\int_0^x\{\chi(x',{\mathcal E})\}\,dx'\right)dx\\
   &=\frac{i\tau^3}{4}\int_0^1\,dx\int_0^x\,dx'\{V\}(x')-
   \frac{\tau^4}{4}\int_0^1(V(x)-V(0))\,dx
   -\frac{i\tau^5}{8}\left(\frac32\,\int_0^1\,dx\int_0^x\,dx'\{V''-V^2\}(x')\right.\\
   &\hspace{7.5cm}
    \left.- \int_0^1\,dx(V(x)-V(0))\,\int_0^x\,dx'\{V\}(x')\right)+O(\tau^6),
\end{align*}
where the operator $\{\cdot\}$ was defined in Lemma~\ref{le:omega-as}.
Substituting~\eqref{omega-as:1} and the result of the last computation
into the definition of $\omega$, we obtain~\eqref{omega-as}. This
completes the proof of Lemma~\ref{le:omega-as}.\qed
\subsubsection{The proof of  Lemma~\ref{derivatives}}
\label{proof:le:der}
We prove the lemma by studying the analytic extensions of the
functions $({\mathcal P_1},{\mathcal P_2})\mapsto C_1({\mathcal
  P_1},{\mathcal P_2})$ and $({\mathcal P_1},{\mathcal P_2})\mapsto
C_2({\mathcal P_1},{\mathcal P_2})$. The proof is divided into
``elementary'' steps.\\ 
{\bf 1.} Following~\cite{MR86k:35142} (chapter II, sections 6 and 5),
we describe the set of the two gap potentials
giving rise to the spectrum $[E_1,E_2]\cup[E_3,E_4]\cup [E_5,\infty[$. \\
Pick $P_1$ and $P_2$ so that $E_2\le P_1\le E_3$ and $E_4\le P_2\le
E_5$. Recall that $R(E)$ is defined in~\eqref{omega-hol}. Let
$\xi\mapsto\lambda_1(\xi)$ and $\xi\mapsto\lambda_2(\xi)$ be the
solutions to the Cauchy problems
\begin{equation}\label{D}
  \begin{cases}
   \D \frac{d\lambda_1}{d\xi}=2\sqrt{-R(\lambda_1)},\\
    \lambda_1(0)=P_1
  \end{cases}
  \quad{\rm and}\quad\quad
  \begin{cases} 
   \D\frac{d\lambda_2}{d\xi}=2\sqrt{-R(\lambda_2)},\\
    \lambda_2(0)=P_2.
  \end{cases}
\end{equation}
The solutions of these equations are smooth real valued periodic
functions of $\xi\in\R$. The range of $\lambda_1$ is $[E_2,E_3]$, and
the one of $\lambda_2$ is $[E_4,E_5]$.  The periods of $\lambda_1$ and
$\lambda_2$ are equal respectively to
\begin{equation*}
  \Xi_1=\int_{E_2}^{E_3}\frac{d\lambda}{\sqrt{|R(\lambda)|}}
  \quad\text{and}
  \quad\Xi_2=\int_{E_4}^{E_5}\frac{d\lambda}{\sqrt{|R(\lambda)|}}.
\end{equation*}
One has $\Xi_1>\Xi_2$. Set 
\begin{equation}\label{x-xi}
x=\int_{0}^\xi (\lambda_2(\xi)-\lambda_1(\xi))\,d\xi.  
\end{equation}
then a two gap potential giving rise to the spectrum
$[E_1,E_2]\cup[E_3,E_4]\cup [E_5,\infty)$ is described by the formula
\begin{equation}
  \label{GLD-1}
    V(x)=-2\left(\lambda_1(\xi)+\lambda_2(\xi)\right)+c_0,
\end{equation}
where $c_0$ is a constant depending only on $(E_j)_{1\leq j\leq 5}$.\\
The potential $V$ is periodic if $\Xi_1$ and $\Xi_2$ are rationally
dependent, i.e., there exists $l$ and $m$ mutually prime such that
$l\Xi_1=m\Xi_2$. As $\Xi_1>\Xi_2$, this can happen
only for $l<m$.  \\
Assume that $l\Xi_1=m\Xi_2$. The period of the potential $V$ is equal
to $\int_{0}^{\Xi} (\lambda_2(\xi)-\lambda_1(\xi))\,d\xi$, where
$\Xi=l\Xi_1=m\Xi_2$. For the periodic operator with the potential $V$,
the gaps $[E_2,E_3]$ and $[E_4,E_5]$ are respectively the $l$-th and
the $m$-gaps. The points $P_1$ and $P_2$ are the projections on the
complex plane of the poles of the Bloch solution $\psi(x,{\mathcal
  E})$. The sheets of the Riemann surface where these poles are are
determined by the choice of the branches of the square roots
in~\eqref{D}. The poles of the Bloch solutions are the parameters
indexing the family of the periodic potentials with the fixed
spectrum. Finally, we note that the potential $V$ satisfies the
relation
\begin{equation}\label{GLD-2}
  V^2(x)-V''(x)=-8\left(\lambda_1^2(\xi)+\lambda_1(\xi)\lambda_2(\xi)
    +\lambda_2^2(\xi)\right)+ c_1\,(\lambda_1(\xi)+\lambda_2(\xi))+c_2.
\end{equation}
where $c_1$ and $c_2$ are constants depending only on 
$(E_j)_{1\leq j\leq 5}$.\\
In the sequel, we assume that
\begin{equation}\label{1-per}
\int_{0}^{\Xi} (\lambda_2(\xi)-\lambda_1(\xi))\,d\xi=1,\quad \Xi_1=2\Xi_2=\Xi,
\end{equation} 
i.e., that $V$ is 1-periodic two gap potential whose the first two
gaps are open.  Note that the conditions~\eqref{1-per} are satisfied
for a suitable choice of the points
$(E_j)_{j=1,\dots 5}$.\\
We complete this discussion with the formulae expressing $C_1$ and
$C_2$ in terms of $\lambda_1$ and $\lambda_2$.
Formulae~\eqref{C0C1},~\eqref{GLD-1},~\eqref{GLD-2} and~\eqref{x-xi}
imply that
\begin{equation*}
  C_1({\mathcal P_1},{\mathcal P_2})=3i F(P_1,P_2) +c_3 G(P_1,P_2)
  \quad\text{and}\quad
  C_2({\mathcal P_1},{\mathcal P_2})=iG(P_1,P_2)
\end{equation*}
where $P_j=\pi_c({\mathcal P_j})$ for $j=\{1,2\}$, $c_3$ is a constant
depending only on $(E_j)_{1\leq j\leq 5}$, and
\begin{equation}
  \label{AB}
  \begin{split}
    F(P_1,P_2)&=\int_0^{\Xi}d\xi(\lambda_2-\lambda_1)(\xi)\int_0^\xi
    d\xi'(\lambda_2^3-\lambda_1^3)(\xi')-
    \frac{1}2 \int_0^{\Xi}d\xi(\lambda_2^3-\lambda_1^3)(\xi),\\
    G(P_1,P_2)&=\int_0^{\Xi}d\xi(\lambda_2-\lambda_1)(\xi)\int_0^\xi
    d\xi'(\lambda_2^2-\lambda_1^2)(\xi')- \frac{1}2
    \int_0^{\Xi}d\xi(\lambda_2^2-\lambda_1^2)(\xi).
  \end{split}
\end{equation}
%
\begin{floatingfigure}[r]{5cm}
  \centering
  \includegraphics[bbllx=71,bblly=590,bburx=219,bbury=735,width=5cm]{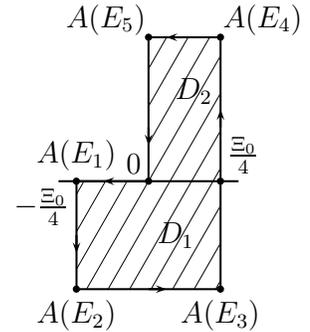}
  \caption{The domain $D$}\label{Abel-map}
\end{floatingfigure}
%
\noindent{\bf 2.} 
To study $\lambda_1$ and $\lambda_2$ for complex $P_1$ and $P_2$,
in~(\ref{D}), we consider the branch of $\sqrt{-R(u)}$ analytic in
$\C_+$ fixed by the condition $\sqrt{-R(u)}<0$ for $u<E_1$.  We
introduce the mapping $A:\C_+\mapsto \C$ defined by the formula
\begin{equation*}
   A(\lambda)=\int_{\infty}^{\lambda}\frac{du}{2\sqrt{-R(u)}},
\end{equation*}
and study its properties.\\
The Christoffel-Schwarz formula (see~\cite{MR819482}) shows that $A$
conformally maps $\C_+$ onto $D$, the domain shown on
Fig.~\ref{Abel-map}. In the same figure, we have shown how
$A(\lambda)$ goes along the boundary of $D$ as $\lambda$ moves along
the real axis from $-\infty$ to $+\infty$. Note that
\begin{equation}
  \label{A:D}
  \hskip-3.5cm\re A(E_1)=\re A(E_2)=-\Xi/4,\quad \re A(E_3)=\re A(E_4)=\Xi/4. 
\end{equation}
\vskip.1cm\noindent Consider the inverse of $A$ on $D$ i.e. the
function $\xi\in D\mapsto\lambda(\xi)\in\C_+$ defined by the relation
$A(\lambda(\xi))=\xi$. It takes real values on the boundary of $D$.
Therefore, using the Riemann-Schwartz symmetry principle, one can
extend it analytically to the strips $S_2=\{0<\im\xi<2\im(A(E_4))\}$
and $S_1=\{2\im(A(E_3))<\im\xi<0\}$.  In result, we obtain a function
satisfying the relations
\begin{gather}\label{lambda:per}
  \lambda(\xi+\Xi/2)=\lambda(\xi)\quad\text{for}\quad \xi\in S_2,\\
  \lambda(\xi+\Xi)=\lambda(\xi)\quad\text{for}\quad
  \lambda(\xi+\Xi/2)=\overline{\lambda(-\overline{\xi})}
  \quad\text{for}\quad\xi\in S_1.
\end{gather}
Discuss the singularities of $\lambda$. Let $\tilde D$ be the
rectangle $\{-\Xi/4\le\re\xi\le\Xi/4, \ \im A(E_3)\le\im\xi\le\im
A(E_4)\}$ cut along the real axis from $-\Xi/4$ to zero. For $\xi \in
\tilde D$, $\lambda$ satisfies the uniform estimates
\begin{equation}
  \label{lambda:rep}
  |\lambda(\xi)+e^{-2\pi i/3}(3\xi)^{-2/3}|\le C\quad\text{and}\quad
  |\lambda'(\xi)+2e^{-5\pi i/3}(3\xi)^{-5/3}|\le C |\xi|^{-1}.
\end{equation}
Here, the branches of the functions $\xi\mapsto(3\xi)^{-2/3}$ and 
$\xi\mapsto (3 \xi)^{-5/3}$ are chosen so that they are both positive  when
$\xi\in \R_+$.\\ 
\noindent{\bf 3.}  For $j\in\{1,2\}$, the solution
$\xi\mapsto\lambda_j(\xi)$ to~\eqref{D} is given by the formulae:
\begin{equation}
  \label{lambdas}
  \lambda_j(\xi)=\lambda(\xi+\xi_j)\quad\text{where}\quad \xi_j=A(P_j).
\end{equation}
From now on, instead of $P_j$, we shall use the parameter $\xi_j$. 
From~\eqref{lambdas}, it is clear, that
$\xi\mapsto\lambda_j(\xi)$ is analytic in the strip $S_j-\xi_j$.\\
\noindent{\bf 4.} Now, we consider the functions $(\xi_1,\xi_2)\mapsto F(\xi_1,\xi_2)$ 
and $(\xi_1,\xi_2)\mapsto G(\xi_1,\xi_2)$. To complete the proof of Lemma~\ref{derivatives},
it suffices to check that, for sufficiently small positive $\delta$,
their partial derivatives satisfy the estimates
\begin{equation}
  \label{part-der}
  |\partial^2_{\xi_1\,\xi_2}F(-\delta,+\delta)-F_0 \delta^{-8/3}|\le C\delta^{-2}
  \quad\text{and}\quad
  |\partial^2_{\xi_1\,\xi_2}G(-\delta,+\delta)|\le C \delta^{-2},
\end{equation}
where $F_0\ne0$ and $C>0$ are independent of $\delta$.\\
Prove the first of these estimates. Representation~\eqref{AB}
and~\eqref{lambdas} and relations~\eqref{lambda:per}
and~(\ref{lambda:rep}) imply that
\begin{equation*}
\begin{split}
  \partial^2_{\xi_1\,\xi_2}&F(-\delta,\delta)
  =-\int_0^{\Xi}(\lambda_2'\lambda_1^3+\lambda_1'\lambda_2^3)d\xi\\
&=-\int_{-\Xi/4}^{\Xi/4}\left[\lambda_2'(\xi)(\lambda_1^3(\xi)+\lambda_1^3(\xi+\Xi/2))+
(\lambda_1'(\xi)+\lambda_1'(\xi+\Xi/2))\lambda_2^3(\xi)\right]d\xi\\
&=-\int_{-\Xi/4}^{\Xi/4}\left[\lambda'(\xi+i\delta)(\lambda^3(\xi-i\delta)+
\overline{\lambda^3(-\xi-i\delta)})+
(\lambda'(\xi-i\delta)-\overline{\lambda'(-\xi-i\delta)})\lambda^3(\xi+i\delta)\right]d\xi.
\end{split}
\end{equation*}
Then, by means of the estimates~\eqref{lambda:rep}, 
we get
\begin{gather*}
  \partial^2_{\xi_1\,\xi_2}F(-\delta,+\delta)=F_0\, \delta^{-8/3}+O(\delta^{-2}),\\
F_0=-2\cdot 3^{-11/3}\,\int_{-\infty}^\infty
\left(2e^{\pi i/3}(t+i)^{-5/3}(t-i)^{-2}-e^{-\pi i/3}(t-i)^{-5/3}(t+i)^{-2}\right)dt,
\end{gather*}
where the branches of the functions $t\mapsto (t\pm i)^{-5/3}$ are
fixed by the condition $\arg((t\pm i)^{-5/3})\to0$ as
$t\to+\infty$. The constant $F_0$ can be computed by means of the
Residue theorem. This gives $F_0=-5i\pi 3^{-11/3} 2^{-2/3}$ which completes the proof
of the first estimate in~\eqref{part-der}. The second one is proved similarly.
This completes the proof of Lemma~\ref{derivatives}.
\qed


%
%
\def\cprime{$'$} 
\def\cydot{\leavevmode\raise.4ex\hbox{.}}

\end{document}